%% file: 1_Manuscript.tex
\documentclass[12pt]{article}
\usepackage[utf8]{inputenc}
\usepackage[fleqn]{amsmath}
\usepackage{amssymb}
\usepackage{graphicx}
\usepackage{geometry}
\usepackage{subcaption}
\usepackage{setspace}
\usepackage{rotating}
\usepackage{pdflscape}
\usepackage{xcolor}
\usepackage{adjustbox}
\usepackage[toc,page]{appendix}
\usepackage{multirow} 
\usepackage{tabularx}
\usepackage{booktabs}
\usepackage{array} 
\usepackage{ragged2e}
\usepackage{longtable}
\usepackage{float}

\usepackage{comment}

\usepackage{natbib}  

\usepackage[colorlinks=true, linkcolor=blue, citecolor=blue, urlcolor=blue]{hyperref}
\usepackage{bookmark} 

\title{\large \textbf{\textit{SUPERSTARS OR SUPER-VILLAINS?} \\ PRODUCTIVITY SPILLOVERS AND FIRM DYNAMICS IN INDONESIA\footnote{\noindent  \textbf{Acknowledgment}: This research has been supported by the Indonesian Endowment Fund for Education (LPDP), which provides the scholarship that enabled the author to pursue PhD program and complete this paper. I am also deeply grateful to my supervisors, Prof. Sai Ding and Dr. Hisayuki Yoshimoto, for their invaluable guidance and support. The author declares no financial relationships or other potential conflicts of interest related to this research.}\\}}
\author{Mohammad Zeqi Yasin \\ \footnotesize Department of Economics, Adam Smith Business School, University of Glasgow \\ \footnotesize Email: \texttt{m.yasin.1@research.gla.ac.uk} \\
\href{https://drive.google.com/drive/folders/1dpwvgxmTH98CeGLyRTfY0FA9HazC8pY1?usp=sharing}{Updated Version}}

\date{\today}

\begin{document}

\maketitle
\begin{singlespace}
\begin{abstract}
\noindent Do industrial ``superstars" help others up or crowd them out? We examine the relationship between the spillovers of superstar firms (those with the top market share in their industry) and the productivity dynamics in Indonesia. Employing data on Indonesian manufacturing firms from 2001 to 2015, we find that superstar exposures in the market raise both the productivity level and the growth of non-superstar firms through horizontal (within a sector-province) and vertical (across sectors) channels. When we distinguish by ownership, foreign superstars consistently encourage productivity except through the horizontal channel. In contrast, domestic superstars generate positive spillovers through both horizontal and vertical linkages, indicating that foreign firms do not solely drive positive externalities. Furthermore, despite overall productivity
growth being positive in 2001-2015, the source of negative growth is
mainly driven by within-group reallocation--evidence of misallocation among surviving firms--notably by domestic superstars. Although Indonesian superstar firms are more efficient in their operations, their relatively modest growth rates suggest a potential stagnation, which can be plausibly attributed to limited innovation activity or a slow pace of adopting new technologies.  
\end{abstract}
\end{singlespace}
\newpage

\section{Introduction}
The declining share of labour income since the 1980s has attracted significant concern from economists. Both developed and developing economies experiencing this decrease have devoted attention to identifying its causes, with a primary focus on the trade-off between physical capital and labour in the production process \citep{barkai2020declining}. Instances of these trade-off causes include the mechanism of lower relative cost of capital compared to labour and technological change \citep{firooz2025automation},  globalization and trade liberalization (\citet{xu2018made}, \citet{damiani2020labour}, \citet{perugini2017globalisation}, \citet{leblebiciouglu2021openness}), as well as the recent hypothesis, the reallocation process to the most productive firms, known as superstar firms \citep{autor2020fall}. Among these causes of labour share decline, the rise of superstar firms might attract the most attention as it integrates other causes, such as how superstar firms can optimally utilize the technological revolution and cause a reallocation among firms in the open economy (\citet{aghion2023creative}, \citet{aghion2019innovation}). 

In the literature, most studies discussing superstar firms focus on advanced economies, such as the US (\citet{ciliberto2021superstar}, \citet{cheng2024exposure}, \citet{ayyagari2024rise}, \citet{firooz2025automation}), Belgium (\citet{amiti2024fdi}, \citet{abraham2020impact}), and European countries \citep{bormans2023productivity}, where the adoption of advanced technologies and automation is well-documented in the data. Moreover, in developed countries, such as the US, industries have become more concentrated in firms with better performance, i.e., superstar firms \citep{autor2020fall}. Since superstar firms mainly utilize high technology in production, such as robots, it has led to a rise in automation \citep{firooz2025automation}. Although the rise of automation through robot utilization requires more skilled workers \citep{firooz2025automation}, it may also lead to a decrease in unskilled labor, which is evidently abundant in many developing countries \citep{Amiti2007c}. This question leaves a significant gap in understanding how superstar firms operate and exert influence in developing economies.

Scrutinizing the relationship between superstar firms and productivity is essential for developing countries. Superstar firms may have direct effects on the economy by generating demand in their supply chains, creating new markets, and offering surpluses for workers' benefits \citep{Ciani2020}. Moreover, superstar firms may also make indirect effects through productivity and knowledge spillovers stemming from intensive innovation, agglomeration, and extending linkages with their trade partners as they connect both domestically and internationally \citep{di2017large}. Evidently, in 2005, superstar firms contributed approximately 1.4\% in Vietnam, 1.5\% in Indonesia, and 0.8\% in Morocco, compared with small and medium-sized enterprises (SMEs), which showed a negative contribution \citep{Ciani2020}. Meanwhile, in Indonesian economy, which is supported by more than 61\% of small enterprises \citep{widita2024spatial}, superstar firms serve as trailblazers for government policy \footnote{For instance, in early 1990s, the Indonesian government introduced the Foster Father (\textit{Bapak Angkat}) program, which encouraged superstar firms to cooperate with smaller firms in providing parts and components in supply chain. Korea also implemented this strategy in the 1970s for heavy and chemical industries; see \citet{pyo2018there}. The program was also implemented simultaneously with Indonesian Regulation No. 176/2009 on Import Duty Exemption on Imported Machinery, which led superstar firms to establish partnerships with domestic firms, as they obtained import duty reductions, provided that at least 30\% of the machinery they utilized was sourced domestically. These rules had been in effect since 2000 under the Decree of the Minister of Finance number 135 of 2000 on Import Duty Relief on the Importation of Machinery, Goods and Materials.}. Moreover, empirical evidence from \citet{cali2025robots} also revealed that there was a significant increase in robot adoption in the Indonesian industry in the early 2010s\footnote{Comtrade data reported that robot imports in 2015 reached more than 50\% in Indonesia.}. It also implies the potential development of superstar firms in Indonesia, consistent with the empirical arguments of \citet{firooz2025automation}, which suggest that superstar firms utilize automation more extensively than non-superstar firms. According to this evidence, it is worthwhile to scrutinize how superstar firms impact productivity in the economy.

This study aims to fill the gap in the nexus between superstar spillovers and productivity in developing economies by examining manufacturing firms in Indonesia using panel data from 2001 to 2015. Indonesia deserves more attention to the issue of superstar firms due to several reasons. First, in terms of macroeconomic evidence, Indonesia has experienced a prolonged episode of economic fluctuation since the 1980s, with 1998 being the most severe period due to the Asian Financial Crisis. In the post-1998 period, the Indonesian manufacturing sector remains the most significant driver of the economy, despite its declining trend over the years\footnote{See Figure \ref{Share Manufacture} in the Appendix.}. However, although the contribution toward Gross Domestic Product (GDP) declines, in terms of productivity, the Indonesian manufacturing sector shows a rebounding trend in the early 2000s after the Asian Financial Crisis in 1998\footnote{See Figure \ref{TFP Macro} in the Appendix.}. It is in line with labour productivity growth relative to 2001 between superstar firms and non-superstar firms, where superstar firms have higher productivity growth from the early 2000s until 2006\footnote{See Figure \ref{Labour Productivity} in the Appendix.}. However, non-superstar firms then caught up and outperformed superstar firms until 2015. This evidence is interesting in showing how the catching-up process occurred.

Another piece of evidence is the dominance of superstar firms in market share. Data from BPS-Statistics of Indonesia from 2001 to 2015 report that superstar firms dominate more than 76\%\footnote{See Figure \ref{Aggregate Share} in the Appendix.}. This evidence suggests that the Indonesian manufacturing industry exhibits significant market concentration, primarily driven by superstar firms. Furthermore, the rise of superstar firms in Indonesia can also be observed in the increasing trend of automation \citep{firooz2025automation}. In the Indonesian case, there has been a rising trend in automation since 2001, dominated by superstar firms, while non-superstar firms have increased their automation level and converged to the level of superstar firms\footnote{See Figure \ref{automation_general} in the Appendix. This stylized fact also supports the data from \citet{cali2025robots}, which shows a significant increase in robot imports in Indonesian manufacturing firms after 2008.}. Moreover, the dynamics of superstar firms may also be demonstrated through the association between automation and market concentration \citep{autor2020fall}. The hypothesis of \textit{``The Rise of Superstar Firms''} in the US from \citet{autor2020fall} is indicated by the decline of labour share, implying a substitution mechanism through automation. In the Indonesian case, the correlation also reveals similar behavior, in which there is a positive correlation between automation and market share in CR20 (the top 20 firms for each subsector), implying that firms with higher automation are associated with higher market share\footnote{See Figure \ref{automation and share nexus} in the Appendix.}.

The Indonesian economy initiated market liberalization in the 1960s, which significantly influenced inward foreign direct investment (FDI), with inward FDI growth exceeding 50\% between 2005 and 2007 \citep{uttama2010foreign}. A major reformation of FDI policy occurred in 1966 under the New Order regime of President Soeharto \citep{Genthner2022}. This reform led to massive structural economic changes compared to the preceding Old Order regime, including deregulation policies in the banking sector, trade, investment, and capital markets, which culminated in Indonesia's membership in the World Trade Organization (WTO) in January 1995. Under the New Order regime, FDI inflows were encouraged, leading to the establishment of production facilities, primarily in the form of multinational corporations (MNCs), to drive high-technology intensification for local firms. However, as suggested by \citet{amiti2024fdi}, FDI in the form of MNCs may not be the only or even the most effective mechanism for generating positive economic spillovers. Their findings suggest that large domestic firms generate similar spillovers while maintaining stronger local linkages. Hence, instead of prioritizing subsidies and incentives for MNCs, there is an essential need to foster a level playing field for all large firms, including domestic superstar firms. In this regard, rather than focusing on FDI spillovers in Indonesia, on which many studies have devoted attention, there is a need to examine how superstar firms, not only MNCs, create spillovers for the Indonesian economy.

In this study, two strategies are employed to examine the relationship between superstar spillovers and firms' productivity. The first strategy is to examine the correlation between the exposure of superstar firms and their productivity. A higher exposure of superstar firms within the sector and region may cause spillovers through the mechanism of demonstration, labour migration, and competition \citep{orlic2018cross}. The demonstration channel takes place through observation and mimicking, by which non-superstar firms aim to copy technology or production management from superstar firms. Labour mobility occurs when non-superstar firms hire workers who have migrated from superstar firms, while the channel of competition occurs when the presence of superstar firms stimulates non-superstar firms to compete in order to become more efficient.

The second strategy is to look at superstar firms' contribution to aggregate productivity, considering firm dynamics in terms of entry and exit behaviour. The idea is to assess whether productivity changes stem solely from within-firm productivity improvements, or are also supported by the entry-exit behaviour of firms. Evidently, data from BPS-Statistics Indonesia from 2001-2015 show that the entry-exit rate of manufacturing firms reached about 8\%, which is essential in determining the productivity composition of the country. Hence, this study decomposes the extent to which superstar firms contribute to aggregate productivity over time, complementing the findings from the correlation test in the first strategy. The standard decomposition from Olley-Pakes \citep{olley1996dynamics} has been extended by \citet{melitz2015dynamic} and \citet{collard2015reallocation} by incorporating the entry and exit of firms. This study extends this strategy to incorporate superstar firms' dynamics.

This study contributes to the literature in three main ways. First, it re-evaluates the literature on productivity spillovers that are mainly claimed to come from foreign firms. In the early 2000s, several studies explored firms with characteristics resembling superstar firms and their spillover effects on productivity. For instance, incoming foreign direct investment (FDI) in the form of multinational companies (MNCs), assumed to be superior to domestic companies \citep{helpman2004export}, was often associated with increased productivity among local firms through spillover mechanisms (see \citet{javorcik2004does}, \citet{Suyanto2009}, and \citet{Sari2016}). These findings are mainly relevant for developing countries with inward FDI's share on world total FDI increasing from 16.69\% in 1990 to 55.47\% in 2014 \citep{sahu2021does}. Based on these statistics, prior studies claimed productivity spillovers from FDI for the domestic economy.

However, whether these spillovers truly come from MNCs needs further investigation. Recent findings from \citet{amiti2024fdi} argue that productivity spillovers from MNCs should be re-assessed, as it is not obvious whether productivity spillovers are truly anchored by MNCs or, additionally, by other types of firms such as large or exporting firms. This hypothesis is plausible, as domestic firms with characteristics such as being large and export-oriented can also create productivity spillovers for their counterparts, since they too are engaged in global value chains like MNCs. Likewise, a finding from \citet{herzer2018long} reveals that FDI might cause negative spillovers on productivity in the long term, implying that a country cannot rely solely on MNCs to stimulate the economy. In this sense, it is essential to redefine which firms' characteristics cause productivity spillovers, namely through superstar firms, which depict more relevant characteristics of top firms, since they are not necessarily MNCs.

Our second contribution stems from the notion of how firm-level productivity dynamics may actually arise from production changes in other firms--particularly those with more intensive adoption of automation, namely superstar firms. We seek to reinforce the findings of \citet{cali2025robots}, which show that superstar firms, empirically found to adopt automation very intensively, can generate spillovers to non-superstar firms. \citet{cali2025robots}'s findings indicate that the net effect of automation in advanced economies is negative, as the potential productivity gains from automation have already been largely exhausted. However, in developing countries such as Indonesia, the effect of automation may be greater, since many factories have not yet maximized the use of robots in their production processes. In this regard, consistent with \citet{firooz2025automation}--who highlights the intensive use of robots by superstar firms--this paper attempts to capture the spillover effects of such firms. Although this study does not directly examine the effect of robot importation by superstar firms due to limited data availability, it does show that productivity improvements in non-superstar firms can occur via spillovers from superstar firms that, as prior evidence demonstrates, have already imported a large number of robots. 

The third contribution is the perspectives by which superstar firms are addressed. We define our definition of superstar firms, which are more fundamental and relevant for the evidence of Indonesia as a developing country. Prior studies have some limitations in classifying superstar firms. For instance, \citet{autor2020fall} defined superstar firms as top-500 firms in the industry, which is arbitrary in the context of Indonesia. Since the number of firms in Indonesia for each year is dynamically changed, it is not possible to employ the quantitative strategy, as suggested by \citet{firooz2025automation}. Likewise, the definition of superstar firms from \citet{amiti2024fdi}, taken from whether firms are either foreign-owned, exporters, or large firms, results in a large number of firms being called superstar, which is no longer relevant with the term ``super'' in this context. This strategy is also similar to the FDI spillovers literature. Conversely, the study of \citet{rowley2024domestic} defined superstar firms too strictly by using the top 3 domestic and foreign firms in terms of export. This strategy is also difficult to capture the spillover effect across different horizons\footnote {For comparison, we report the number of superstar firms from this literature for the context of Indonesia in the Appendix.}. 

The findings of this study are as follows. First, we capture positive spillovers from both horizontal and vertical channel on TFP level and growth, implying a higher share of superstar firms within province and subsector causes non-superstar firms more productive. The positive relationship of superstar spillovers in horizontal direction implies that horizontal linkages on the TFP may stimulate competition from the outset, putting pressure on non-superstar firms and leading to lower productivity; firms unable to survive this pressure may exit the market, leaving only those with a certain level of productivity. Likewise, we found a positive association between vertical channels, both backward spillovers (from whom superstar firms purchase) and forward spillovers (to whom superstar firms sell), on the TFP level and growth. It implies that acting as suppliers to superstar firms or purchasing intermediate inputs from superstar firms improves the productivity growth of non-superstar firms. 

In terms of heterogeneity of the superstar firms by distinguishing them into foreign and domestic superstars, we found that foreign superstars consistently show positive effects on non-superstar productivity, except for horizontal channels, showing an insignificant effect. A plausible reason is the technology protection provided by their parent company, which hinders spillovers from demonstration and imitation channels. However, domestic superstar shows positive results for entire channels, implying that domestic superstars generate spillovers more than foreign superstars do. These findings are robust across different specification strategies, such as medium- to large-sized samples, alternative instrumental variables, and different alternative productivity measurements. 

Moreover, the results from the decomposition strategy show that although overall TFP growth is positive in 2001-2015, the source of negative aggregate productivity growth is mainly driven by within-group reallocation, which implies misallocation within survivors in the markets. Interestingly, the negative growth of the survivor group is mainly driven by domestic superstar firms. More specifically, superstar firms in Indonesia exhibit slower growth compared to their non-superstar counterparts. Accordingly, this finding contrasts with the ``Rise of Superstar Firms" hypothesis commonly observed in advanced economies, as in \citet{autor2020fall} and \citet{amiti2024fdi}. Although Indonesian superstar firms are more efficient in their operations, their relatively modest growth rates point to a potential stagnation. Such stagnation could be attributed to limited innovation activity or a slow pace in adopting new technologies. While productivity spillovers to non-superstar firms are evident, they may largely stem from the existing wide TFP gap between superstar firms and their competing, supplying, and purchasing partners. 

\subsection{Related Literature}
Our study is related to several studies on superstar firms such as \citet{autor2020fall}, \citet{amiti2024fdi}, \citet{firooz2025automation}, \citet{ciliberto2021superstar}, \citet{cheng2024exposure}, and \citet{rowley2024domestic}. Study of \citet{autor2020fall} aims to investigate the cause of labour share declining behaviour in the US and show that the industry with the highest increase in concentration shows the largest decline in labour share. This decline stems mainly from between-firm reallocation, which occurred in most firms with the largest sales concentration. \citet{amiti2024fdi} aims to reveal whether by having networking with superstar firms will increase firm’s productivity in Belgium in 2002-2014. The findings show that firms are embarking on a strong relationship with superstar firms. Moreover, the relation with other types of superstars, large firms or exporters, also stimulates TFP growth for non-superstar firms, implying that a firm does not necessarily need to channel with foreign-owned superstar firms to stimulate TFP\footnote{Similar design of firm-to-firm behaviour to capture superstar spillovers is also shown in \citet{alfaro2022effects} for Costa Rica evidence, while \citet{li2024fdi} captures the supplier-customer network from firms-to-countries evidence.}.

In terms of exporter superstar, the study of \citet{ciliberto2021superstar} shows that the strong competitive effect significantly encourages a firm's decision to export. Without this strong competitive effect, superstar firms are more likely to export by 53.2\%, implying that competition discourages benefits and export participation of the firm. Meanwhile, \citet{rowley2024domestic} compared the different granularities of foreign and domestic exporter superstars and revealed that foreign and domestic superstar granularity shows a positive impact on market concentration and depends on the knowledge gap between superstars and non-superstars. If the gap is absent, the granularity effect might be weaker. Another issue of superstars is connected with the robot development, as in \citet{firooz2025automation}. \citet{firooz2025automation} shows that the declining price of robots and automatic machines leads to better accessibility of firms in the US, which in turn improves their labour productivity but raises industry concentration. \citet{firooz2025automation} defined superstar firms as firms in the top 1\% in terms of sales and employment share and suggested a proportional level of robot subsidy to mitigate markup distortions and improve welfare by stimulating automation investment.

The following discusses our model setup, while Section 3 outlines the data and methodology. Section 4 presents our findings and discussion. Section 5 concludes the study and presents some policy implications.

\section{The Model Setup}
The theoretical model of our study embraces several prior seminal papers on heterogeneous firm models, such as \citet{brock1972models}, \citet{smith1974optimal}, \citet{hopenhayn1992entry}, \citet{melitz2003impact}, and \citet{melitz2008market}\footnote{The tractability of this theoretical model is explained in the Appendix.}. The notion of a heterogeneous firm model with spillovers was proposed in a strand of literature, such as from \citet{anwar2019firm}, who postulated the idea for FDI spillovers based on the cutoff criteria as in \citet{melitz2003impact}. However, the model introduced by \citet{anwar2019firm} generalized the behaviour of spillovers. Hence, we introduce the heterogeneous model with superstar firms by adjusting the generalized FDI spillovers model of \citet{anwar2019firm}. The model is based on the constant elasticity substitution as follows:
\begin{equation}
    q=\Theta p^{1/{\rho-1}}
\label{CES}
\end{equation}
where $q$ is the total demands, $p$ denotes the price, $\Theta$ represents a given aggregate demand, and $\rho$ is the parameter of the consumer utility function where $
\rho \in (0,1)$. In this model, when a non-superstar firm enters the market, it possesses and explores its capability ($\lambda$) with a continuous distribution in the density function $g(\lambda)$ where $\lambda \in (0,+\infty)$. 

The composition of cost can be decomposed from the production function as follows:
\begin{equation}
    q=\varphi.L
\label{ProductionFunctionCost}
\end{equation}
where $q$ is total outputs, $\varphi$ is average productivity of input $L$, and $L$ is an index of inputs. The nexus of left-side and right-side in equation \ref{ProductionFunctionCost} is that the total output is not only determined by the level of inputs $L$, but also by the average productivity of inputs $L$. From equation \ref{ProductionFunctionCost}, we may obtain the information on the inputs utilization from $L=\frac{q}{\varphi}$. We obtain the information on variable cost by multiplying it by the cost of inputs $L$, $w$. Hence, we obtain the total cost composition as follows:

\begin{equation}
    TC=\underbrace{f}_{\text{fixed cost}}+\underbrace{w\frac{q}{\varphi}}_{\text{variable cost}}
\label{TotalCost}
\end{equation}
where $TC$ denotes total cost, $f$ is the fixed cost incurred by each domestic firm, and variable cost consisting of $w$ as the price of input $L$, $q$ as the total outputs, and $\varphi$ as the firm's productivity. By following \citet{anwar2019firm} who endogenized $\varphi$ to capture spillover behaviour, we arrange: 
\begin{equation}
    \varphi=\lambda e^{\alpha \gamma} c(Z)
\label{ProductivityFormulaS}
\end{equation}
Where $\varphi$ is the productivity, $\alpha$ and $\gamma$ represent the productivity spillovers from superstar firms and the degree to which superstar firms create spillovers (mostly the share of sales), respectively, and $Z$ is a vector of firm characteristics as control. In equation \ref{ProductivityFormulaS}, the value of $\alpha$ captures whether the spillovers are positive ($\alpha>0)$), negative ($\alpha<0)$), or have no effects ($\alpha=0)$), for non-superstar firms. In this regard, the productivity of non-superstar firms is determined by three components: capability level $\lambda$, the exponential value of spillovers $\alpha$, and the share of superstar firms $\gamma$, as well as the firm's characteristics. 

We can also distinguish spillovers into some channels, namely horizontal and vertical (backward and forward) direction. Horizontal spillovers occur within a subsector involving competition and demonstration, such as mimicking. Meanwhile, vertical spillovers occur across subsectors through the networking with suppliers (backwards) and purchasers (forward). Equation \ref{ProductivityFormulaS} can then be extended into:

\begin{equation}
    \varphi=\lambda e^{\alpha HSpill}e^{\tau BSpill}e^{\psi FSpill} c(Z)
\label{ProductivityFormulaS_Extended}
\end{equation}

We can then arrange the profit function as follows:
\begin{equation}
    \pi=p.q-(f+w\frac{q}{\varphi})
    \label{ProfitFunction}
\end{equation}
Where $\pi$ is the profit. Under a monopolistic market, the first order condition (FOC) of equation \ref{ProfitFunction} requires marginal revenue to equal marginal cost. We re-arrange equation \ref{CES} into: $q=\Theta p^{1/{\rho-1}} $ and $p=\frac{q^{\rho-1}}{\Theta^{\rho-1}}$, we then obtain:
\begin{equation}
TR = \frac{q^\rho}{\Theta^{\rho-1}}
\end{equation}

Hence, we obtain Marginal Revenue (MR)$=\rho p$

Meanwhile, using the function from cost in equation \ref{TotalCost}, we obtain Marginal Cost (MC)$=\frac{w}{s}$. Hence, we obtain the pricing rules under $MR=MC$ as:

\begin{equation}
        p=\frac{MC}{\rho}=\frac{w}{\rho s}
    \label{FOCMRMC}
\end{equation}

We then apply the pricing rule in equation \ref{FOCMRMC}, demand function in \ref{CES}, and average productivity of inputs in equation \ref{ProductivityFormulaS} for optimal profit ($\pi^*$) function in equation \ref{ProfitFunction} as follow\footnote{The optimization process is explained in the appendix}:

\begin{equation}
\begin{split}
 \pi^*=(1-\rho)w^{\frac{\rho}{\rho-1}}\rho^{\frac{\rho}{1-\rho}}\lambda^{\frac{\rho}{1-\rho}} e^{\frac{\alpha HSpill\rho}{1-\rho}}e^{\frac{\tau BSpill\rho}{1-\rho}}e^{\frac{\psi FSpill\rho}{1-\rho}} c^{\rho/1-\rho}\Theta-f
\end{split}
\label{ProfitOptimalFunction}
\end{equation}

According to equation \ref{ProfitOptimalFunction}, we identify that $\pi^*$ is a monotonically increasing function of the firm capability ($\lambda$). As firm capability converges to zero, profit will converge to $-f$. In this regard, there is a need for a cut-off of firm capability to prevent firms from obtaining negative profit. Hence, in the condition of non-negative profit non-superstar firms ($\pi=0$), we may obtain the optimal firm capability cut-off ($\lambda^*$) as follows:
\begin{equation}
\begin{split}
\lambda^*=f^{\frac{1-\rho}{\rho}}(1-\rho)^{\frac{\rho}{1-\rho}}w^{-1}e^{-\alpha HSpill}e^{-\tau BSpill}e^{-\psi FSpill}c\Theta^{\frac{\rho}{1-\rho}}
\end{split}
\label{CutoffCapability}
\end{equation}

According to equation \ref{CutoffCapability}, the cutoff capability of the market depends on the degree of the superstar firm's exposure ($HSpill$, $BSpill$, and $FSpill$). If positive horizontal spillover occurs ($\alpha>0$), an increase of $HSpill$ leads to lower cut-off capability level, implying that the market is getting less strict in terms of productivity performance, which in turn enables non-superstar firms that were not originally capable for surviving in the industry to re-enter. It may cause aggregate productivity to decrease afterwards. Meanwhile, if negative spillovers occur ($\alpha<0$), a higher share of superstar firms in the industry leads to a higher level of cut-off capability, which causes crowding out effects for less productive firms. 

The channel of vertical spillovers is similar. If positive vertical spillovers occur (either $\tau >0$ or $\psi>0$), an increase of $BSpill$ or $FSpill$ (i.e. superstar firms purchase more intermediate inputs from other subsectors, or sells more to other subsectors), the lower the cutoff becomes because superstar firms may become increasingly dependent on other sectors (their product sales are dominated by sales to other sectors rather than to final consumers). This situation leads to less strict productivity requirements, so any firm can potentially become a supplier to superstar firms since superstar firms require more suppliers, while existing suppliers may not be able to meet superstar firms' demands. As a result, more firms can enter the market and may become suppliers or purchasers of superstar firms. It will then lower overall productivity because superstar firms still have to purchase intermediate goods from suppliers with lower productivity performance. 

Therefore, we obtain (logged of) expected productivity from equation \ref{ProductivityFormulaS_Extended} given the capability level above the cut-off as follows:
\begin{equation}
\begin{split}
        E(ln(s)|\lambda>\lambda^*)=\alpha HSpill +\tau BSpill+ \psi FSpill+ \text{ln c} + \frac{\int ln\lambda g(\lambda)d\lambda}{\int_{\lambda^*}^\infty g(\lambda)d\lambda}
\end{split}
\label{ExpectedProductivity}
\end{equation}

According to equation \ref{ExpectedProductivity}, there are two channels of superstar spillovers to occur. First, the direct effect from $\alpha$, $\tau$, and $\psi$, and the second channel is through the cutoff level $\lambda$. We can then estimate the marginal effects with respect to the superstar firm's share, as follows:

\begin{equation}
  \frac{\partial E(ln(s)|\lambda>\lambda^*)}{\partial Spill}=\underbrace{\alpha+\tau +\psi}_{\text{Direct Effect}} + \underbrace{\frac{\int ln\lambda g(\lambda)d\lambda}{[\int_{\lambda^*}^\infty g(\lambda)d\lambda]^2}g(\lambda^*)\frac{\partial \lambda^*}{\partial Spill}}_{\text{Indirect Effect}}
\label{DirectIndirectSpillovers}
\end{equation}

Where $Spill$ denotes $HSpill$, $BSpill$, or $FSpill$. According to equation \ref{DirectIndirectSpillovers}, there is a contradictory nexus between direct and indirect effects. If spillovers positive occurs ($\alpha>0$, $\tau>0$, $\psi>0$) the change of capability is negative ($\frac{\partial \lambda^*}{\partial HSpill}<0$, $\frac{\partial \lambda^*}{\partial BSpill}<0$, $\frac{\partial \lambda^*}{\partial FSpill}<0$), so indirect effect is negative, \textit{vice-versa}. The intuition of this opposite direction is that at the initial entry of superstar firms, the direct effect of spillover enables non-superstar firms--both survivors and those that have exited the market-- to identify benchmark firms that serve as a reference for their own production process. In this context, an increase in the share of superstar firms within the sector ($HSpill$) accompanied by the rise in productivity may indicate that non-superstar firms are directly responding through mechanisms such as demonstration, labour migration, and competitive pressure, as suggested by \citet{orlic2018cross}. Meanwhile, an increase in the share of superstar firms across sectors ($BSpill$ and $FSpill$) may indicate that non-superstar firms are directly responding by entering the market and becoming suppliers/purchasers. 

However, there also exists an indirect channel of productivity spillovers. When the superstar firm's share increases and generates positive direct effects, the resulting shift can lower the capability cutoff ($\lambda$), thereby enabling a greater number of firms to enter the market. It implies that the increase in superstar firm share indirectly contributes to the lowering of the entry threshold. In the presence of such potentially conflicting direct and indirect effects, the initial magnitude of spillovers $\alpha$, $\tau$, and $\psi$ becomes crucial, as it determines the extent of the shift in the cutoff and consequently the change in aggregate productivity. 

Positive spillovers may lead to a surge of less productive firms into the market, thereby reducing aggregate productivity due to a decline in the productivity threshold. Conversely, negative spillovers (rivalry effect and market stealing from superstar firms) may initially reduce productivity from non-superstar firms directly--possibly get them out of the market--but simultaneously raise the cutoff level, ensuring that only more capable firms survive, which could eventually increase aggregate productivity.

\section{Data and Methodology}
\subsection{Data}
This study utilizes the survey of Indonesian manufacturing firms from \textit{Statistik Industri}, henceforth referred to as \textit{SI}, published by BPS-Statistics from 2001 to 2015. The survey of SI was embarked in 1975 and selects firms with 20 or more workers \citep{Marquez-Ramos2020}. The survey is at the level of firm/establishment/plant, so SI will use the terms ``plant" and ``establishment" interchangeably to refer to a firm\footnote{Although some firms may have more than one factory, \citet{Marquez-Ramos2020} reports that less than 5\% of Indonesian factories belong to multi-factory firms. The BPS-Statistics, represented by the field agents, aims to increase the compliance rate by visiting each non-respondent \citep{Marquez-Ramos2020}. It also ensures that a firm may end its production activity. Nonetheless, the firm's response rate shows a decreasing trend over years.}. \citet{Marquez-Ramos2020} report that response rate was about 74\% in 2004, about 63\% in 2011, but it dropped into 47\% in 2017. For the years 2001-2015, the total observation reached 356,057 with a varying number of firms in each year. Specifically, 2003 has the smallest number of firms with 20,310 firms, and 2006 has the largest number of firms with 29,468 firms.  

The BPS-Statistics, known as Badan Pusat Statistik, provides the questionnaire for establishments to fill out by themselves. In general, a questionnaire in \textit{SI} consists of several basic information, such as firm identification code, the International Standard Industrial Classification of All Economic Activities (ISIC), and production value. It also covers different information depending on the year, such as the information on innovation and research activity that is available in 2011 but not in other years. Other information is also available, such as ownership (public, private, or foreign), export status, total assets, electricity utilization, fuel consumption, output, expenses, and labour. 

There are several adjustments to the industrial classification code in this data. The data of 2001-2005 follow the industrial classification code from The Indonesian Standard Industrial Classification (KBLI) 2000, the data of 2006-2009 refer to the KBLI 2005 and International Standard of Industrial Classification (ISIC) Rev. 3 (1990). Moreover, the years 2010-2015 refer to ISIC Rev. 4 (2008) and KBLI 2009. In this regard, the concordance is required to merge the datasets to ensure comparability of the firms. This study  refers to KBLI 2009 in analysing sectoral firm-panel behaviour in the datasets. Another important adjustment is that some firms are located in a province with a limited number of firms, causing a riskier condition to detect the firm. In this regard, BPS-Statistics merely uses 3-digit or 2-digit classification for these firms, while the concordance also needs wise adjustment. 

Another adjustment is the regional code, for which some regions (province, regency, district, or village) are merged and possess a new identification. In this regard, it is important to check the consistency of the code over the years. Likewise, some firms have no province code as there is a limited number of firms in that code. There is no alternative, unlike the 5-digits to be 3-digits, for this case. 

Most importantly, the information on total fixed assets in 2006 is not available. Some studies implement interpolating strategies, such as \citet{Amiti2007b}. In this study, we apply an interpolating strategy from \citet{Amiti2007b} for several variables, such as capital, labour, materials, and energy, by averaging the values from one year before and one year after (i.e., summing the values from the previous and subsequent year and dividing by two). In addition, a more advanced method was used specifically for the capital variable, as this variable is not available in 2006. This strategy involved estimating capital based on the lagged value of total output, labour, materials, and energy to obtain the relevant coefficients. The residuals were also predicted in this process.

Subsequently, the capital value was calculated by multiplying each coefficient by the lagged value of its corresponding variable (total output, labour, materials, and energy) and adding the residual. The inclusion of the residual component is necessary to avoid bias \citep{enders2010}. This strategy aims to estimate the degree of correlation between capital and the input/output variables, rather than capturing a causal relationship \citep{riandy2024dynamics}. Furthermore, following \citet{Sari2016}, we use lagged values for all variables on the right-hand side to ensure that the values from the previous period are associated with the current value of capital, thus enabling us to predict current capital.\footnote{For example, in 2006, all firms lacked data on fixed assets. Therefore, values from other periods are used to predict the missing value in 2006 using the estimated coefficients. More specifically, the values of total output, labour, materials, energy, and the residual in 2006 will be associated with the value of fixed assets in 2007, and so on.} The distribution of logged capital before and after interpolation is summarized in the Appendix in Figure \ref{Capital Before-After Interpol}.

Another data used in this study is the Input-Output Table of Indonesia in 2010. The Input-Output (I-O) Table aims to capture the relation among the sectors in the economy. In Indonesia, BPS-Statistics has released the I-O Table in 1971, 1975, 1980, 1985, 1990, 1995, 2000, 2005, and 2010. The I-O Table was released every 5 years according to the economic structure and technology utilization of the economic sectors at that time. The I-O table is then merged and adjusted based on the 3-digit ISIC from the \textit{SI} datasets. However, some subsectors are not classified elsewhere. In this regard, we merge these subsectors altogether; for instance, Computers (262) \& Accessories and Communication Equipment (263) are merged into Computers and Communication Equipment. Finally, we have 51 3-digit subsectors to analyze. 

We also use sectoral and regional datasets for instrumental variables. For sectoral datasets, we use tariff data from World Integrated Trade Solution (WITS) of World Bank in 2001-2015. The tariff data is taken from the 3-digit subsector in manufacturing, specifically for Most-Favoured Nation (MFN), with a simple average. Meanwhile, for regional datasets, we use road density measured from the ratio of length of road (country, province, and regency, in kilometer) to the size of the province (in kilometer squared), as suggested by \citet{Rodriguez-Pose2013a}\footnote{The data of length of road is only available for 2008-2015, but the size is available for 2001-2015 with some minor interpolation. Hence, we use the data of 2008 for the length of road for the years 2001-2007, assuming the length of road did not change before 2008, as suggested by \citet{Rodriguez-Pose2013a}}.

\subsection{Variables}
This study employs some variables, which are classified into two types: production function variables and productivity determinants variables. Production function variables consist of value-added, total workers, raw materials (intermediate inputs), and total fixed assets, and are used to estimate total factor productivity (TFP). We also include total (gross) outputs as an indicator to measure market share. Some variables are in monetary value, such as total outputs, capital, and raw materials. Hence, we deflate these variables using the Wholesale Price Index in 2-digit ISIC with the year 2000 as a base year. 

The second type is the determinants of TFP, consisting of spillover variables and control variables. The control variables are dummies for superstar firms, foreign-owned, and exporters. Dummy of superstar firms refer to whether a firm a superstar firm, based on the top 5\% total outputs share within 3-digits subsectors. The dummy of foreign-owned refers to the capital ownership of the firms. If a firm is owned by at least 10\% foreign ownership, it means that the firm is foreign-owned or multinational. This cutoff is also employed by prior studies such as \citet{Sari2016} and \citet{Suyanto2009}. The dummy variable for exporter refers to the condition whether a firm exports its outputs \footnote{In the \textit{SI}, there is information on the degree to which a firm exports its outputs. Still, this information has high missing values and is not available for the entire years of 2001-2015.}. Some ratio variables are also included, namely imported intensity (the ratio of imported materials to the total materials) and market concentration from the Herfindahl-Hirschman Index (HHI). We also include absorptive capacity, measured by the ratio of labour cost per worker.

\subsubsection{Superstar Firms Definition}
This study proposed an indicator of superstar firms, namely the share of total outputs produced (gross outputs), to be in the top 5\% in the three-digit ISIC, according to the stylized facts that this cutoff shows remarkable market share domination by more than 76\%. To ensure the invariant market share, we use the median of the share within the period of firms observed. Furthermore, a firm should have been in the market for more than 10 years. Moreover, some firms may not consistently appear in the top 5\% of the output share throughout the entire period of observation. For example, a firm that is observed from 2001 to 2005 may only be in the top 5\% in 2001 and 2002. In such cases, we define a firm as a superstar if it belongs to the top 5\% in more than 90\% of the years during which it is observed\footnote{Our findings are not sensitive to lower thresholds such as 75\%. Moreover, our results remain robust when we exclude firms that are in the top 5\% within sectors but do not last more than 10 years or are not in the top 5\% in at least 90\% of the observed years}. 

The definition of superstar firms in this study shall not use arbitrary number as in \citet{autor2020fall}, capture too many firms as in \citet{amiti2024fdi}, or capture too strict number of firms as in \citet{rowley2024domestic}\footnote{For the description of the number of superstar firms, we selected the top 5\% of all firms without distinguishing subsectors. However, in the correlation analysis, we selected the top 5\% from each subsector to ensure sufficient observation from a dynamic perspective, i.e., survivor-entry-exit. Consequently, the number of firms in the table remains as stated, but in the inferential analysis, the total number is necessarily higher.}. The criteria for identifying superstar firms from our definition, with 1\% and 5\% cutoffs, are deliberately stringent, resulting in a very limited selection of firms. Although the stricter approach ensures that the selected firms genuinely represent ``superstar" firms, rather than merely ``star" firms, the dynamic of entry and exit of superstar firms with a too strict cut-off, such as 1\%, might be extremely limited. Hence, the cutoff of 5\% is then used for further analysis.

\subsubsection{Superstar Spillovers and Its Economic Intuition}
There is a strand of literature discussing how the process of superior firms creates externalities for non-superior firms. Prior study from \citet{javorcik2004does} introduced the spillover process from superior foreign-owned firms, while more recent ones, such as \citet{amiti2024fdi}, introduced spillovers from superstar firms. In general, the process through which spillovers occurred from superior firms (either superstar firms or multinational firms) is similar, namely, how non-superior firms attempt to mimic them through various channels. The only difference of this mechanism is that superstar spillovers are not bound to foreign-owned firms, as in FDI spillovers, but domestic firms can also generate spillovers, as suggested by \citet{amiti2024fdi}. In the context of superstar and non-superstar firms, spillovers occur when the presence of superstar firms increases the productivity of non-superstar firms. Inward superstar firms may stimulate non-superstar firms by producing more efficiently. 

In this study, we examine three types of superstar spillovers. The first one is Horizontal Spillovers, capturing the degree to which superstar firms dominate the market, shown by the share of outputs produced within the three-digit ISIC, province, and year. 
    \begin{equation}
        HSpill_{kjt}=\frac{\sum_{i \in k \in j} DProvi_{j} \times DSubsector_{k} \times DSuperstar_{it}\times Outputs_{it}}{\sum_{i\in k \in j} Outputs_{it}}
        \label{HSpill}
    \end{equation}

Where $HSpill_{kjt}$ denotes Horizontal Spillovers of three-digits ISIC subsector $k$ in province $j$ in year $t$, $DProvi_{j}$ denotes dummy of province $j$, $DSubsector_{k}$ denotes dummy of subsector $k$, $DSuperstar_{it}$ denotes dummy of superstar firms, $Outputs_{it}$ is the total outputs produced of firm $i$ in year $t$. The numerator in Equation \ref{HSpill} captures the total outputs produced by superstar firms located in province $j$ and operating in subsector $k$ during year $t$. The dummy variables $DProvi_j$, $DSubsector_k$. and $DSuperstar_{it}$ ensure that only outputs from superstar firms in the specified province and subsector are counted. These values represent the production dominance of superstar firms within that local market. The denominator is the total outputs of all firms, both superstar and non-superstar, in the same province and subsector, providing a benchmark against which the superstar share is measured. A higher $HSpill$ means that superstar firms are more dominant in that local market. 

The economic intuition of Horizontal Spillovers is that superstar spillovers capture the degree to which superstar firms create externalities for non-superstar firms. The mechanism works under the mechanism of the Cournot model, as in \citet{shen2021productivity}, where in the beginning, superstar firms dominate market share, and non-superstar firms can then react by adjusting costs more efficiently. Some non-superstar firms may fail to adjust, forcing them to exit from the markets (crowding-out effects). However, as in \citet{melitz2003impact}, some firms may be only able to produce for domestic markets if they are efficient enough, while other non-superstar firms with higher efficiency can serve for export markets. In this regard, we capture the degree to which superstar firms dominate markets from the share of outputs produced by superstar firms in certain subsectors and provinces. This proxy is suitable as a higher share of outputs produced by superstar firms in a subsector and a province implies that superstar firms dominate the market and are able to impose a high markup \citep{aghion2023creative}. 

The second type of spillover is Backward Linkage Vertical Spillovers (which are firms purchasing from), capturing the degree to which a firm in a subsector supplies intermediate inputs for superstar firms. In this case, superstar firms are in the downstream sector while non-superstar firms are in the upstream sector. 
 
        \begin{equation}
        BSpill_{kjt}=\sum_k b_{kl}\times HSpill_{kjt}
        \label{BSpill}
    \end{equation}
where $BSpill_{kjt}$ denotes backward linkage (who are firms buying from), $b_{kl}$ denotes the input–output matrix coefficient from the Input-Output Table of Indonesia in 2010, that captures the amount of intermediate output used from industry $l$ to produce
one unit of output in the downstream industry $k$. The value of $BSpill_{kjt}$ measures, to a certain degree, the derived demand from superstar firms in subsector $k$ for subsector $l$. A higher value of $BSpill_{kjt}$ captures a higher demand of intermediate inputs from superstar firms in subsector $k$ to the subsectors in the upstream industry, regardless it is superstar or non-superstar firms. 

The economic intuition of the Backward spillovers strategy is that when superstar firms operate in different subsectors with non-superstar firms, they may allow spillovers to take place, as it will benefit them through backward linkage \citep{amiti2024fdi}. Suppose technology diffusion occurred from superstar firms in the downstream sectors to the non-superstar firms in the upstream sectors. In that case, superstar firms will obtain better quality intermediate inputs. At the same time, non-superstar firms in the upstream will also upgrade their efficiency and productivity due to the stringent standards for being superstar firms' suppliers and demand security \citep{amiti2024fdi}. Some superstar firms even provide mentoring to their suppliers to ensure that the required standards are met. This mentoring serves as a mechanism through which technology and knowledge diffusion can occur in different subsectors (downstream vs upstream sectors).

The third spillover is the Forward Linkage Vertical Spillovers (who are firms selling to), depicting the degree to which non-superstar firms purchase intermediate inputs from superstar firms. 
 
        \begin{equation}
        FSpill_{kjt}=\sum_k b_{km}\times HSpill_{kjt}
        \label{FSpill}
    \end{equation}
where $FSpill_{kjt}$ denotes forward linkage (who are firms selling to). $b_{km}$ denotes the input–output matrix coefficient that captures the amount of intermediate output sold for industry $m$ from upstream industry $k$. A higher degree of $FSpill_{kjt}$ implies a higher intermediate input of superstar firms in the subsector $k$ sold for subsectors $m$ in the downstream. The economic intuition of this strategy is that non-superstar firms may also benefit from purchasing intermediate inputs from superstar firms, as they gain access to higher-quality and more efficient inputs-- representing a forward linkage. Additionally, superstar firms may offer supplementary services that would not be available if non-superstar firms were to import intermediate inputs instead, as suggested by \citet{javorcik2004does} for multinational companies' evidence. Moreover, the ability of non-superstar firms to import may be limited, making it more advantageous for them to source inputs from domestic firms.

\subsubsection{Total Factor Productivity}
The variable of Total Factor Productivity (TFP) is the part of the production function. Total production may consist not only of proportional input utilization, but also of the degree to which the firm benefits from TFP. The seminal paper of \citet{olley1996dynamics} introduced more robust estimates for productivity coming from unobserved shocks in the production function\footnote{The mechanism of the \citet{olley1996dynamics}'s TFP calculation is explained in the Appendix.}. Meanwhile, \citet{levinsohn2003estimating} introduced an alternative for the proxy variable. Rather than using investments that are prone to costly adjustment, \citet{levinsohn2003estimating} used intermediate inputs as the proxy for productivity shocks. The strategy is similar to \citet{olley1996dynamics}, namely by providing that there is a strictly increasing association between intermediate inputs and productivity, implying that more productive firms allocate more intermediate inputs for production\footnote{See empirical studies such as \citet{olper2017imported} for imported inputs.}. This monotonicity assumption is then used to invert the productivity equation and show consistent parameters. The log of outputs as a function of the log of inputs and the shocks in the standard Cobb-Douglas production function can be arranged as follows:
\begin{equation}
    y_{it}=\beta_{l}l_{it}+\beta_k k_{it}+\beta_{r} r_{it}+\varphi_{it}+\varepsilon_{it}
    \label{CobbDouglas}
\end{equation}

Where $k_it$ is capital proximate from fixed assets such as buildings, land, and other equipment. $r_{it}$ denotes raw materials, $\varphi_{it}$ denotes total factor productivity. Compared to \citet{olley1996dynamics} who used investment as a shock to productivity, there are fewer drawbacks when we use raw materials in the context of Indonesia. As suggested by \citet{rovigatti2018theory}, data for investments may be largely omitted. Evidently, this is also relevant for our data, where the data from 2006 for capital is not available. Meanwhile, raw material is relevant under the monotonicity assumption, which imposes higher material associates to higher productivity for all relevant capital \citet{levinsohn2003estimating}.

In the first stage, we estimate equation \ref{CobbDouglas} using \citet{levinsohn2003estimating}'s strategy to obtain expected value of outputs ($\hat{y}_{it}$) and estimate for $\varepsilon_{it}$, which is arranged as:

\begin{equation}
    \hat{y}_{it}=\beta_k k_{it}+\beta_{l}l_{it}+\varphi_{t}(m_{it},k_{it},z_{it})
    \label{ExpectedValueofY}
\end{equation}

Where $z_{it}$ is exogenous control variables affecting $\varphi_{it}$. Then we can calculate TFP ($\varphi_{it}$) by subtracting $\hat{y}_{it}$ with all components in the right side of equation \ref{ExpectedValueofY}, as follows:

\begin{equation}
    \varphi_{it}=\hat{y}_{it}-\beta_k k_{it}-\beta_{l}l_{it}
\end{equation}

where $\varphi_{it}$ Total Factor Productivity of firm $i$ in time $t$ in the logged-form. Other studies demonstrates the law of motion to capture the effect of prior period's productivity and some incorporate control variables in affecting current value of $\varphi_{it}$, i.e. $\varphi_{it}=g(\varphi_{it-1},z_{it-1})+\xi_{it}$ (\citet{bournakis2022productivity}, \citet{ackerberg2015identification})\footnote{In this study, we do not impose any control variables, such as $z_{it}$ in TFP estimates. It is because we estimate the production function to calculate TFP in each three-digit subsector separately as in \citet{Amiti2007b}. It is not possible to disaggregate the subsector into more specific digits, such as 4 to 5, as this may lead to unreliable statistical outcomes. Moreover, some firms have no three-digit classification due to the limited plants in the province where the firm is located, requiring BPS-Statistics to use a more general sectoral classification. In this case, we group the subsectors into ``Others" based on each two digits with relatively similar technology}. The descriptive statistics of the above-mentioned variables are reported in Table \ref{DescriptiveStatistics}.

\subsection{Empirical Strategy}
\subsubsection{The Nexus of Superstar Spillovers and Total Factor Productivity}
In this study, after we estimate TFP from the first and second stage using \citet{levinsohn2003estimating} strategy, we then arrange the empirical specification for superstar firms on the TFP. We look at the association between the share of superstar firms within the subsector and province (horizontal spillovers) and across the subsector (vertical spillovers), shown by the total output share of superstar firms within the 3-digit ISIC. We test all these spillovers on the TFP for both level ($\varphi$) and growth ($\Delta \varphi$), and arrange the equations as follows:

\begin{equation}
        \varphi_{it}=\beta_{0}+\beta_{Spill} Spill_{kjt}+\beta_{Z}Z_{it}+\varepsilon_{it}
        \label{TFPmodel_baseline}
    \end{equation}

    \begin{equation}
        \Delta\varphi_{it}=\beta_{0}+\beta_{Spill} Spill_{kjt}+\beta_{Z}Z_{it}+\varepsilon_{it}
        \label{TFPgrowthmodel_baseline}
    \end{equation}

Where $\varphi_{it}$ denotes Total Factor Productivity (TFP) from \citet{levinsohn2003estimating} in log-form, while $\Delta\varphi_{it}$ is the TFP growth for which the TFP level in each time $t$ is compared to the initial period $t1$ ($\varphi_{i,t}-\varphi_{i,t1}$), $Spill_{kt}$ is the spillover variables consisting of $HSpill$, $BSpill$, and $FSpill$. We split these three types of spillovers to capture average treatment effects (ATE) from each exogenous instrument. 

Intuitively, how horizontal ($HSpill$) and vertical ($BSpill$ and $FSpill$) spillovers correlate with TFP consists of two mechanisms. First, a higher spillover associated with a higher TFP for all firms (superstar and non-superstar) indicates an overall positive correlation between a superstar firm's exposure to the productivity development. If $\beta_{Spill}>0$, an increase in the superstar firm's share within the subsector and province is associated with an increase in the firm's productivity in general. This hypothesis may work under the mechanism of demonstration, labour migration, and market competition \citep{orlic2018cross}. The second mechanism is by isolating the pure effects solely for non-superstar firms. The extent to which superstar firms in certain subsectors and provinces create spillovers should address their association only with non-superstar firms. The positive magnitude of $\beta_{Spill}$ from the first mechanism might overstate the true correlation magnitude, as it shows both superstar and non-superstar inter-linkage. In this regard, we limit the nexus of spillovers and TFP only for non-superstar firms\footnote{This strategy has been implemented by prior studies such as \citet{Suyanto2009} and \citet{yasin2023spillover}.}. The notation of $Z_{it}$  denotes the set of firm heterogeneous control variables, namely dummy of foreign-owned, and exporter, as well as import intensity, market concentration from Herfindahl-Hirschman Index (HHI), and absorptive capacity. 

We estimate equation \ref{TFPmodel_baseline} and  \ref{TFPgrowthmodel_baseline} using two-stage least squares (2SLS) with fixed effects in the industry, region, island, and year. All results are estimated using robust standard errors and clustered at the firm level. Moreover, we also test the superstar firm's heterogeneity, i.e., foreign-owned and domestic superstars. Some interaction terms between spillovers and firm heterogeneity in control variables are also examined for robustness tests. 

The estimation of each spillover is conducted separately due to the measurements of Backward and Forward Spillovers that also stem from Horizontal Spillovers. Theoretically, spillover processes can be interpreted from multiple perspectives, two of the most commonly studied being horizontal (within-sector and within-province) and vertical (across sectors) spillovers. Each of these spillover channels is associated with distinct mechanisms and thus relies on its own set of instrumental variables. To ensure clarity in identifying the effects of each spillover type, and given that vertical spillovers are mechanically constructed as deterministic functions of horizontal spillovers (as shown in Equation \ref{BSpill} and \ref{FSpill}), which may induce multicollinearity, a separate strategy is more appropriate. This approach offers a more transparent interpretation of each channel's effect and avoids potential identification issues arising from their structural correlation. 

\subsubsection{Potential Endogeneity}
Prior studies on FDI spillovers have treated spillovers indicators as exogenous, typically by measuring the share of superior firms within a sub-sector (see \citet{Sari2016}, \citet{spithoven2023productivity}, and \citet{bournakis2021spillovers}). Intuitively, the share of superstar firms located in a province may also be determined by sectoral and regionally specific factors, such as import tariffs, national shocks, and regional infrastructure. As a result, it may lead to a misleading interpretation of the true effects of productivity spillovers  \citep{bournakis2022productivity}. 

While prior studies on FDI Spillovers do not address endogeneity concerns, we aim to construct a proxy for endogenous superstar firm presence. We identify several potential endogeneities for using the share of superstars as a spillover proxy.  First, a higher share of superstar firms in subsector $k$ in province $j$ may be associated with higher productivity, suggesting that spillovers occur from superstar to non-superstar firms. However, it is also possible that superstar firms increase their share in response to productivity improvements among non-superstar firms, raising concerns about reverse causality. Moreover, the decision of firms to expand their production in a certain province may also be determined by sectoral-specific dynamics at the national level as well as province-specific characteristics. 

Another potential endogeneity is the selection bias. We aim to capture the effect of horizontal and vertical spillovers from superstar firms, specifically on the non-superstar firms. Meanwhile, for being superstar firms, there is potential selection criteria which may cause selection bias if we directly isolate non-superstar firms in the sample\footnote{However, evidently, in our datasets, more than 92\% of superstar firms have existed since the beginning of the period. It implies the dominant pre-determined non-superstar samples.}. Hence, we use inverse probability weighting (IPW) as robustness to check whether the selection issue occurred if we only include non-superstar firms in the estimation\footnote{IPW approach can reduce bias due to self-selection by eliminating the correlation between observed/unobserved factors and non-superstar firms \citep{petersen2024inverse}. Another way to tackle the selection problem is by using Heckman selection criteria (\citet{heckman1976common} and \citet{heckman1979sample}) to determine whether a firm is non-superstar. However, Heckman selection does not accommodate other endogeneity problems such as reverse causality between spillovers and productivity.}.

In addition to applying standard approaches like fixed-effects, which already mitigate potential endogeneity bias \citep{Amiti2007b}, we implement an alternative estimation using instrumental variables (IV). It is well-known that selecting instrumental variables is difficult, notably the mechanism by which spillovers in our proxy have different dimensions. Hence, we refer to prior studies of spillovers such as \citet{du2014fdi} and \citet{xu2012productivity} in selecting the instrumental variables. In our design, the instrumental variables should correlate with spillovers but not directly with non-superstar TFP level and growth. Most importantly, the IV should not correlate with error terms from the second stage. Some prior studies on spillovers solely use the lag of endogenous spillovers under the assumption that the delayed reaction of productivity from non-superstar firms (see \citet{njikam2019productivity} and \citet{barrios2011spillovers}). However, the lag of endogenous spillovers may not be a strong instrument if there is a persistent dynamic error process, causing sample bias and imprecision \citep{blundell1998initial}.

\subsubsection{Bartik Instruments: Output Growth and Tariff}
In the construction of our instrumental variables, we apply the Bartik instrument, known as Shift-Share IV, originally introduced by \citet{bartik1991benefits}, popularized by \citet{blanchard1992regional}, and formally evaluated by \citet{goldsmith2020bartik} and \citet{borusyak2025practical}. The design of our Bartik-IV is different across spillover dimensions. First, we design Bartik-IV from labour-based and Output Growth for Horizontal Spillovers ($HSpill$). As suggested by \citet{Amiti2007b}, we construct the instrument using the initial share of unskilled workers of superstar firms of subsector $k$ in province $j$  in the baseline year, 2001, combined with the subsequent national output growth of sector $k$, excluding the related province, as follows.

\begin{equation}
    LabSh^{S}_{jk,2001}=\frac{\sum_{k\in\mathcal{K} } Unskilled^{S}_{jk,2001}}{\sum_{k\in\mathcal{K}} Unskilled_{jk,2001}}
    \label{Share Bartik Unskilled}
\end{equation}

\begin{equation}
G^{-j}_{kt} = 
\frac{
    \sum_{m \in \mathcal{M} \setminus j} Y_{m,k,t} - \sum_{m \in \mathcal{M} \setminus j} Y_{m,k,t-1}
}{
    \sum_{m \in \mathcal{M} \setminus j} Y_{m,k,t-1}
}
\label{Shift Bartik Output}
\end{equation}

\begin{equation}
    LabBartikIV^{S}_{kt}=\sum_{k \in \mathcal{K}} LabSh^{S}_{j,k,t=2001}\times G^{-j}_{kt}
    \label{Bartik Equation Labour}
\end{equation}

Where $LabSh^{S}_{j,k,2001}$ is the share of superstar firms' unskilled workers in province $j$ from subsector (3-digits) $k$ in 2001, $G^{-j}_{kt}$ denotes the growth of outputs of subsector $k$ in national level excluded subsector $k$ in province $j$, so $m \in \mathcal{M}$ denotes national level excluding related subsector and province. $LabBartikIV^{S}_{kt}$ is the exogeneous Bartik Instrument, an instrument for $HSpill$. We follow a shift-share approach that endogenizes spillovers variation while also ensuring the exogeneity of the instrument--a key requirement for instrumental variable validity. Specifically, we implement a \textit{leave-one-out} strategy \citep{goldsmith2020bartik}, whereby the 3-digit subsector in province $j$ of interest is excluded when aggregating sectoral growth at the national level. 

The economic intuition of this Bartik-IV is to capture how the initial superstar share in province $j$ and subsector $k$ determines the extent to which the production expansion of superstar firms from subsector $k$ in a given provinces is influenced by the initial share of unskilled labour employed by these superstar firms\footnote{In the Indonesian manufacturing data, unskilled workers are assumed as workers for production activities, while unskilled workers are for non-production activities. These arguments are assumed by \citet{amiti2012trade} and \citet{matsuura2023foreign} under the evidence that in 2006 data, about 10\% of non-production workers are university graduates, and 63\% are high-school graduates. Meanwhile, 1\% of production workers have graduated from university, and 42\% have completed high school.}. The use of unskilled workers as a component of the $LabBartikIV$ share is intended to capture how the proportion of workers directly involved in production activities can determine whether a superstar firm will choose to expand or not, given a sector-specific shock at the national level. If subsector $k$ experiences a shock in the form of high national-level growth, which is exogenous to firms in subsector $k$ located in provinces $j$, then subsector $k$ in province $j$ with a high initial share of unskilled workers becomes a target for expansion by these superstar firms. This expansion will lead to an increase in the superstar firms' output share, which is captured by the proxy for Horizontal Spillovers. As suggested by  \citet{trefler2004long} and \citet{Amiti2007b}, unskilled workers influence the propensity of an industry to become organized. In other words, the extent to which subsector $k$ in province $j$ is organized by superstar firms, through efforts to dominate market share, is determined by the initial condition of unskilled labour in that province and subsector. For the robustness test, we construct this labour-based $LabBartikIV$ using both skilled and unskilled workers.

We design another Bartik-IV using the share of superstar from outputs-based (gross outputs) as the share component and the change of tariff (Most Favored Nation, MFN) for vertical spillovers (backward and forward). Although prior studies have found a significant effect of tariffs on a firm's productivity (\citet{Amiti2007b}, \citet{Gupta2023}, \citet{zhang2021overcoming}), our model argues that there is a strong potential for superstar firms to be more affected by tariff shocks at the subsector level. To ensure this assumption, rather than using tariff level as a direct IV, we design Bartik-IV with tariff change as the shift component and the initial share of superstar as the weight component.

\begin{equation}
    OutSh^{S}_{jk,2001}=\frac{\sum_{k\in\mathcal{K} } Y^{S}_{jk,2001}}{\sum_{k\in\mathcal{K}} Y_{jk,2001}}
    \label{Share Bartik Output}
\end{equation}

\begin{equation}
\Delta Tariff_{k,t} = Tariff_{k,t} - Tariff_{k,t-1}
\label{Shift Bartik Tariff}
\end{equation}

\begin{equation}
    TarrBartik IV^{S}_{kt}=\sum_{k \in \mathcal{K}} OutSh^{S}_{j,k,t=2001}\times \Delta Tariff_{k,t}
    \label{Bartik Equation Tariff}
\end{equation}

where $OutSh^{S}_{j,k,2001}$ is the share of superstar firms outputs in province $j$ from subsector (3-digits) $k$ in 2001, $TarrBartik IV^{S}_{kt}$ denotes the change of tariff of subsector $k$ in national level. $TarrBartik IV^{S}_{kt}$ is the exogeneous Bartik Instrument, an instrument for $BSpill$ and $FSpill$. 

The intuition of $TarrBartik IV^{S}_{kt}$ as the IV for $BSpill$ and $FSpill$ is that the level of share of superstar firms in the initial period may be subject to the decision of firms to purchase from the upstream sectors or sell to the downstream sectors. When the superstar share is small and tariff decreases (tariff change is negative), superstar firms may enhance the connection with their suppliers in the upstream sectors and purchasers in the downstream sectors. When a superstar firm decides to expand its production, the change of import tariffs on intermediate inputs becomes a key determinant of whether the firm will proceed with such expansion and further increase its market connection. As superstar firms are typically more internationally exposed compared to non-superstar firms, for example, through larger import activities, a higher import tariff may discourage superstar firms from expanding, leading to a decline in their output and supply-chain domestic market networks.

In this framework, tariffs are hypothesized to be negatively associated with the market of superstar firms. This mechanism, in turn, has implications for the productivity dynamics of non-superstar firms. Specifically, if tariffs are high and rising, leading to a reduction in superstar firms' market share, and if negative spillovers from superstar firms are present, then the productivity of non-superstar firms may improve, conditional on their import intensity (which is therefore controlled for in our empirical specification). Conversely, if tariffs decline and negative spillovers persist, superstar firms may further dominate the market, potentially suppressing the productivity of non-superstar firms within the same sector and province through the rivalry and market-stealing phenomenon. In the empirical study, for instance, \citet{du2014fdi} found that tariffs affect the magnitude and direction of spillovers. Specifically, they revealed that tariff changes following China's accession to the World Trade Organization (WTO) strengthened FDI spillovers, particularly backward spillovers\footnote{Some studies have also used tariffs as instrumental variables, such as \citet{pane2023role} and \citet{tsionas2023productivity}. For the robustness test, we use Road Density for the IV of $BSpill$ and $FSpill$. Road Density is measured from the total length of the country, province, and regency divided by the province's size (in square kilometers). Prior study of \citet{Rodriguez-Pose2013a} examined the effect of road density on the firm decision to firm's decision to export. It implies that road density can influence not only whether superstar firms choose to locate in a particular province, but also the extent of their market share within that province. We use road density as the robustness rather than the main IV due to the fact that road density merely captures province variation. Meanwhile, $TarrBartikIV$ captures both sectoral and provincial variation.}.

\subsubsection{The Dynamic Olley-Pakes Decomposition}
In addition to investigating superstar spillovers, we also aim to capture the dynamic of superstar firms' productivity overall. The idea is that whether productivity changes solely stems from within-firm productivity improvements, or is also supported by the entry-exit behaviour of the firms. In this regard, we refer to the notion of Dynamic Olley-Pakes Decomposition (DOPD) that was introduced by several papers such as \citet{baily1992distribution}, \citet{griliches1995firm}, and \citet{foster2001aggregate}. The latest decomposition was proposed in \citet{melitz2015dynamic} by elaborating firms' entry and exit behaviour into the model of Olley-Pakes Decomposition from \citet{olley1996dynamics} and \citet{collard2015reallocation}, decomposing TFP based on technology group. The decomposition notion was proposed in the mechanism by which firms are heterogeneous and thus contribute to their aggregate productivity growth to a different extent \citep{Karagiannis2018}. In this study, we employ such a decomposition in the application of superstar and non-superstar groups. We first look at the standard Olley-Pakes Decomposition (henceforth OP Decomposition) arranged as follows:
\begin{equation}
 \begin{split}
    \Phi_t=\bar{\varphi_t}+\sum_i (Sh_{it}-\bar{Sh_t})(\varphi_{it}-\bar{\varphi_t}) \\
    = \bar{\varphi_t}+ cov(s_{it},\varphi_{it})
    \end{split}
    \label{BasicOlleyPakes}
\end{equation}
where $\bar{\varphi_t}$ denotes unweighted firm productivity mean ($\bar{\varphi_t}=\frac{1}{n}\sum^{n}_{i=1} \varphi_{it}$) and $\bar{Sh}_t$ is the mean of market share ($\bar{Sh_t}=\frac{1}{n}\sum^{n}_{i=1} Sh_{it}$). The level of productivity growth over time ($\Delta\Phi$) is obtained by looking at the change of unweighted firm productivity mean ($\Delta \varphi_t$) and the change of covariance ($\Delta cov(s_{it},\varphi_{it}) $). The component of $\Delta \varphi_t$ captures the shifts in the productivity distribution, showing that a firm's productivity improves over time. Meanwhile, the component of $\Delta cov(s_{it},\varphi_{it}) $ captures the process through which more dominant firms, shown from higher market share, are more productive, which thus contributes to their aggregate productivity. We prove this assumption in the Results section that market share is positively correlated with productivity. 

The same decomposition can also be applied by a different group, as introduced by \citet{collard2015reallocation} for technology group evidence in the US. In this case, we apply for the group of superstar ($S$) and non-superstar ($NS$) firms, noted by group $\psi$. This strategy enables the changes in productivity between superstar and non-superstar firms (within-firms) and substantial reallocation across firms to be captured. In this regard, we may note the market share of each group $\psi$ denoted by $Sh_t(\psi)=\Sigma_{i \in \psi} Sh_{it}$. Hence, we may denote group-specific aggregate productivity as $\Phi_t(\psi)$. while $\bar{\varphi}_t(\psi)$ as the average productivity within group $\psi$. By referring to the OP Decomposition, we may obtain a static decomposition for within superstar-nonsuperstar as follows: \\
\\
\textbf{DEFINITION 1:} \textit{The Static Olley-Pakes Decomposition}

\begin{equation}
\begin{split}
        &\Phi_t= \sum_{\psi \in S, NS} Sh_t(\psi) \Big (\bar{\varphi}_t(\psi)+ cov_t(\psi)\Big ) \\
        & = \sum_{\psi \in S, NS} Sh_t(\psi) \Big (\bar{\varphi}_t(\psi)+\sum_i (Sh_{it}(\psi)-\bar{Sh}_t(\psi))(\varphi_{it}(\psi)-\bar{\varphi_t}(\psi))\Big ) \\
        &= Sh_t(\psi=S) \Bigg \{\bar{\varphi}_t(\psi=S)+\sum_{i \in \psi=S} [Sh_{it}(\psi=S)-\bar{Sh}_t(\psi=S)][\varphi_{it}(\psi=S)-\bar{\varphi_t}(\psi=S)]\Bigg \}+  \\
        &Sh_t(\psi=NS) \Bigg \{\bar{\varphi}_t(\psi=NS)+\sum_{i \in \psi=NS} [Sh_{it}(\psi=NS)-\bar{Sh}_t(\psi=NS)][\varphi_{it}(\psi=NS)-\bar{\varphi_t}(\psi=NS)]\Bigg \} 
\end{split}
\label{AugmentedOlleyPakes}
\end{equation}

The concept of decomposition in equation \ref{BasicOlleyPakes} was also extended by \citet{melitz2015dynamic} by incorporating entry and exit firms' behavior (henceforth MP Decomposition). The standard form of MP Decomposition from \citet{melitz2015dynamic} is as follows:
\\
\\
\textbf{DEFINITION 2:} \textit{Dynamic Olley-Pakes Decomposition}

\begin{equation}
    \Phi_1=Sh_{Sur_1}\Phi_{Sur_1}+Sh_{Ex_1}\Phi_{Ex_1}=\Phi_{Sur_1}+Sh_{Ex_1}(\Phi_{Ex_1}-\Phi_{Sur_1})
    \label{MPDecomt1}
\end{equation}

\begin{equation}
    \Phi_2=Sh_{Sur_2}\Phi_{Sur_2}+Sh_{En_2}\Phi_{En_2}=\Phi_{Sur_2}+Sh_{En_2}(\Phi_{En_2}-\Phi_{Sur_2})
    \label{MPDecomt2}
\end{equation}
where $Sh_{G_t}=\Sigma_{i \in G}Sh_{it}$ represents the aggregate market share of a group $G$ of firms and we define $\Phi_{Gt}=\Sigma_{i \in G}(Sh_{it}/Sh_{Gt})\varphi_{it}$ as that group's aggregate (average) productivity. The group of $Sur$, $Ex$, and $En$ denotes Survivors, Exiters, and Entrants. Survivors refer to firms that exist in both $t=1$ and $t=2$. For instance, in terms of $t_1=2001$ compared with $t_2=2002$, a firm is a survivor if it exists in 2001 and 2002. When we compare non-respective behaviour, such as 2001 and 2003, a firm is called a Survivor of 2003 if it exists in 2001, 2002, and 2003. If it did not exist in 2002, we would exclude this firm as a Survivor in 2003. Instead, we may capture this type of firm as Exiters in 2002. Meanwhile, Exiter firm is if a firm exists in $t=1$ but does not exist in $t=2$, while it is Entrant if the firm exists in $t=2$ but does not exist in $t=1$. In this regard, we may obtain equation identity based on the share for two periods as $Sh_{Sur_1}+Sh_{Ex_1}=Sh_{Sur_2}+Sh_{En_2}=1$. From equations (8) and (9), we can find the growth rate as follows:

\begin{equation}
\begin{split}
    \Delta \Phi=\Phi_{Sur_2}-\Phi_{Sur_1}+Sh_{En_2}(\Phi_{En_2}-\Phi_{Sur_2})+Sh_{Ex_1}(\Phi_{Sur_1}-\Phi_{Ex_1}) \\
    = \underbrace{\underbrace{\Delta \varphi_{Sur}}_{\text{plant improvements}} + \underbrace{\Delta cov_{Sur}}_{\text{within reallocation}}}_{Survivors}+\underbrace{\underbrace{Sh_{En_2}(\varphi_{En_2}-\varphi_{Sur_2})}_{\text{plant improvements}}+\underbrace{Sh_{En_2}(cov_{En_2}-cov_{Sur_2})}_{\text{between reallocation}}}_{Entrants}+ \\
    \underbrace{\underbrace{Sh_{Ex_1}(\varphi_{Sur_1}-\varphi_{Ex_1})}_{\text{plant improvements}}+\underbrace{Sh_{Ex_1}(cov_{Sur_1}-cov_{Ex_1})}_{\text{between reallocation}}}_{Exiters}
     \end{split}
     \label{MPDecomGrowth}
\end{equation}

The component of $\Phi_{Sur_2}-\Phi_{Sur_1}$ shows the contribution of survivor firms consisting of productivity improvement from its own and market share reallocation. Meanwhile, the component of $Sh_{En_2}(\Phi_{En_2}-\Phi_{Sur_2})$ shows the contribution of entering firms, and $Sh_{Ex_1}(\Phi_{Sur_1}-\Phi_{Ex_1})$ captures the exiting firm dynamics.

\section{Results}
\subsection{Stylized Facts}
We first report the stylized facts of superstar firms in Indonesia by presenting the Total Factor Productivity (TFP) level and growth estimation over time and across firms' characteristics. According to Figure \ref{LogofTFovertime}, the TFP level of superstar firms is relatively higher than that of non-superstar firms, implying that superstar firms, as the market leader within the sector, operate production more efficiently than non-superstar firms. However, the TFP level trend for non-superstar firms converges with the level of superstar firms after 2008. Meanwhile, TFP growth, measured by the change in TFP level relative to 2001, shows the opposite. The growth of non-superstar' TFP since 2002 is tied with superstar firms, but they outweigh superstar firms in 2010, implying that non-superstar firms' productivity grows faster since this year \footnote{This evidence also shows the catching-up behaviour of less productive firms towards more productive firms, as postulated by development studies such as \citet{barro1992convergence}.}.

We also report the distribution of superstar firms on average from 2001 to 2015, in Figure \ref{Superstar Distribution}. According to Figure  \ref{Superstar Distribution}, on average, from 2001 to 2015, Java Island dominates the number of superstar firm presence, notably West Java, which possesses more than 100 superstar firms from various subsectors. Meanwhile, provinces outside Java with a large number of superstars are dominated by Sumatra and Kalimantan.

\subsection{Results and Discussion}
The following result shows the relationship of superstar spillovers to the productivity level and growth. To isolate the pure correlation on non-superstar firms, we also examine the relationship of spillovers exclusively on non-superstar firms. Moreover, the nexus is captured for both TFP level and growth, as well as the interaction with control variables to reveal its robustness. 

First, Table \ref{Main Results Horizontal} shows the results from Horizontal Spillovers. We capture a positive relationship between horizontal spillovers and TFP level ($\varphi$) and growth ($\Delta\varphi$). The results are consistent across different designs, namely interaction with controls, i.e., exporters and foreign-owned. It implies that a higher share of superstar firms within a province and a subsector makes non-superstar firms more productive. A plausible reason for this finding may stem from the spillovers through the automation production process that becomes pervasive among not only superstar firms but also non-superstar firms, based on the stylized facts in Figure \ref{automation_general}. If we interpret this direction with the results in the First-Stage, reported in Table \ref{First-Stage Main Horizontal} in the Appendix, it implies that a higher share of superstar firms in certain province and subsector shocked by the industrial national growth leads to the higher share of the superstar firms, which in turn enhance non-superstar firms productivity level and growth within the sector and province.

Meanwhile, in the case of vertical spillovers from backwards ($BSpill$), the results are reported in Table \ref{Main Results Backward}, and Table \ref{First-Stage Main Backwards} for the First-Stage estimation. We capture a positive correlation between Backward spillovers and productivity for both level and growth. It suggests that when non-superstar firms establish a supply connection by acting as suppliers for superstar firms, their productivity level and growth are higher. Meanwhile, in the first stage, tariffs negatively affect the degree to which superstar firms establish backward channels, suggesting that a high share of superstar firms within a province, followed by a higher tariff rate--a positive tariff change--discourages superstar firms from expanding their market shares. Conversely, when the share is low and followed by a lower tariff, it stimulates superstar firms in connecting with their suppliers, which in turn stimulates productivity enhancement for non-superstar firms in the upstream sectors.

The results from Backward channels show a similar direction to forward ($FSpill$) linkages in Table \ref{Main Results Forward} and Table \ref{First-Stage Main Forwards} in the Appendix for the First-Stage estimates. The results show that being a buyer of intermediate inputs from superstar firms is associated with a better productivity level and growth. Accordingly, superstar firms act as stimulators that enhance the equality of outputs in the upstream and downstream industries. By becoming suppliers to superstar firms, non-superstar firms are compelled to raise their production standards, while they may gain high-quality inputs by purchasing intermediate materials from superstar firms.

We then test superstar-heterogeneous characteristics by more specifically redefining superstar firms into two groups: foreign-owned and domestic superstars. In this strategy, we may capture whether superstar spillovers occur from foreign enterprises. In this strategy, foreign superstar refers to superstar firms with foreign capital ownership of more than 10\%. Meanwhile, for domestic superstar, we specifically define as non-foreign-owned superstar firms (capital ownership from foreign is less than 10\%). In this test, both foreign and domestic superstars are non-exporters to isolate the real effect of superstar \footnote{This two definitions of superstar firms are also used by \citet{amiti2024fdi}.}. The results are reported in Table \ref{Heterogeneous Superstar} for the second-stage, while the first-stage is reported in the Appendix in Table \ref{Heterogeneous Superstar FS}.

According to Table \ref{Heterogeneous Superstar}, we found that there are robust effects of superstar firms across different types of ownership. First, foreign superstars consistently show positive effects on non-superstar productivity, except for the horizontal channel. It implies that competing with foreign superstar firms has no significant effects on non-superstar firms. However, domestic superstar shows positive results for entire channels, implying that domestic superstars generate spillovers more than foreign superstars do. We also report the estimation results for the standardized $HSpill$, $BSpill$, and $FSpill$ on both TFP level and growth in Figure \ref{Standardized Spillovers Coefficient} in the Appendix. The results show that horizontal spillovers indeed have smaller effects compared to vertical spillovers (both backward and forward). Similarly, when we break down the results by heterogeneous superstar firms, we find that domestic superstars have a larger effect than foreign superstars (see Figure \ref{Heterogeneous Superstar Spillovers} in the Appendix). 

Our finding is consistent with \citet{amiti2024fdi} arguing that there are heterogeneous relationships between the type of superstar firms, not necessarily the foreign firm, that cause positive spillovers. When domestic firms are also involved in the international markets by being exporters, they may obtain global technological knowledge, which in turn possibly causes spillovers for non-superstars. Although in early 2000, not many firms in Indonesia could afford automation, except for foreign superstar firms that have gained access to advanced technology from their parent company, after the implementation of the Indonesian rule number 176/2009 about Import Duty Exemption on the Imported Machinery, there is an indication that it may encourage the productivity of non-superstar firms. Recently, in early 2022, the Indonesian government, through Ministerial Regulation (\textit{Peraturan Pemerintah}) Number 1 of 2022, also promulgated a regulation requiring large or superstar firms to establish cooperation with smaller firms, specifically Small and Medium Enterprises (SMEs). This regulation also complements previous foreign investment regulations, such as the Negative Investment List, which specifically regulates foreign entry into Indonesia since early 2000, as discussed by \citet{Genthner2022}. This finding aligns with our findings, which indicate that acting as suppliers for superstar firms (domestic and foreign) enhances the productivity of non-superstar firms. 

Moreover, superstar firms may also function as a dating agency, facilitating connections between non-superstar firms and other firms in downstream industries (both superstar or non-superstar). Once a non-superstar firm meets the high production standard required to become a superstar supplier, it builds a strong portfolio, making it more attractive to other superstar firms or non-superstar firms in the downstream sectors. Moreover, being a supplier to a superstar firm requires a company to focus and specialize in a specific product category. For instance, in packaging, superstar firms prefer suppliers that are solely dedicated to packaging production, ensuring that the quality standards remain exceptionally high. 

The automation process in superstar firms, which leads to an increase in their output share and creates rivalry effects for non-superstar firms, is an inevitable phenomenon. Superstar firms drive production through automation; if non-superstar firms are unable to keep up with this production speed, their market share will decline. In this context, if the government intervenes by supporting technology adoption for non-superstar firms, the strategy will be quite challenging due to the limited availability of skilled workers in these firms. On the other hand, superstar firms can receive an incentive in the form of an import tax discount from importing advanced machines if they collaborate with non-superstar firms. Although the government has promulgated this regulation, major obstacles remain--superstar firms struggle to find smaller firms that can meet their demands, which, in turn, may merely be recognized as a barrier to entry for superstar firms. 

Meanwhile, a plausible reason for the insignificant effects of horizontal spillovers ($HSpill$) from foreign superstars is the protection of technology from the parent company to which foreign superstars are affiliated. Although non-superstar firms can imitate production technology from foreign superstar firms, they cannot solely mimic the technology to produce similar outputs if the parent company of these foreign superstar firms protects their technology. In this regard, vertical spillovers from horizontal channels cannot occur. This mechanism is different in the case of domestic spillovers. Domestic superstar and non-superstar firms may already be familiar with local markets, stimulating fair competition and encouraging non-superstars' productivity improvements.

We then look at the results from productivity decomposition to determine whether productivity dynamics are determined by the entry-exit behaviour of the firms. We first show the assumption that there is a positive association between share and productivity. There are three components, namely correlation from covariance of OP Decomposition in each 2001-2005, 2006-2010, and 2011-2015, which are reported in the Appendix in Figure \ref{CorrelationProof}. According to Figure \ref{CorrelationProof}, it is proven that there is a positive association between TFP and market share in the first two panels. Specifically, the assumption of \citet{olley1996dynamics} by which an increase in share correlates positively with the TFP is proven and captures market reallocation into more productive firms. 

We then reveal the productivity decomposition in our study. We explore the contribution for 3 time windows (2015, 2010, and 2005) with 2001 as a base year.  First, the result from the static OP Decomposition with superstar and non-superstar groups, excluding entry-exit behaviors, is reported in Table \ref{Static OP}. According to Table \ref{Static OP}, the average overall TFP growth is positive for all time windows, contributed mainly from plant improvements, while the Reallocation component mainly contributes negatively to the aggregate TFP change in 2001-2015. It implies that TFP change is mainly supported by the improvements within firms, while the reallocation process occurred towards less productive firms. 

If we compare the components between Superstar and Non-superstar, it is clear that the aggregate TFP growth of non-superstar firms outweighs superstar firms, with plant improvements being the dominant component for both groups. This decomposition corroborates the results in Figure \ref{LogofTFovertime}. This finding also implies that although superstar firms possess a higher TFP level, indicating that they are more efficient, they do not necessarily grow faster than non-superstar firms.

Furthermore, we present the results from the Dynamic OP Decomposition, a-la \citet{melitz2015dynamic}, with the extension for heterogeneous superstar firms (foreign and domestic)\footnote{The robustness test for Dynamic OP Decomposition of superstar firms is reported in \ref{Robustness DOPD} by referring to \citet{amiti2024fdi}'s definition.}. The results are reported in Table \ref{ContributionDynamicOP}. According to Table \ref{ContributionDynamicOP}, there is a consistent finding with the static decomposition where the reallocation process contributes negatively towards aggregate productivity, notably in the period of 2010 and 2015. However, in this dynamic approach, we can address that the reallocation within survivors that causes worsened productivity, implying the market share is reallocated towards survivors who are less productive. This finding also indicates misallocation among survivor firms. Meanwhile, the reallocation between survivors and exiters-entrants gains positive drivers.

If we look at the components by group of General Superstar, it can be seen that non-superstar firms outperform in the productivity improvement component for 2001-2005 and 2001-2015 time windows. Meanwhile, superstar firms experience a severe negative within-group reallocation component for all time windows, which is much higher in magnitude compared to the non-superstar group. It indicates a negative contribution from the survivors within the superstar group. An intriguing result is shown by the Heterogeneous Superstar group, where the negative reallocation within survivors is mainly driven by the domestic superstar with negative growth for all time windows, although the domestic superstar grows faster than the foreign superstar in terms of plant improvements. On the other hand, reallocation from the exiters and entrants components shows a better contribution for superstar firms, although only in 2001-2005 do superstar firms outperform non-superstar firms, and even then, only by a moderate margin. 

We also report the dynamic decomposition from the overall group, superstar, and non-superstar for 2001-2015 in Figure \ref{DOPD Decomposition}. According to Figure \ref{DOPD Decomposition}, in panel (i), it is obvious that in 2001-2015, the largest drivers of the productivity growth stem from Plant-Improvements from survivor firms. More specifically, Plant-Improvements within superstar survivors are relatively stable between 2007 and 2012, although the components of Within-Reallocation are negative. It implies that there is reallocation into less productive firms within superstar survivors during 2007-2015. Meanwhile, the group of non-superstar firms is also driven mostly by Plant-Improvements, where some periods show a negative contribution from Within-Reallocation.

The finding that superstar firms have higher TFP level compared to non-superstar firms (as shown in Figure \ref{LogofTFovertime} and Table \ref{Main Results Horizontal}, Table \ref{Main Results Backward}, and Table \ref{Main Results Forward}), and generate positive spillover effects for non-superstar firms but exhibit lower growth rates than non-superstar firms, indicates that superstar firms in Indonesia behave differently from the ``Rise of Superstar Firms" hypothesis observed in advanced economies, as discussed by \citet{autor2020fall} or \citet{amiti2024fdi}. While Indonesian superstar firms indeed operate more efficiently than non-superstar firms, their lower growth suggests stagnation among these firms. This stagnation may reflect a lack of innovation or limited adoption of new technologies within Indonesian superstar firms. Although positive productivity spillovers occur, it may simply be a consequence of the already substantial TFP gap between superstar firms and their non-superstar competitors, suppliers, and purchasers (as seen in the considerable TFP gap at the level presented in Figure \ref{LogofTFovertime}).

Furthermore, when superstar firms account for more than 75\% of market share and experience TFP growth, this evidence may also contribute to the premature de-industrialization discussed by \citet{rodrik2016premature}. In line with \citet{autor2020fall}, superstar firms may be regarded as ``winners-take-all"; however, in the Indonesian context, market-dominating superstar firms do not outperform in terms of growth--in fact, what is observed could be characterized as ``The Fall of Superstar Firms". This stagnation or even decline in TFP growth among superstar firms may be one of the reasons why, at the aggregate national level, the manufacturing subsector's contribution has declined significantly: the market leaders themselves are stagnating or regressing in terms of productivity growth. 

Considering other characteristics, such as automation and market share (see Figure \ref{automation_general}), \citet{autor2020fall} notes that a rise in superstar firms is typically associated with increased market concentration. Similar to Indonesian evidence, figure \ref{automation_general} shows that an increase in automation--indicating a rise in superstar firms-- is also associated with an increase in market concentration. It further supports the finding that Indonesian superstar firms primarily seek to stimulate market competition within the sector and province, sourcing and selling from non-superstar suppliers in upstream and downstream sectors, thereby resulting in positive backward and forward spillovers. At the same time, although superstar firms may not be engaging in innovation, shown by their slower productivity growth, their rate of automation adoption has increased and is relatively higher than that of non-superstar firms. It may explain why their TFP level remains higher compared to non-superstar firms.

The findings also imply that although we observe an increasing trend of TFP growth and levels over 2001-2015, the decomposition result shows that the market mechanism failed to work well due to misallocation, even though firm-level upgrading improved. It also corroborates the evidence that the share of manufacturing output in GDP decreases, as the increase in TFP levels and growth was not accompanied by the reallocation of production toward the most productive firms, i.e., misallocation. Meanwhile, regarding positive spillovers, although spillovers from firm superstar to non-superstar occurred, they only affected within-plant improvement and do not necessarily determine how market misallocation unfolded. In this sense, TFP level and growth might have been even lower without these spillovers, given that misallocation worsened aggregate productivity. Therefore, while TFP did increase, it failed to support structural upgrading of manufacturing by raising its share of GDP.

Furthermore, the findings from heterogeneous superstar decomposition that reveal negative results for domestic superstar firms imply misallocation within domestic superstar survivor firms in Indonesia. This finding is plausible when the contribution of state-ownership to the domestic superstar firms reaches about 13\% on average in 2001-2015, which is 4 times higher than foreign superstar firms that possess about 3\% ownership from Indonesian central and regional governments. Among these domestic superstar firms, about 4\% are state-owned enterprises (SOE) and 7\% are regionally-owned enterprises (ROE)\footnote{SOE and ROE with foreign-ownership more than 10\% are less than 1.5\% in 2001-2015}. In this regard, there is any possibility that the misallocation occurred mainly from this SOE and ROE\footnote{Prior studies, such as \citet{han2021subsidies} and \citet{bach2019state} have also found economic distortion due to misallocation of state-owned enterprises in China, while \citet{de2024productivity} scrutinize misallocation of Indonesia, Vietnam and Malaysia and found the distortion due to misallocation which hinders them to grow}. 

Moreover, the findings that misallocation occurs among domestic surviving firms imply that domestic firms in Indonesia, possibly including SOEs, have not experienced technological progress due to privileges from the government, such as better access to the financial system \citep{zhao2019chinese}. In this context, although they hold a large market share of sales and operate efficiently (so they become a superstar), thereby creating positive spillovers for non-superstar firms, they are reluctant to technology-upgrade. Consequently, their TFP growth stagnates and is even slower than that of non-superstar firms. When misallocation occurs, resources are not reallocated from less productive superstar firms to more productive ones, but rather the opposite, due to distortion.

\subsection{Robustness Tests}
We conduct several strategies for additional robustness tests. First, we resample the observation into two groups, namely medium and large firms, based on the definition of BPS-statistics. Medium firms are those firms with workers less than 100, while large firms possesses at least 100 workers. This robustness test also aims to reveal homogeneous treatment effects of each instrument on the endogenous spillovers. The results are reported in the Appendix in Table \ref{Robustness Large Medium}. According to Table \ref{Robustness Large Medium}, the results are relatively consistent with the main estimation, where there are positive spillovers from all channels.  

Another strategy for robustness testing is to use alternative instrumental variables. We re-design $LabBartikIV$ from equation \ref{Bartik Equation Labour} by changing the share component into the initial share of all workers. We use this IV for $HSpill$. Meanwhile, we use the average of Road Density for the IV of $BSpill$ and $FSpill$. The results are reported in Table \ref{Robustness IV Horizontal}, \ref{Robustness IV Backward}, and \ref{Robustness IV Forward}. According to these tables, all results are relatively consistent, where we capture positive effects of spillovers from all dimensions. 

Another robustness test is for heterogeneous superstar spillovers from the results in Table \ref{Heterogeneous Superstar}. In this result, we test the robustness by using a different productivity indicator, namely the simple ratio of value added to the total workers (in log form). The results are reported in Table \ref{Heterogeneous Superstar Simple}. According to Table \ref{Heterogeneous Superstar Simple}, the results are consistent where both foreign and domestic superstars generate positive spillovers for non-superstars, except for horizontal spillovers from foreign superstars. 

\section{Conclusion and Further Development}
We provide empirical evidence that firms dominating the market exhibit key characteristics of superstar firm behaviour. In this regard, we aim to reveal the relationship between superstar spillovers and productivity and look at the contribution of superstar firms to productivity. Our findings capture a positive relationship between productivity level and growth on the horizontal spillovers and vertical spillovers, implying that a higher share of superstar firms within a province and subsector causes non-superstar firms to be more productive and leads them to grow faster. In the case of vertical spillovers from backward and forward linkages, the results indicate a positive relationship, suggesting that when non-superstar firms establish supply connections by acting as suppliers or purchasers, their productivity is higher and more rapid. In terms of the decomposition strategy, the results show that the negative aggregate productivity growth that occurred is mainly driven by within-group reallocation, which implies misallocation within survivors in the markets. 

Our study leaves at least three policy implications. First, the strategy for superstar firms to bridge with non-superstar firms is essential in stimulating TFP growth. With the results showing that becoming suppliers to foreign superstar firms increases productivity level and growth, government policy should focus on improving the upstream and downstream sectors. This strategy could include initiatives such as providing training for companies that have not yet become superstar suppliers and developing strategies to help them meet higher industry standards. By enhancing the capabilities of these firms, the government can facilitate their integration into the supply chains of superstar firms, ultimately boosting overall industrial productivity. Furthermore, the training program aims to equip workers with expert skills in technology. It is inevitable that superstar firms may replace repetitive tasks in their plants with machines to maintain consistent production volumes. However, they still require skilled workers to operate advanced technology, which, in turn, necessitates that human resources keep up with technological advancements. Additionally, it is also inevitable that superstar firms are more likely to be involved in the global economic frontier, leading them to provide high-quality intermediate inputs for the downstream industry in the country. However, it is essential to maintain the price remains affordable for the downstream industry due to the fact that superstar firms might impose high markups.

\newpage

\input{1_Manuscript.bbl}
\newpage
\begin{appendices}

\section{Figures and Table}
\renewcommand{\thefigure}{A-\arabic{figure}}
\setcounter{figure}{0}

\input{Stylized_Facts}

\input{Optimal_Profit_Function}

\input{Heterogeneous_Firm_Model}

\input{TFP_Full_Description}

\input{Distribution_of_capital}

\input{TFP_across_different_production_function}

\input{Correlation_of_TFP_and_Share}

\begin{figure}[htpb]
    \centering
        \centering
        \includegraphics[width=0.65\textwidth]{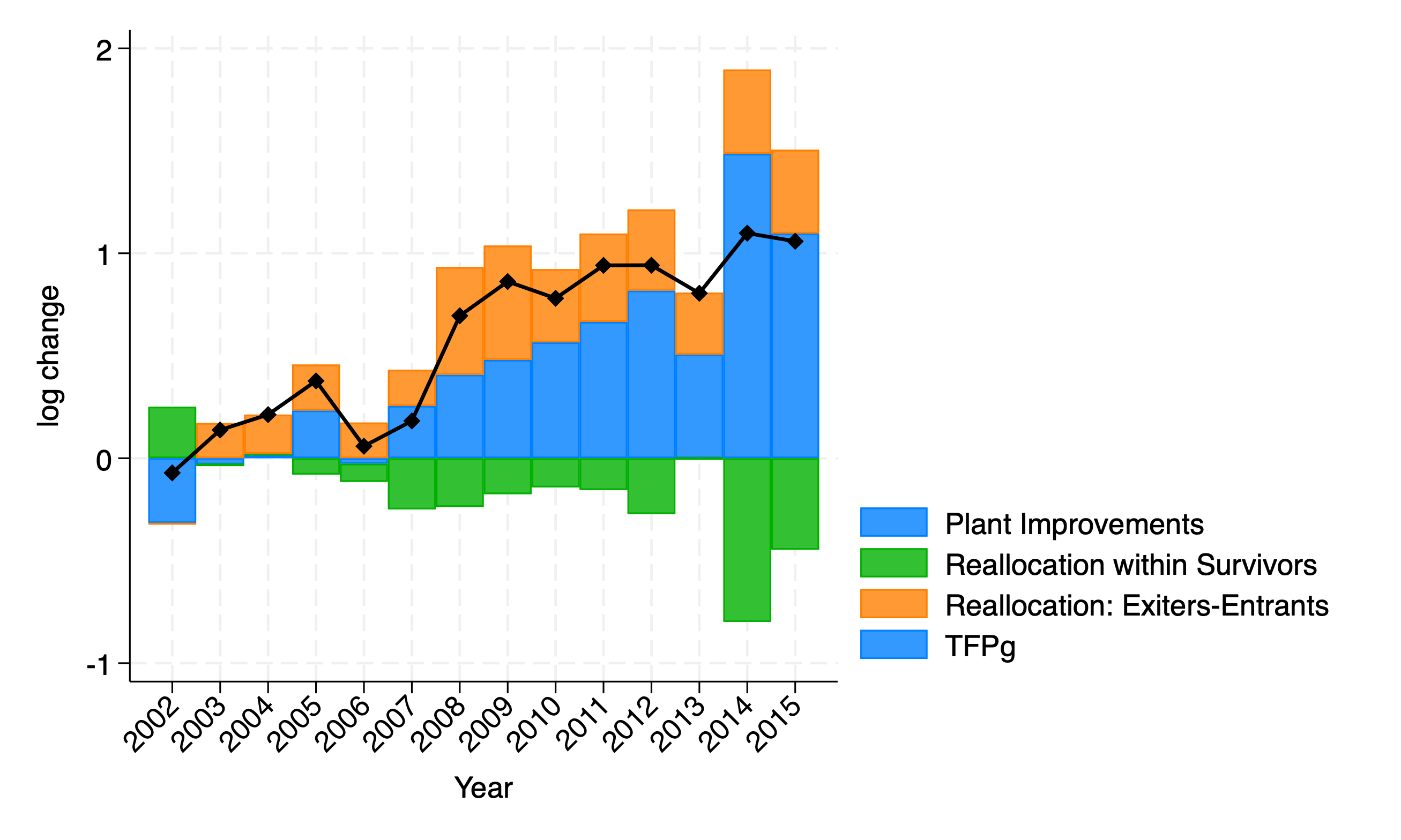}
       \\
        i. Overall \\
        \includegraphics[width=0.65\textwidth]{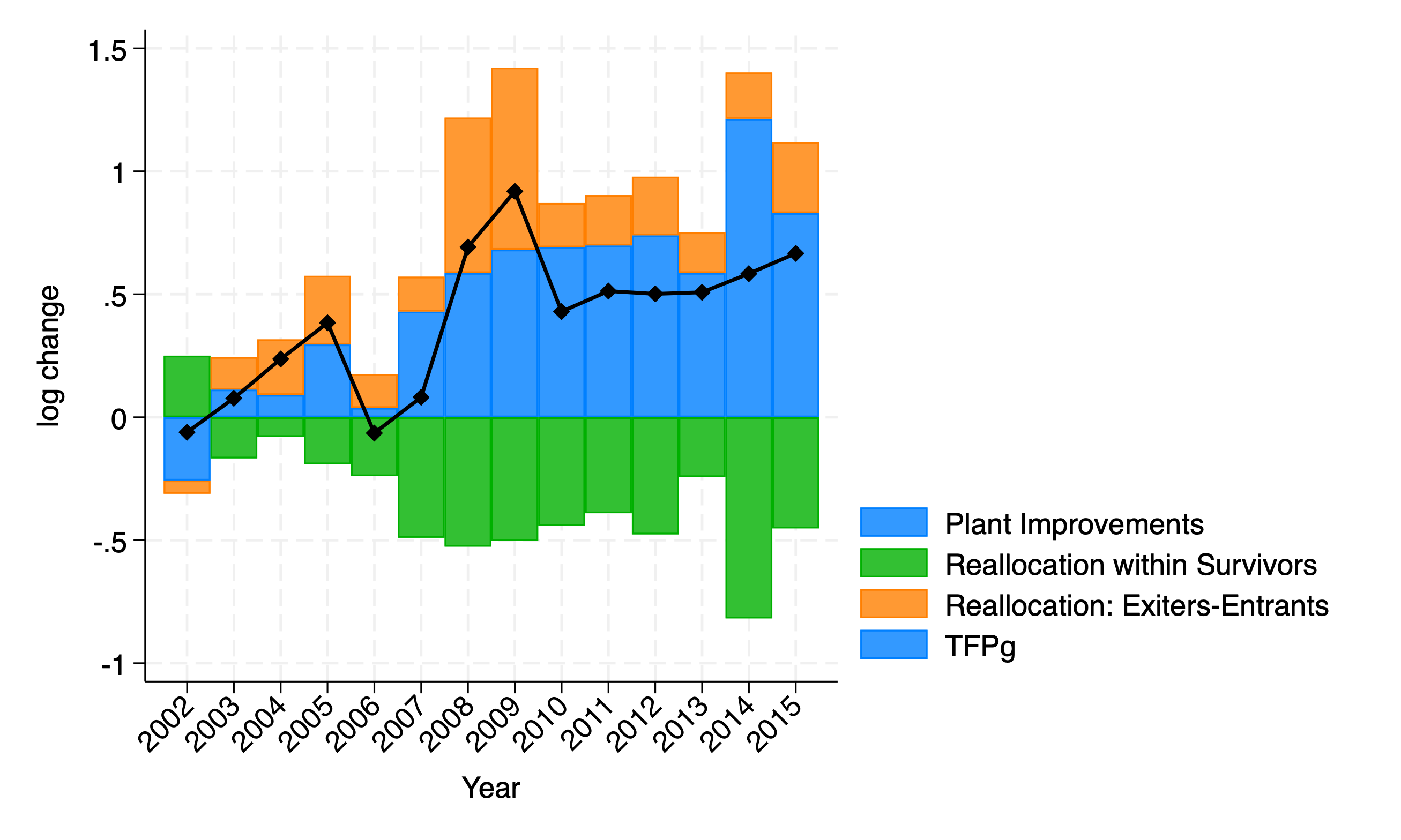}
        \\
        ii. Superstar \\
        \includegraphics[width=0.65\textwidth]{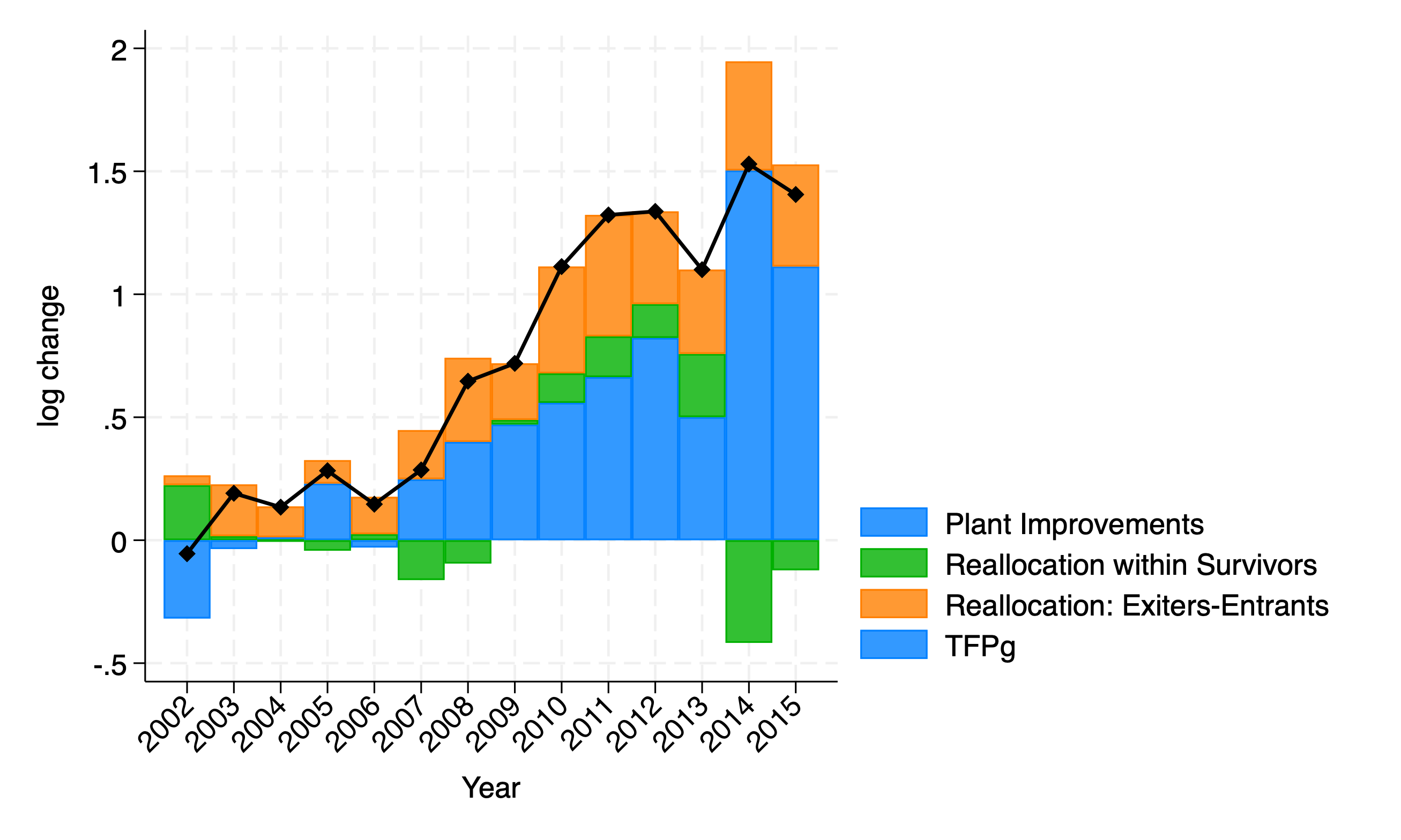}
        \\
        iii. Non-superstar \\
    \caption{Productivity Growth Decomposition}
    \label{DOPD Decomposition}
\end{figure}

\setcounter{table}{0}
\renewcommand{\thetable}{A-\arabic{table}}

    \input{Literature_of_Superstar_Firms}

\input{Descriptive_Statistics}

\input{Main_Horizontal}

\input{Main_Backward}

\input{Main_Forward}
\input{Heterogeneous_Spillovers}

\input{Static_OP}

\input{DOPD}

\input{Elasticity_2_digits}

\input{Elasticity_3_digits_1}

\input{Elasticity_3_digits_2}

\input{First_Stage_Horizontal}

\input{First_Stage_Backward}

\input{First_Stage_Forward}

\input{Main_Horizontal_IPW}

\input{Main_Backward_IPW}

\input{Main_Forward_IPW}

\input{First_Stage_Horizontal_with_IPW}

\input{First_Stage_Backward_with_IPW}

\input{First_Stage_Forward_with_IPW}

\input{Heterogeneous_Spillovers_FS}

\input{Robustness_Split_Large_Medium}

\input{Robustness_Horizontal_-_IV}
\input{Robustness_Backward_-_IV}

\input{Robustness_Forward_-_IV}

\input{Robustness_Heterogeneous_-_Simple_Productivity}

\input{DOPD_Robustness_Amiti}

\end{appendices}

\end{document}

%% file: Stylized_Facts.tex
Stylized Facts of Indonesian Economy

\begin{figure}[H]
    \centering
        \includegraphics[width=0.6\linewidth]{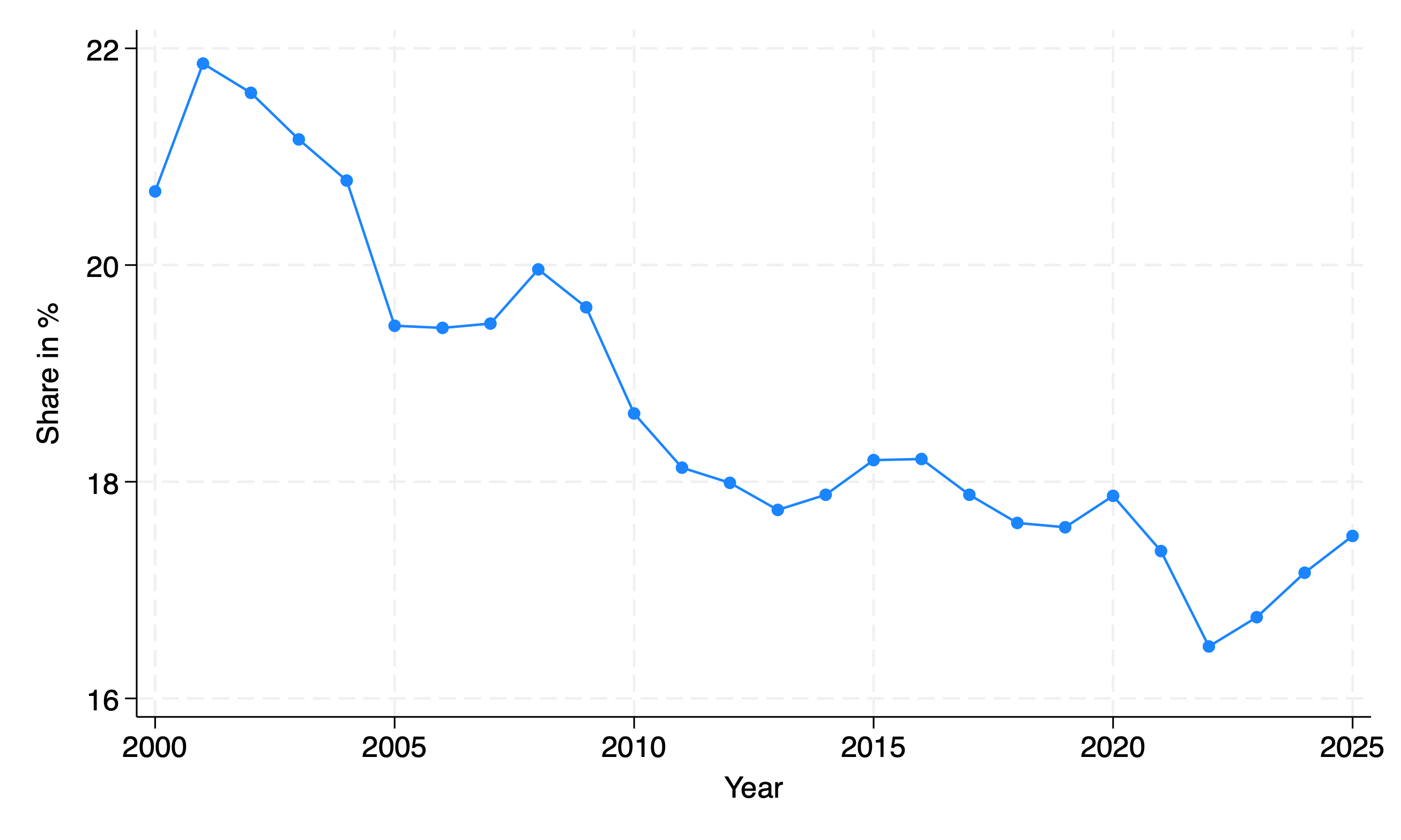}
    \caption{Share of Manufacturing Sector to GDP}
    \label{Share Manufacture}
    \tiny
    Source: BPS-Statisctis
\end{figure}

\begin{figure}[H]
    \centering
        \includegraphics[width=0.6\linewidth]{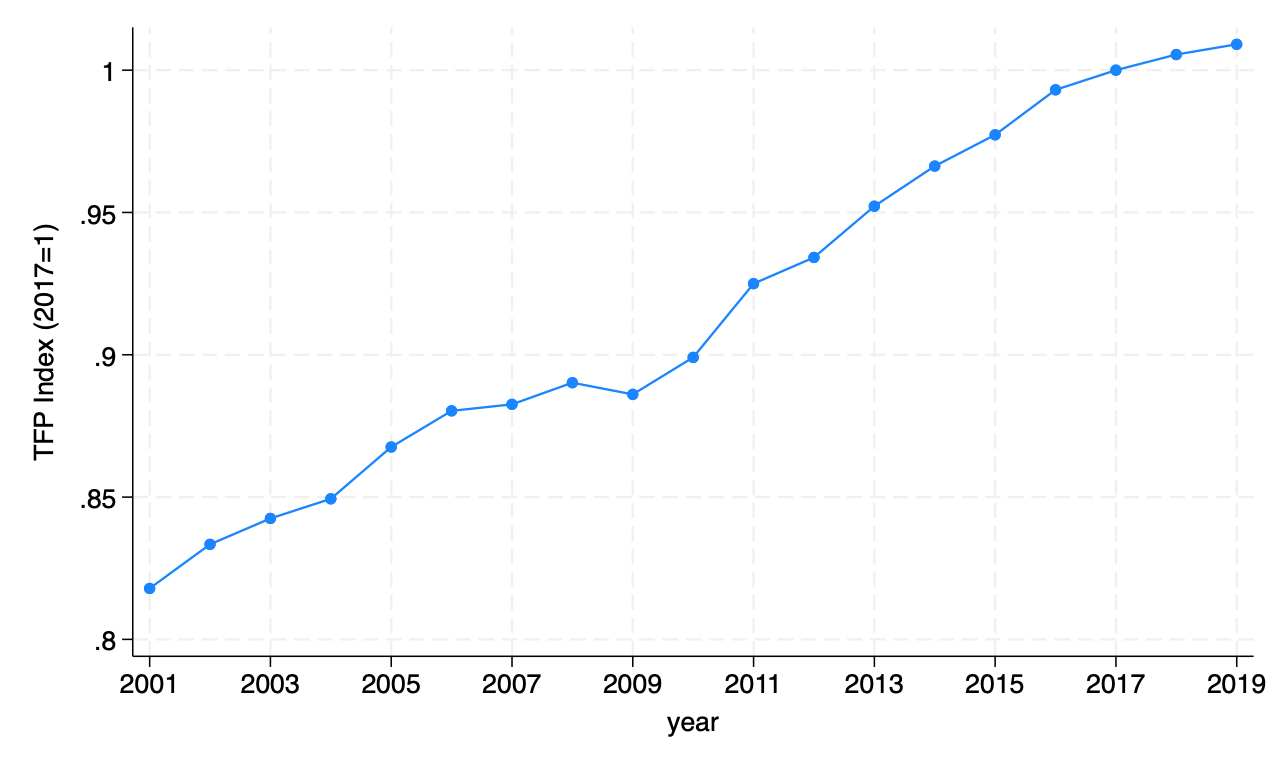}
    \caption{Total Factor Productivity Index Relative to 2017}
    \label{TFP Macro}
    \tiny
    Source: \citet{QT5BCC_2023}
\end{figure}

\begin{figure}[H]
    \centering
        \includegraphics[width=0.6\linewidth]{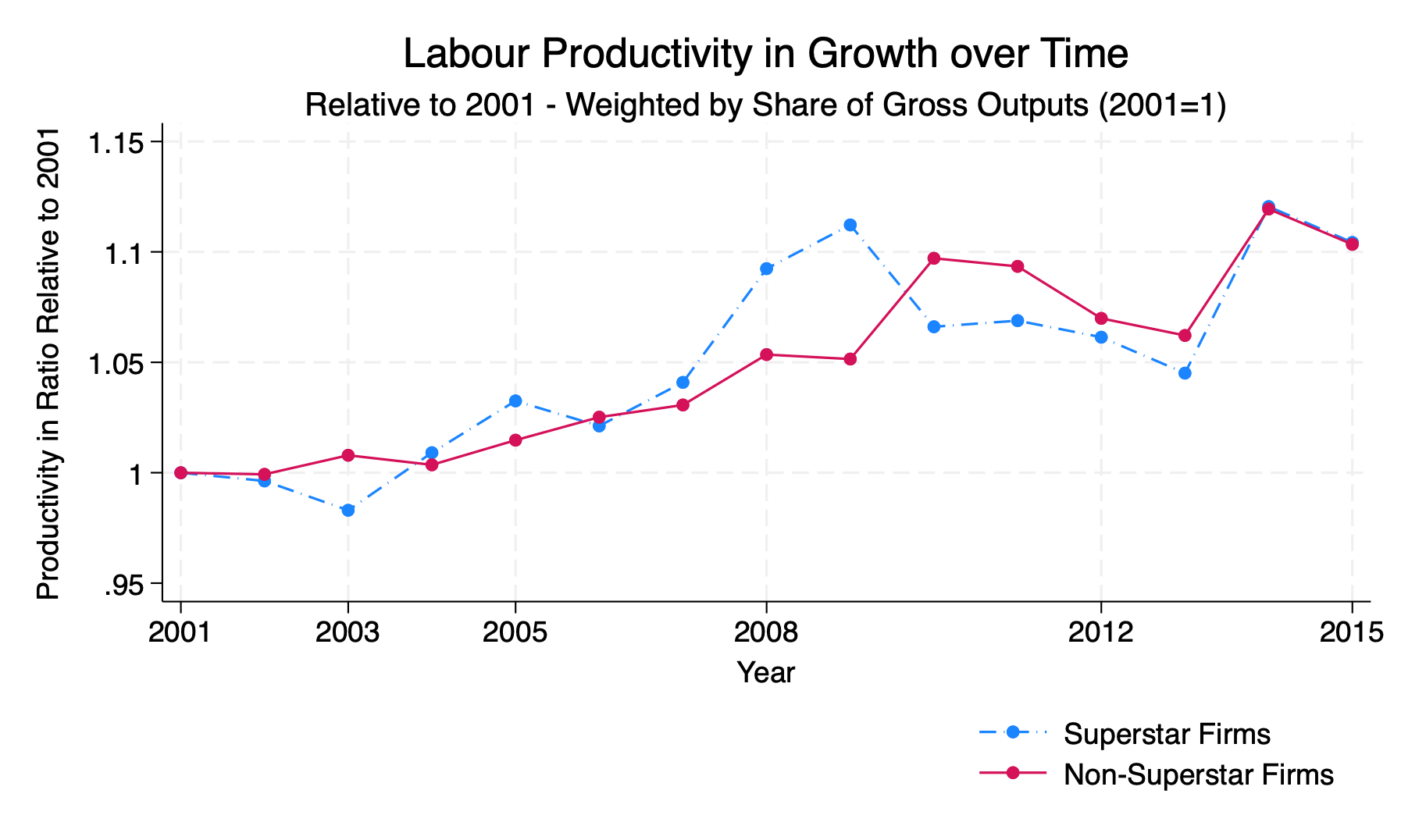}
    \caption{Labour Productivity Growth Relative to 2001}
    \label{Labour Productivity}
    \tiny
    Source: BPS Statistics Indonesia
\end{figure}

\begin{figure}[H]
    \centering
        \includegraphics[width=0.6\linewidth]{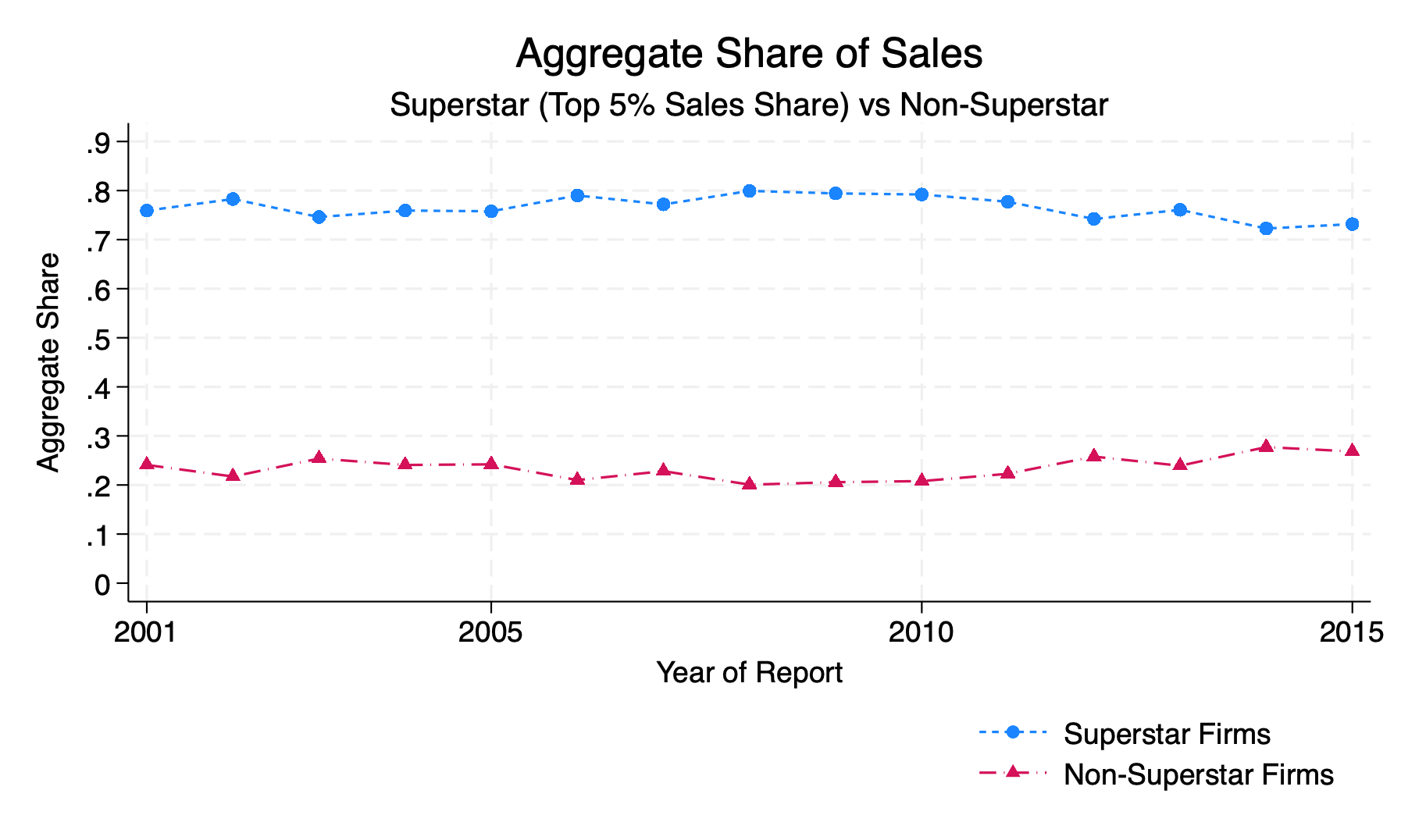}
    \caption{Aggregate Share of Sales: Superstar and Non-Superstar Firms}
    \label{Aggregate Share}
    \tiny
    Source: BPS Statistics Indonesia
\end{figure}
\begin{figure}[H]
        \centering
        \includegraphics[width=0.7\linewidth]{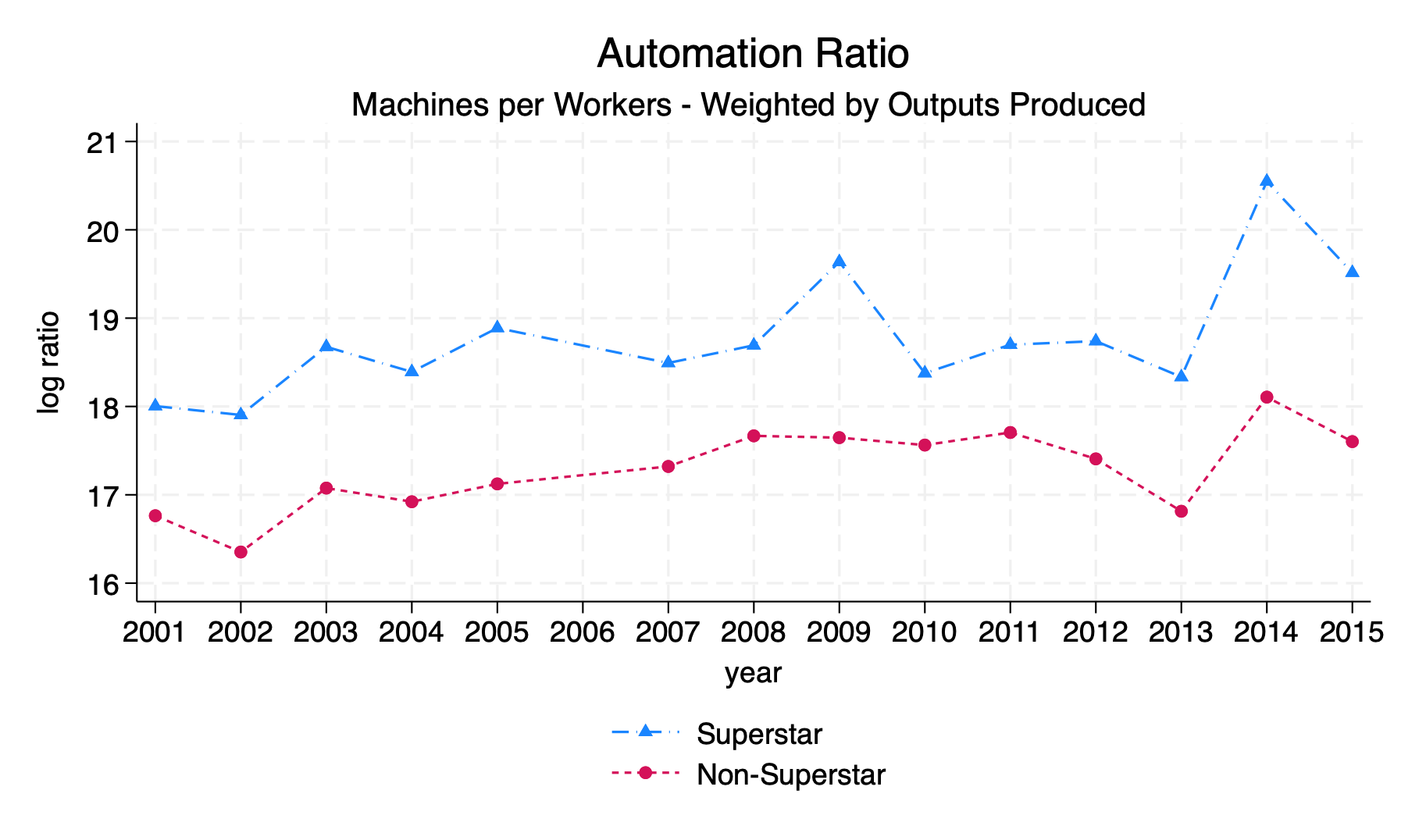}
    \caption{Automation between Superstar and Non-superstar}
    \label{automation_general}
    \tiny
    Source: BPS-Statistics
\end{figure}

\begin{figure}[H]
    \centering
        \includegraphics[width=0.49\linewidth]{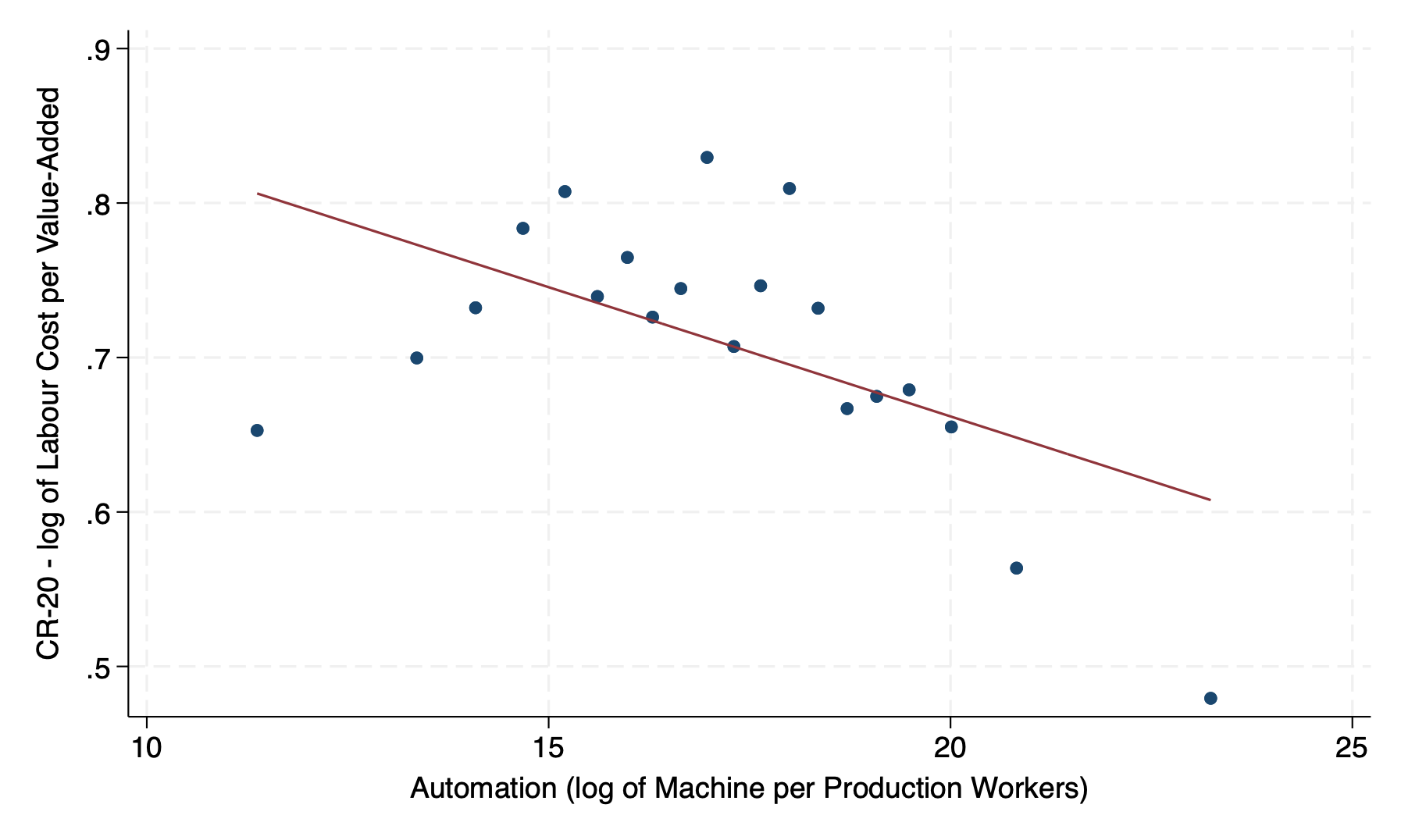}
             \includegraphics[width=0.48\linewidth]{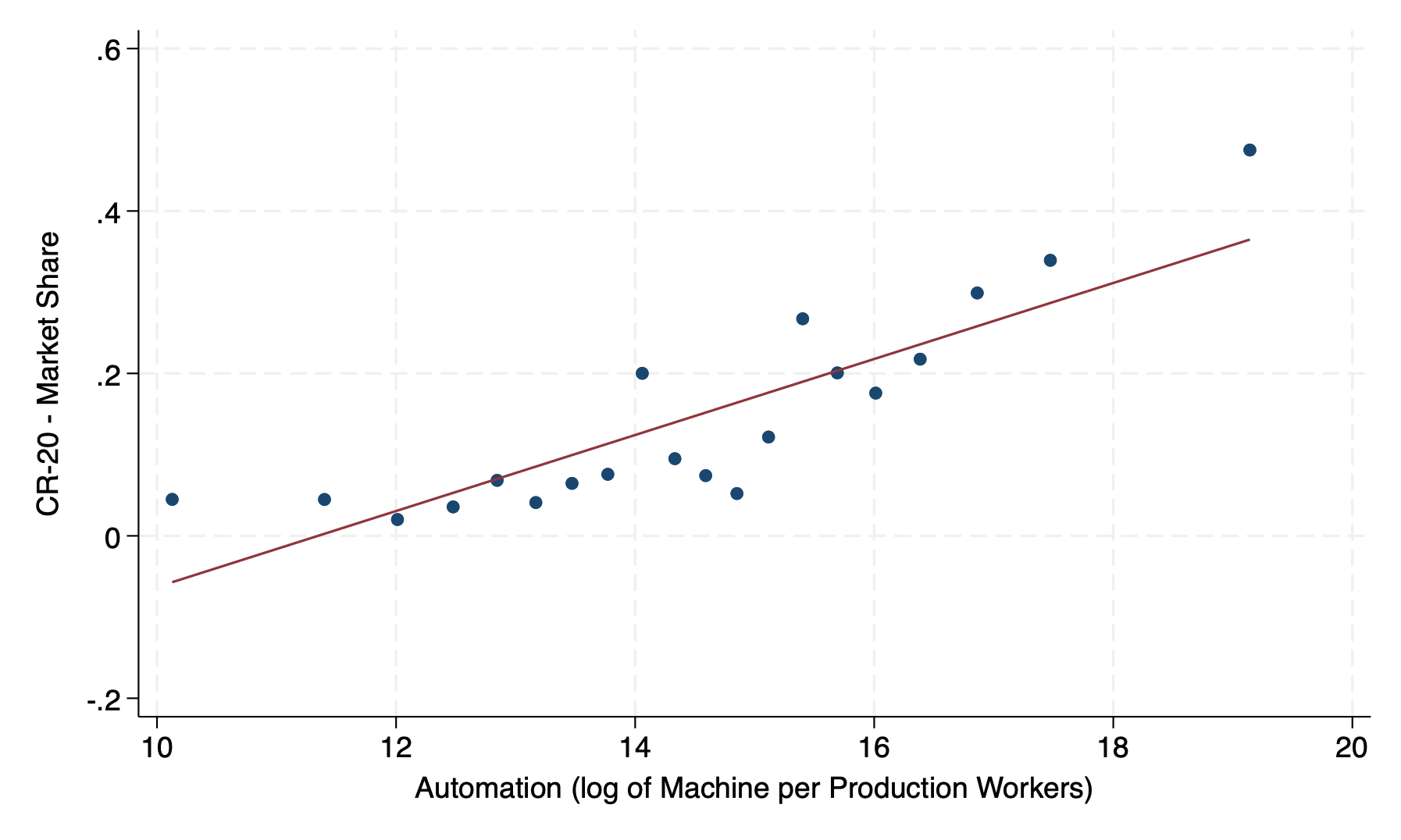}
    \caption{Correlation between Automation and Labour and Market Share}
    \label{automation and share nexus}
    \tiny
    Source: BPS-Statistics
\end{figure}

\begin{figure}[H]
    \centering
     \includegraphics[width=0.49\textwidth]{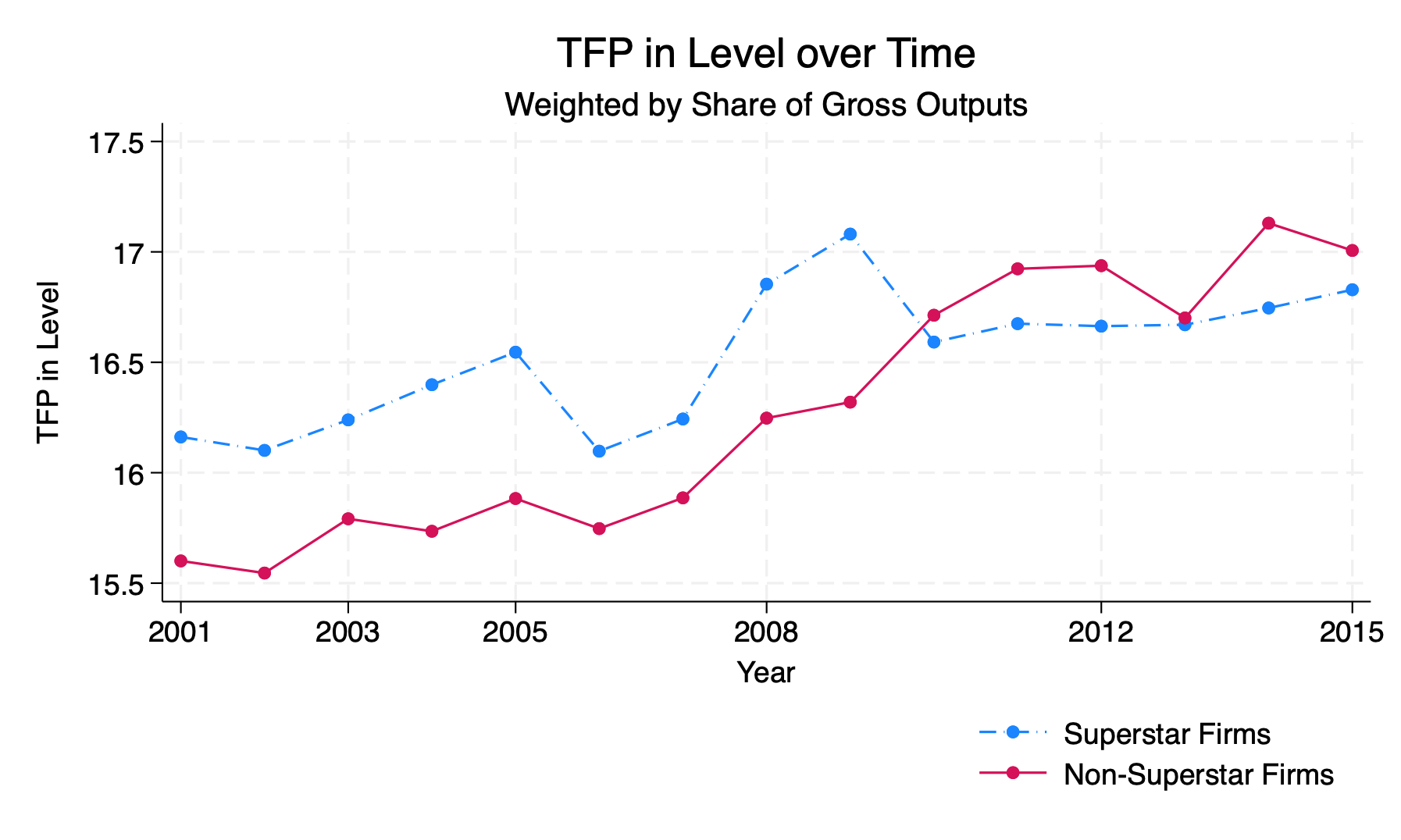}
    \includegraphics[width=0.48\textwidth]{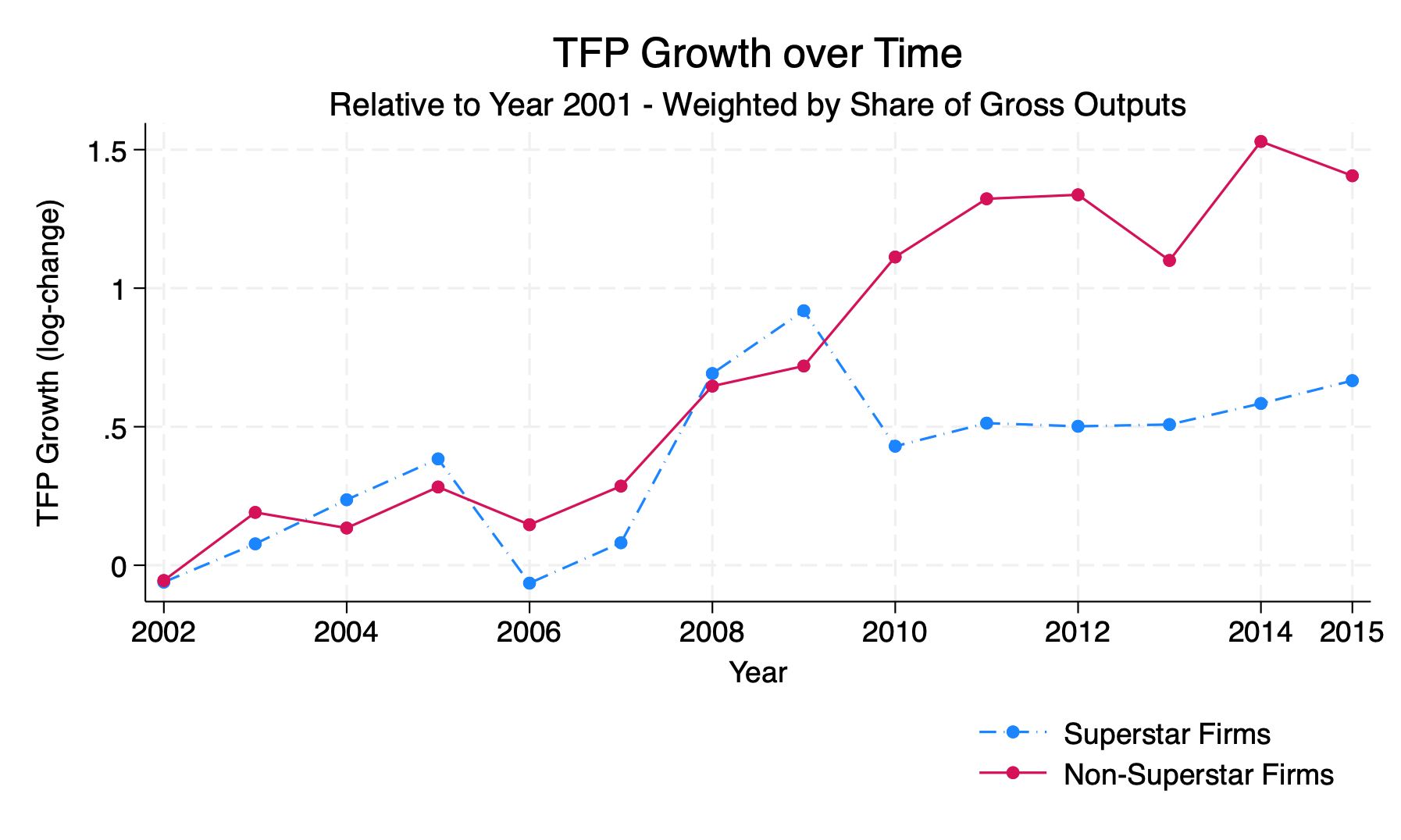}
    \caption{TFP Level and Growth over Time}
    \label{LogofTFovertime}
\end{figure}

\begin{figure}[H]
    \centering
     \includegraphics[width=1\textwidth]{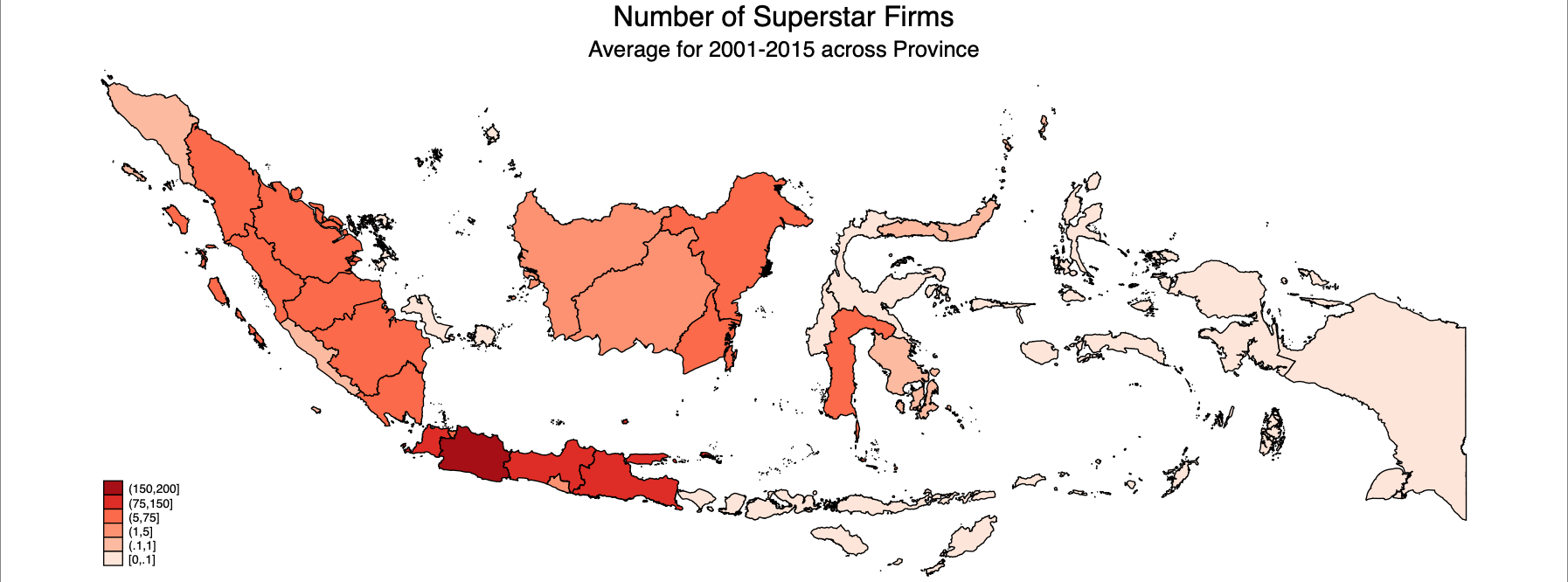}
    \caption{Superstar Distribution}
    \label{Superstar Distribution}
\end{figure}

\begin{figure}[H]
    \centering
     \includegraphics[width=1\textwidth]{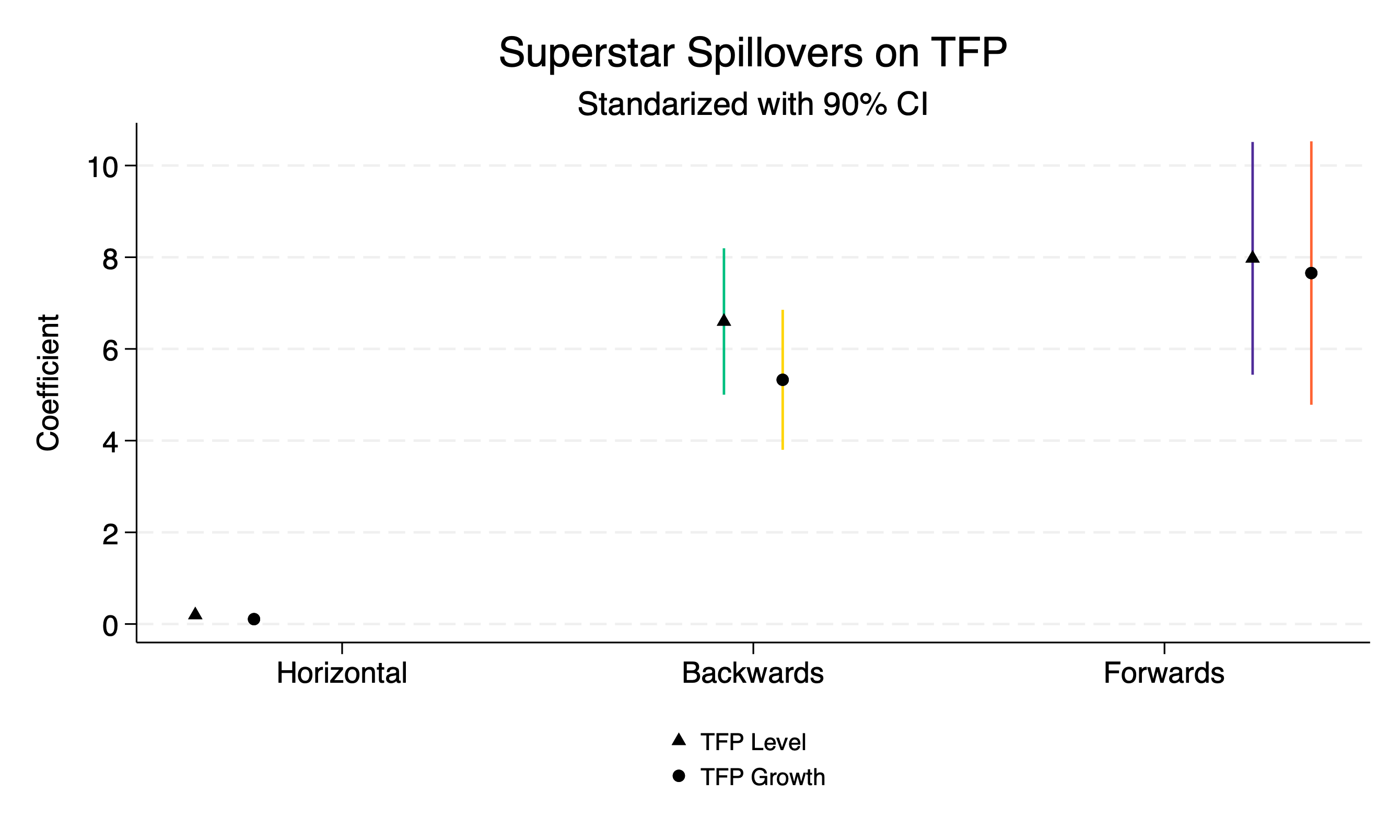}
    \caption{Superstar Spillovers with Standardized Coefficients (90\% CI)}
    \label{Standardized Spillovers Coefficient}
\end{figure}

\begin{figure}[H]
    \centering
     \includegraphics[width=1\textwidth]{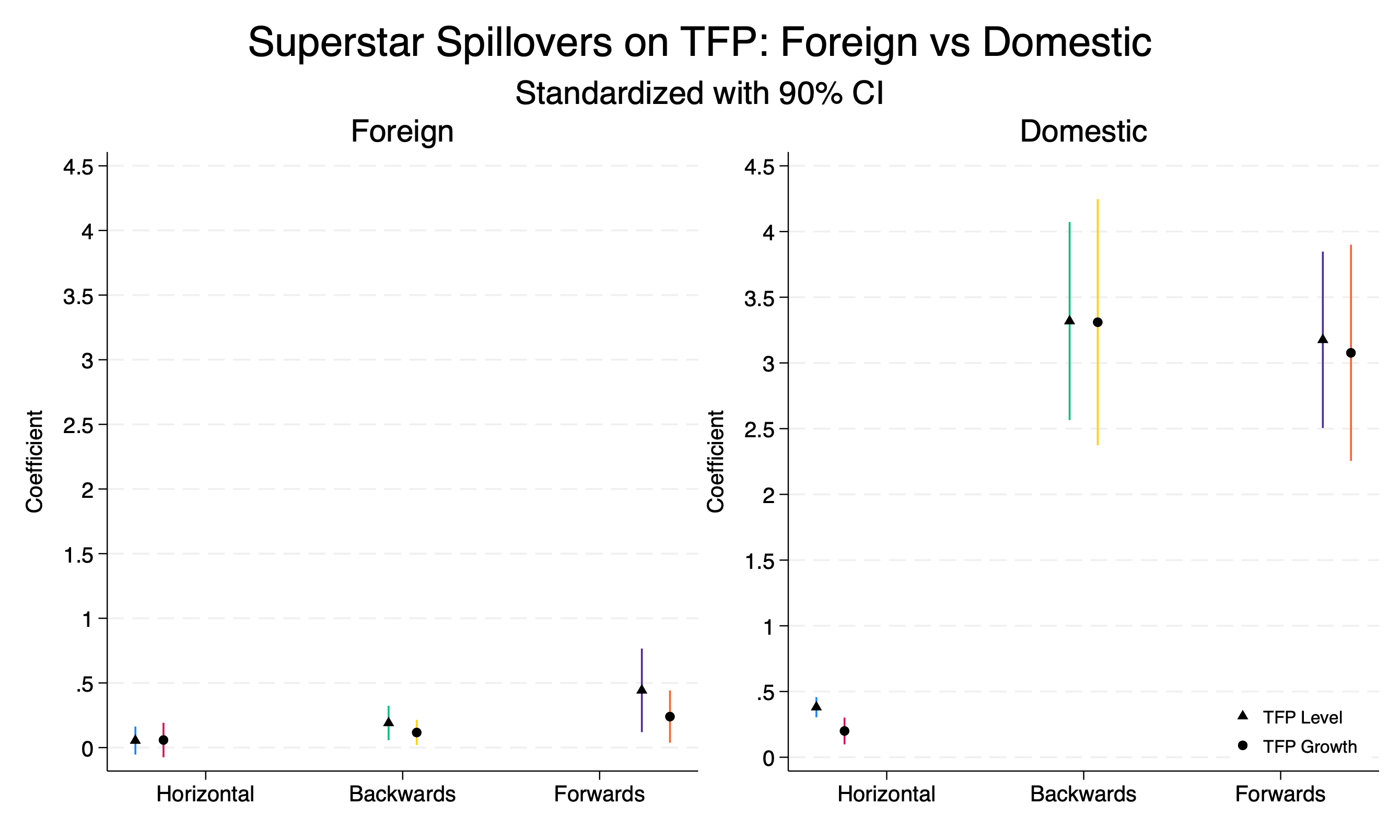}
    \caption{Superstar Spillovers across Heterogeneous Firm}
    \label{Heterogeneous Superstar Spillovers}
\end{figure}

%% file: Optimal_Profit_Function.tex
\section{Optimal Profit Function}
\begin{equation}
\begin{split}
 &  \pi^*=\Big(\frac{w}{\rho \varphi}\Theta \Big[\frac{w}{\rho \varphi}\Big ]^{1/\rho  -1}\Big )-\Big(f+w\frac{\Theta \Big[\frac{w}{\rho \varphi}\Big ]^{1/\rho  -1}}{\varphi}\Big) \\
 & \textcolor{white}{\pi^*}=\frac{w}{\rho \varphi}\Theta \Big[\frac{w}{\rho \varphi}\Big ]^{1/\rho  -1}-f-\frac{w}{\varphi}\Theta \Big[\frac{w}{\rho \varphi}\Big ]^{1/\rho  -1} \\
 & \textcolor{white}{\pi^*}=\frac{w}{\varphi}\Theta \Big[\frac{w}{\rho \varphi}\Big ]^{1/\rho  -1}\Big (\frac{1}{\rho}-1\Big ) - f  \\
 & \textcolor{white}{\pi^*}= \frac{w}{\varphi}\Theta \Big[\frac{w}{\rho \varphi}\Big ]^{1/\rho  -1}\Big (\frac{1-\rho}{\rho}\Big ) - f  \\
 & \textcolor{white}{\pi^*}= \frac{w^{\rho/\rho-1}}{\varphi^{\rho/\rho-1} \rho^{\rho/\rho-1}} \Theta (1-\rho)-f \\
 & \textcolor{white}{\pi^*}= (1-\rho)w^{\rho/\rho-1}\rho^{\rho/1-\rho}\varphi^{\rho/1-\rho}\Theta-f \\
 & \textcolor{white}{\pi^*}= (1-\rho)w^{\rho/\rho-1}\rho^{\rho/1-\rho}\lambda^{\rho/1-\rho}e^{\rho\alpha\gamma/1-\rho}c^{\rho/1-\rho}\Theta-f
\end{split}
\label{ProfitOptimalFunction_full}
\end{equation}

%% file: Heterogeneous_Firm_Model.tex
\section{Heterogeneous Firm Model Theory} 

\subsection{Demand}
Given preferences of a consumer from Constant Elasticity Substitution (CES) utility function over a continuum of goods denoted by $\omega$:
\begin{equation}
    U=\Bigg [ \int_{\omega \in \Omega} q(\omega)^\rho d \omega \Bigg]^{\frac{1}{\rho}}
\end{equation}

\noindent where $\Omega$ denotes the set of available goods in the market, $q(\omega)$ denotes the number of varieties $\omega$, $\rho$ denotes the substitution parameter between varieties where $0<\rho<1$ and related to elasticity substitution $(\sigma=\frac{1}{1-\rho}>1)$. In this case, a higher $\sigma$ associates to the consumer preference on high varieties. This consumer behaviour can then be modeled in aggregate good $Q \equiv U$ correlated with aggregate price as in \citet{dixit1977monopolistic}:

\begin{equation}
    P=\Bigg [ \int_{\omega \in \Omega} p(\omega)^{1-\sigma} d \omega \Bigg ]^{\frac{1}{1-\sigma}}
\end{equation}

\noindent where $P$ is the aggregate price. We then can use this aggregate to derive the optimal consumption and expenditure decision for individual varieties through:

\begin{equation}
    q(\omega)=Q\Bigg[\frac{p(\omega)}{P} \Bigg]^{-\sigma} 
    \label{Optimal Consumption}
\end{equation}

\begin{equation}
    r(\omega)=R\Bigg[\frac{p(\omega)}{P} \Bigg]^{1-\sigma} 
    \label{Expenditure Decision}
\end{equation}

\noindent where $q(\omega)$ is the optimal consumption and $r(\omega)$ denotes the expenditure decision so $R=PQ=\int_{\omega \in \Omega} r(\omega) d\omega$ is the aggregate expenditure.

\subsection{Production}

This section explains how the productivity determines price, output, revenue, and profit in the model. Given a continuum firms choosing individually to produce a different variety of $\omega$, production requires only one factor, labour, assumed in-elastically supplied at its aggregate level $L$ capturing the size of the economy. Meanwhile, technology is assumed as fixed and is represented by marginal cost with a fixed overhead cost. Then, the linear function of output is obtained as a function of labour employed: $q:l=f+\frac{q}{\varphi}$, where $\varphi$ is the firm-heterogeneous productivity levels and $f$ is the strictly positive fixed-cost. Each firm faces a residual demand curve with constant elasticity $\sigma$ and thus chooses the indifferent profit maximizing markup $\sigma/(\sigma-1)=1/\rho$ which then yields the pricing rules as follows:

\begin{equation}
    p(\varphi)=\frac{w}{\rho \varphi}
    \label{Price Function}
\end{equation}

where $w$ denotes wage rate normalized to one. Hence, we may obtain the firm profit function as:

\begin{equation}
    \pi(\varphi)=r(\varphi)-l(\varphi)=r(\varphi)-f-\frac{q}{\varphi}
    \label{Profit Function Simple}
\end{equation}

As in \citet{dixit1977monopolistic}, variable profit also incorporates to \ref{Profit Function Simple} and is defined as the revenue fraction that is not used for variable cost, i.e. $\frac{r(\varphi)}{\sigma}$ where $\frac{1}{\sigma}$ is the fraction of revenue to offset $f$, hence we may obtain another profit function as:

\begin{equation}
    \pi(\varphi)=\frac{r(\varphi)}{\sigma}-f
    \label{Profit Function Dixit}
\end{equation}

Moreover, $r(\varphi)$ and $\pi(\varphi)$ also depend on the aggregate rice and revenue according to \ref{Optimal Consumption} and \ref{Expenditure Decision}, hence we may obtain:

\begin{equation}
\begin{split}
       & r(\omega)=R\Bigg[\frac{p(\omega)}{P} \Bigg]^{1-\sigma} \\\
       & \textcolor{white}{r(\omega)}=R\Bigg[\frac{\frac{w}{\rho \varphi}}{P} \Bigg]^{1-\sigma} \\\
       & \textcolor{white}{r(\omega)}=R\Bigg[\frac{\frac{1}{\rho \varphi}}{P} \Bigg]^{1-\sigma} \\\
       & \textcolor{white}{r(\omega)}=R\Bigg[{\rho \varphi}P \Bigg]^{\sigma-1} \\\
\end{split}
    \label{Expenditure Decision with Productivity}
\end{equation}

\begin{equation}
\begin{split}
       & \pi(\varphi)=\frac{r(\varphi)}{\sigma}-f\\\
       & \textcolor{white}{\pi(\varphi)}=\frac{R}{\sigma}\Bigg[{\rho \varphi}P \Bigg]^{\sigma-1} -f\\\
\end{split}
    \label{Profit with Productivity}
\end{equation}

According to equation \ref{Expenditure Decision with Productivity}, \ref{Price Function}, \ref{Profit with Productivity}, we may conclude that a higher $\varphi$, that means a more productive firm, will associate to a larger size, i.e. larger revenue and output, impose a lower price, and is more profitable than less productive firm.

\subsection{Firm Entry and Exit}
In the dynamic setting of heterogenous firm model, firms should make initial sunk investment cost to enter the markets, i.e. $f_e>0$ denoting a fixed-entry cost. Firms then take into account their initial productivity parameter $\varphi$ from a common distribution $g(\varphi)\in (0,\infty)$ with its continuous cumulative distribution $G(\varphi)$. In the time at which a firm enters the market, it may decide to produce or not produce. If it does, it belongs to a constant probability of shock enforcing them to exit, denoted by $\delta$. Under the assumption of time-invariant productivity level, its optimal per period profit level also remains constant, by excluding $f_e$. An entering firm with productivity $\varphi$ will exit if this profit is negative. In contrast, it remains profitable as long as they stay in the market and do not get bad shock to exit. Hence, we may arrange this condition under the profit discounted with probability to survive $(1-\delta)$ or if it is without time discounting, we may obtain as follows:

\begin{equation}
     v(\varphi)=\text{max} \Bigg \{0,\sum^{\infty}_{t=0}(1-\delta)^t \pi(\varphi)\Bigg \} =\text{max} \Bigg \{0,\frac1\delta \pi(\varphi)\Bigg \}
    \label{Firm Value Function}
\end{equation}

\noindent where $\pi(\varphi)$ depends on the $R$ and $P$ based on the equation \ref{Profit with Productivity}. Hence, we may denote $\varphi^*=inf\{\varphi:v(\varphi)>0\}$ as the lowest productivity level of producing firms, known as cut-off level, to ensure the firms to stay in the markets. When a firm with $\varphi<\varphi^*$, this firm will exit immediately and does not produce. The exit process will not affect the equilibrium productivity distribution $\mu(\varphi)$ under the assumption that there is no correlation between subsequent firm exit and productivity. Instead, $\mu{(\varphi)}$ is determined by the initial productivity draw conditional on sucessful entry, as follows:

\begin{equation}
\mu(\varphi) =
\begin{cases}
\frac{g(\varphi)}{1-G(\varphi)} & \text{if } \varphi\geq \varphi^* \\
0  & \text{if otherwise}
\end{cases}
\label{Distribution of Productivity}
\end{equation}

where $1-G(\varphi)$ is the ex-ante probability of successful entry and represents the level under which firm's productivity enables to survive. Equation \ref{Distribution of Productivity} also captures how the initial distribution, i.e. $g(\varphi)$, changes into equilibrium distribution $\mu(\varphi)$ by solely existing firms in the markets with $\varphi>\varphi^*$. We then can obtain aggregate productivity level, $\tilde{\varphi}$, as follows:

\begin{equation}
     \tilde{\varphi}(\varphi)=\Bigg [\frac{1}{1-G(\varphi^*)} \int^{\infty}_{\varphi^*} \varphi^{1-\sigma} g(\varphi)d\varphi \Bigg ]^{\frac{1}{\sigma-1}}
    \label{Aggregate Productivity Level}
\end{equation}

where equation \ref{Aggregate Productivity Level} defines the average survivors productivity in the equilibrium.

%% file: TFP_Full_Description.tex
\section{Total Factor Productivity from Olley-Pakes}
In the \citet{olley1996dynamics}, we may arrange the standard Cobb-Douglas production function for a panel setting as follows:

\begin{equation}
    y_{it}=\beta_0 + \beta_X X_{it}  + \varepsilon_{it}
    \label{ProductionFunctionFirst}
\end{equation}

Where $y_{it}$ denotes total outputs, either gross outputs or value-added, for firm $i$ in year $t$. $X_{it}$ denotes set of adjustable and dynamic inputs over time. These variables are transformed into a log form. Meanwhile, $\varepsilon_{it}$ denotes error terms capturing Hicks neutral productivity shocks. Under Ordinary Least Squared (OLS) properties, we impose a strict exogeneity assumption, namely $E(\varepsilon|X)=0$ and $cov(\varepsilon_i,X_i)=0$ . However, there is simultaneity problem when contemporaneous correlation between inputs and shocks occurred ($E(\varepsilon|X) \neq 0$ and $cov(\varepsilon_i,X_i)\neq0$) in a way that the decision to allocate inputs ($X_{it}$), such as for labour and capital, are determined by shocks captured by $\varepsilon_{it}$, which in turn causes endogeneity \citep{levinsohn2003estimating}. We demonstrate how this simultaneity problem occurred. First, we may arrange a production function with two input settings, namely labour and capital, as follows:

\begin{equation}
    y_{it}=\beta_0 + \beta_l l_{it} + \beta_k k_{it} + \varepsilon_{it}
    \label{Cobb-Douglas Standard}
\end{equation}

Where $y_{it}$ denotes the value-added of firm $i$ in year $t$, inputs are divided into a freely variable, namely the number of workers employed ($l_{it}$), and the state variable capital ($k_{it}$) (also known as quasi-fixed input). Freely variable is the variable assumed free from adjustment costs \footnote{Later studies have assumed that labour can also be a non-free variable as there are adjustment costs for training and costs of hiring, see \citet{ackerberg2015identification}}. Meanwhile, the state variable is the dynamic variable prone to adjustment costs. We may obtain estimated parameters of $\beta_l$ for $\hat\beta_l$ and $\beta_k$ for $\hat\beta_k$ from \ref{Cobb-Douglas Standard} from least-squared mechanism as follows:
\begin{equation}
    \begin{split}
        & y_{it}=\beta_0 + \beta_l l_{it} + \beta_k k_{it} + \varepsilon_{it} \\
        & \hat \beta_l= \beta_l + \frac{var(k).cov(l,\varepsilon)-cov(l,k).cov(k,\varepsilon)}{var(l).var(k)-cov(l,k)^2} ,\\
        & \hat \beta_k= \beta_k + \frac{var(l).cov(k,\varepsilon)-cov(l,k).cov(l,\varepsilon)}{var(l).var(k)-cov(l,k)^2} ,\\
    \end{split}
    \label{ProofOLSBias}
\end{equation}

where $var(a)$ and $cov(a,b)$ denote variance and covariance sample between $a$ and $b$. There will be biases if we use OLS based on these properties. First, we know that the denominator in \ref{ProofOLSBias} is always positive since $var(l)>0$ and $var(k)>0$, and the bias comes from the numerator. Suppose only labour reacts to the shocks, for example. In that case, more labour is hired in response to a productivity shock, so $cov(l,\varepsilon)>0$, and capital is not correlated with labour, we then know that $\hat \beta_l$ is biased up, albeit $\hat \beta_k$ is unbiased. However, if labour is correlated with capital and it has a strong association with productivity shocks, $\hat \beta_l$ may capture an overestimated magnitude as most shocks are trapped by labour, while  $\hat \beta_k$ will be underestimated. Some studies tackled this issue by providing instrumental variable (IV) strategies, such as setting up input prices and the lagged value of input uses as instruments. However, input prices at the firm level are rarely observed. Hence, \citet{olley1996dynamics} introduced a novel approach, namely utilizing investment as a proxy of productivity shocks. We now set the production function by disentangling the productivity shocks from $\varepsilon$ as follows:
\begin{equation}
    y_{it}=\beta_0+\beta_l l_{it}+\beta_k k_{it} + \varphi_{it}+\eta_{it}
\end{equation}

where $\varphi_{it}+\eta_{it}=\varepsilon_{it}$ from \ref{Cobb-Douglas Standard}, specifically $\varphi_{it}$ captures productivity shock affecting decision rules, while $\eta_{it}$ has no impact for firm's decision. We then arrange the function of investment as follows:

\begin{equation}
    i_{it}=i_{it}(\varphi_{it},k_{it})
     \label{InvestmentFunction}
\end{equation}

where $i_{it}$ denotes investment of firm $i$ in time $t$. This function stems from the empirical evidence that more productive firms with higher capital are associated with higher investments\footnote{See, for example, \cite{siliverstovs2016r} for research and development (R\&D) investment evidence}. In other words, more positive productivity shocks in the present time imply that more investments will occur later, which ultimately accumulates capital. Under monotonicity between these three variables, we may invert the function into:
\begin{equation}
    y_{it}=\beta_l l + \phi(i_{it}, k_{it})+\eta_{it}
    \label{InvertedOP}
\end{equation}

where $\phi(i_{it}, k_{it})=\beta_0 + \beta_k k_{it}+\varphi_{it} (i_{it}, k_{it})$. The equation \ref{InvertedOP} can be estimated using various methods, such as polynomial degree $n-$th or instrumental variables\footnote{See, for example, \citet{olley1996dynamics} for fourth-degree polynomial and \citet{loecker2012markups} for Generalized Methods of Moments (GMM) estimates.}. We then set up \ref{InvertedOP} in terms of conditional expectation as follows:

\begin{equation}
    E(y_{it}|i_{it}, k_{it})=\beta_lE(l_{it}|i_{it},k_{it})+\phi_{it}(i_{it},k_{it})
    \label{InvertedOP_expectation}
\end{equation}

Since $\eta_{it}$ is uncorrelated with $l_{it}$ and $k_{it}$, we can subtract \ref{InvertedOP_expectation} from \ref{InvertedOP} yielding:

\begin{equation}
    y_{it}-E(y_{it}|i_{it}, k_{it})=\beta_l\Big (l_{it}-E(l_{it}|i_{it},k_{it})\Big)+\eta_{it}
    \label{InvertedOP_expectation2}
\end{equation}

Under the strict exogeneity assumption, again, $\eta_{it}$ is mean independent of $l_{it}$, so we may use the OLS strategy without an intercept to estimate $\beta_l$. Some studies define this process as the first stage estimate, such as \citep{ackerberg2015identification}. 

Meanwhile, in the second step, we need to identify the coefficient of capital ($\beta_k$) in more comprehensive ways. Since capital affects productivity twice, i.e., from the investment channel causing capital accumulation and its own channel, a more complete model is needed \citep{levinsohn2003estimating}. Some studies assume that productivity follows a first-order Markov process minus its change from last period ($\xi_{it}$), i.e. $E(\varphi_{it}|\varphi_{it-1})=\varphi_{it}-\xi_{it}$, while at the same time capital does not respond directly to the change of it. Hence, we may arrange the output net of labour's contribution ($y*$) as:
\begin{equation}
    y^*_{it}=y_{it}-\beta_l l_{it}=\beta_0+\beta_k k_{it}+\underbrace{E(\varphi_{it}|\varphi_{it-1})+\xi_{it}}_{\varphi_{it}}+\eta_{it}
\end{equation}

Hence, we may obtain consistent $\beta_k$ by regressing $y^*_{it}$ on $k_{it}$ under assumption of orthogonality of $\xi_{it}$ and $\eta_{it}$ on $k_{it}$. This approach might be more applicable under limited information of price for IV strategy.

%% file: Distribution_of_capital.tex
\begin{figure}[H]
    \centering
        \centering
        \includegraphics[width=0.9\textwidth]{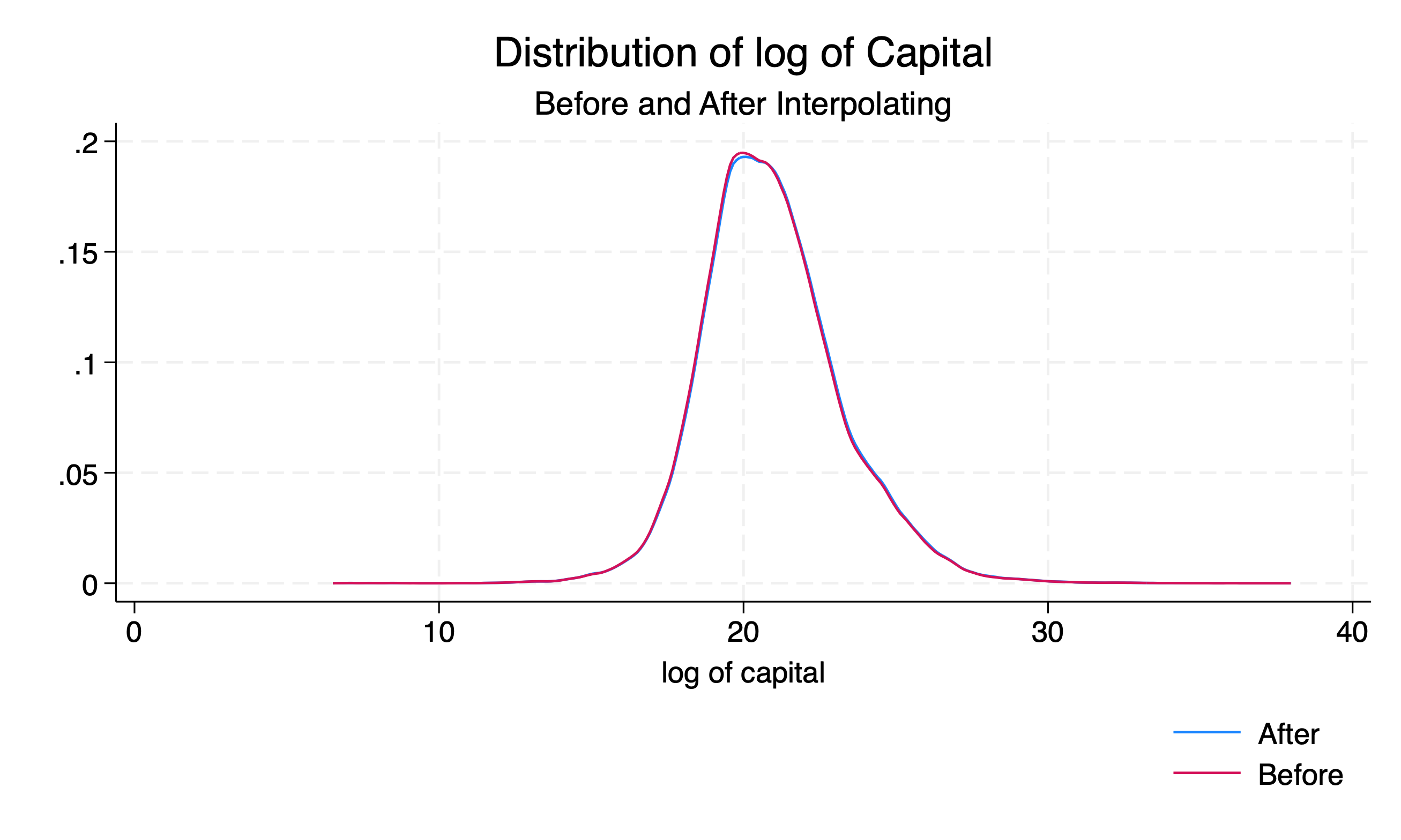}
    \caption{Distribution of Log of Capital: Before and After Interpolation}
    \label{Capital Before-After Interpol}
\end{figure}

%% file: TFP_across_different_production_function.tex
\begin{figure}[H]
    \centering
        \centering
        \includegraphics[width=0.9\textwidth]{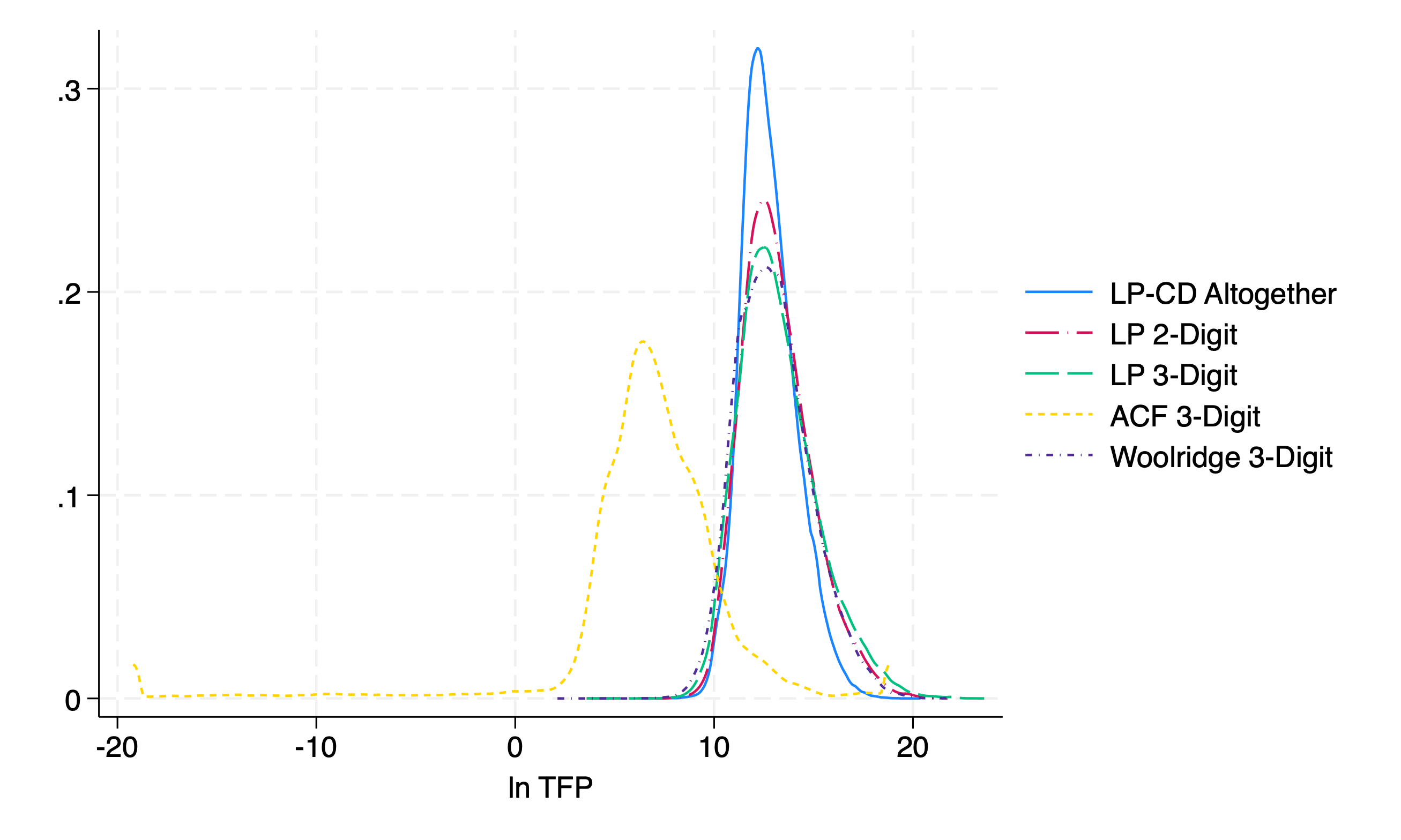}
         \includegraphics[width=0.9\textwidth]{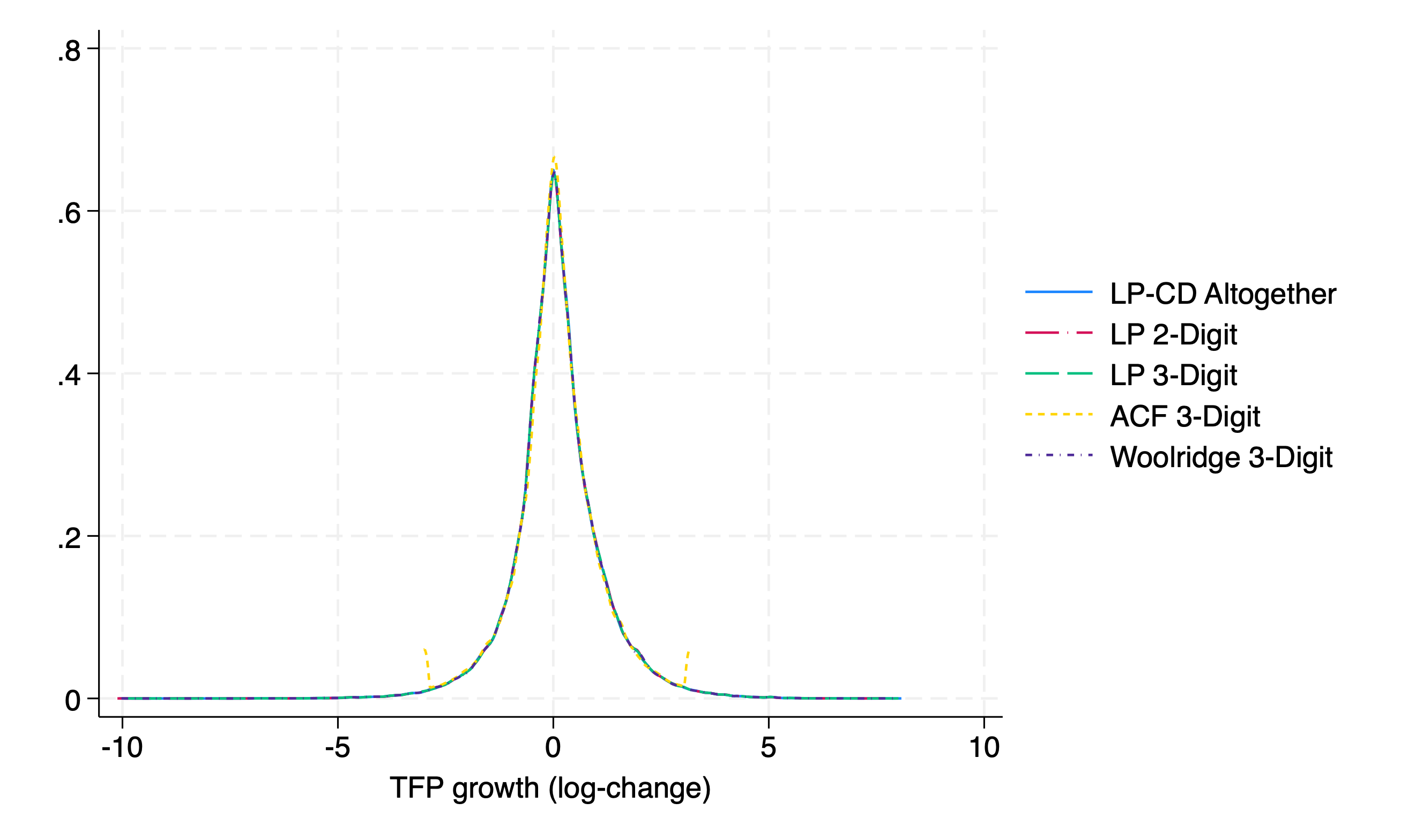}
    \caption{TFP level and Growth across Methodologies}
    \label{TFP level and Growth across method}
\end{figure}

%% file: Correlation_of_TFP_and_Share.tex
\begin{figure}[htpb]
    \centering
        \centering
        \includegraphics[width=0.55\textwidth]{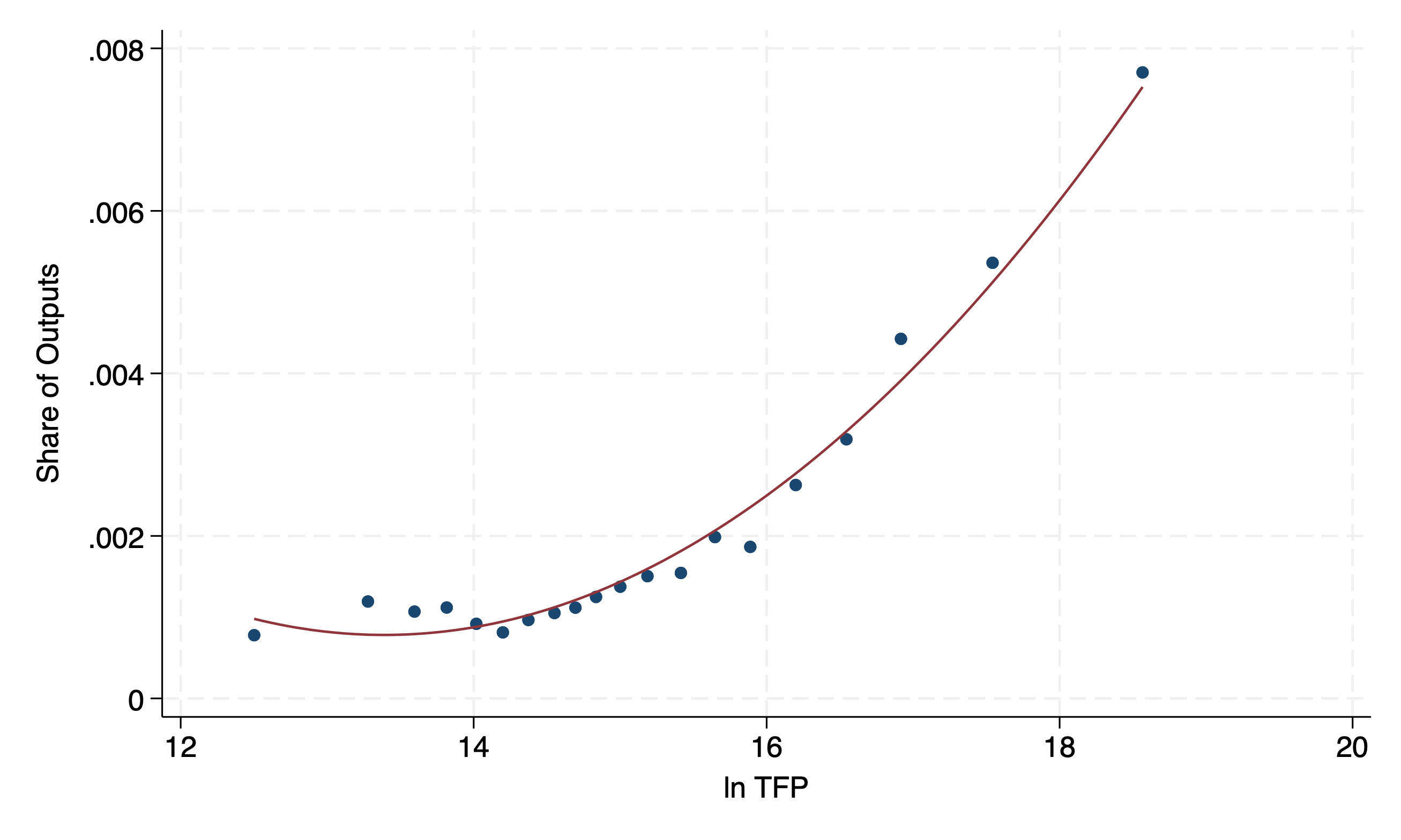}
       \\
        i. 2001-2005 \\
        \includegraphics[width=0.55\textwidth]{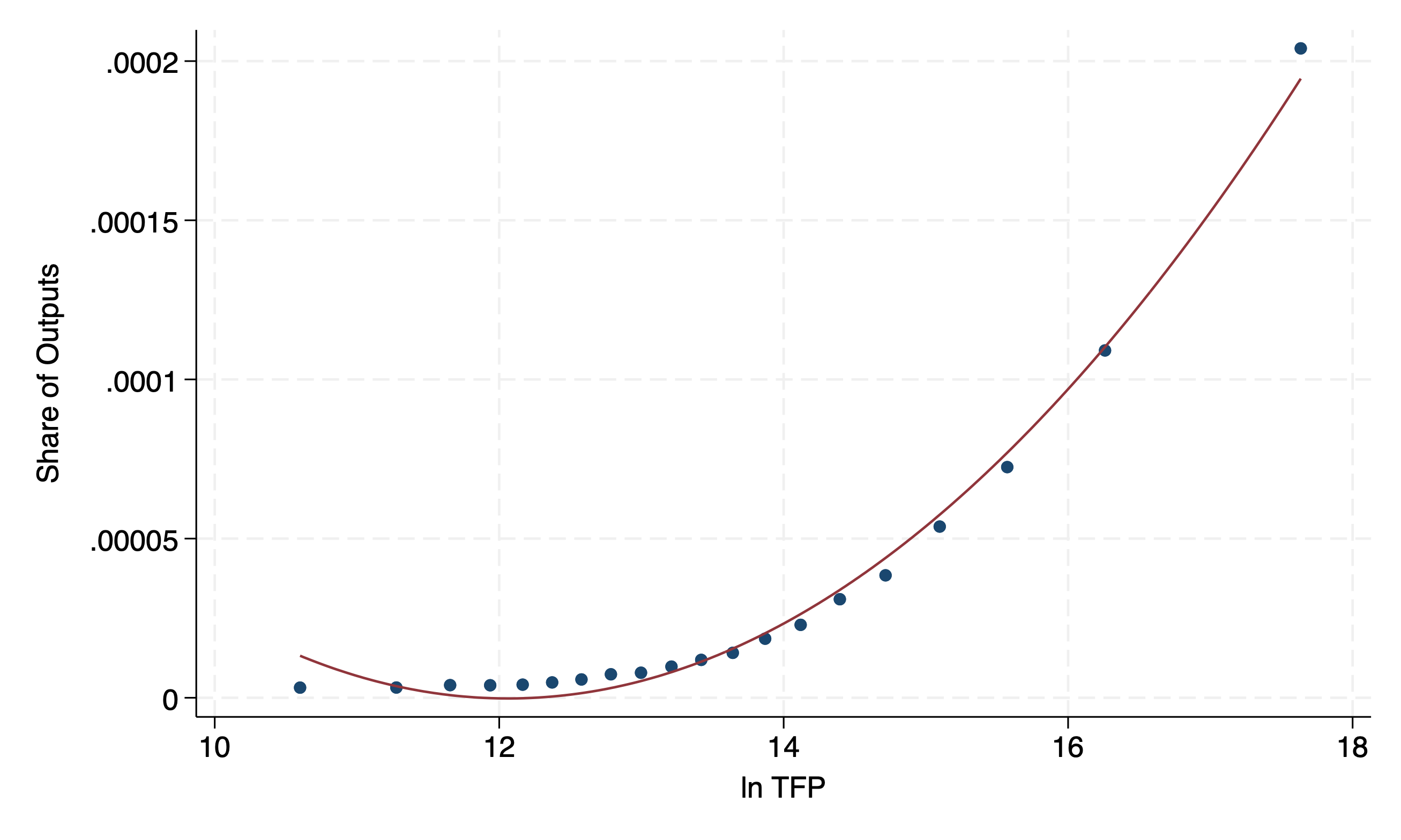}
        \\
        ii. 2006-2010 \\
        \includegraphics[width=0.55\textwidth]{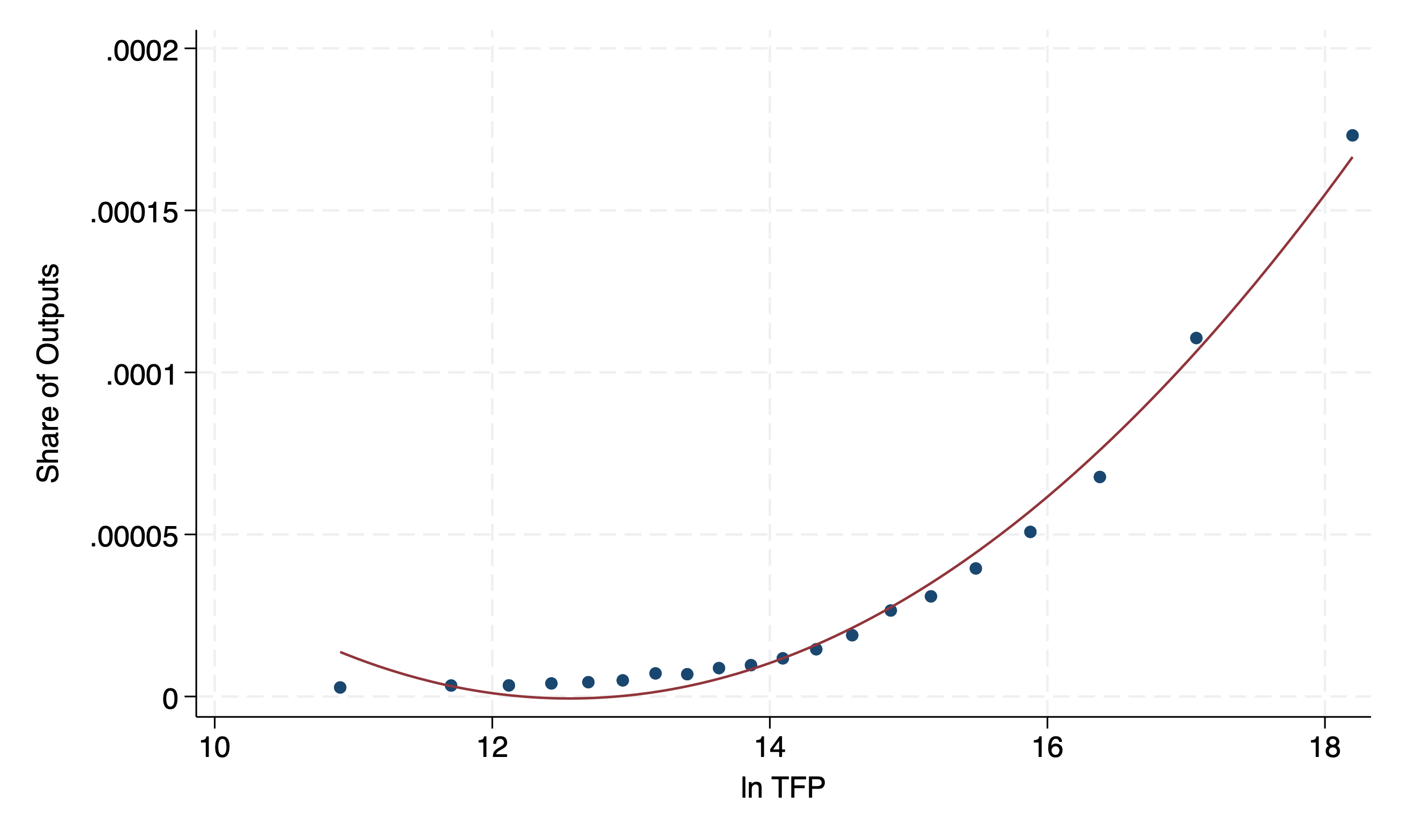}
        \\
        iii. 2011-2015 \\
    \caption{Correlation between TFP and Share of Outputs}
    \label{CorrelationProof}
\end{figure}

%% file: Literature_of_Superstar_Firms.tex
\begin{table}[htbp]
\centering
\caption{Summary of Studies on Superstar Firms}
\resizebox{\textwidth}{!}{
\begin{tabular}{p{3cm} p{4cm} p{2.5cm} p{6cm}}
\toprule
\toprule
\textbf{Study} & \textbf{Criteria} & \textbf{Ratio} & \textbf{Notes} \\
\midrule
\citet{amiti2024fdi} & Foreign Firm (cutoff 10\% capital ownership) & 7.39\%-10.1\% & Less limited number of superstar firms and similar with traditional definition of FDI spillovers \\
& Exporter (cutoff 10\% percentile exports) & 8.47\%-13.76\% & Less limited number of superstar firms and export proportion data is not entirely available \\
& Large Firm (cutoff 10\% sales) & 4.68\%-21.13\% & Less limited number with the cutoff 10\% \\
\citet{autor2020fall} & Top 500 Sales & 0.12\%-11.95\% & The number of firms can be fluctuated in each year \\
\citet{cheng2024exposure} & Markup (cutoff top 5\% and positive growth) & 1.52\%-3.45\% & Limited to positive growth \\
\citet{firooz2025automation} & Sales (cutoff top 1\%) & 1\% & Limited Number \\
& Employment Share (cutoff top 1\%) & 1\% & Not suitable as larger employment share may indicate less capital intensive \\
\citet{rowley2024domestic} & Top 3 Export for Domestic and Foreign & $<$0.1\% & Strictly limited number and export proportion data is not entirely available \\
\bottomrule
\bottomrule
\end{tabular}
}
\label{SummarySuperstarStudies}
\end{table}

%% file: Descriptive_Statistics.tex
\begin{table}[htbp]
\centering
\caption{Descriptive Statistics}
\resizebox{\textwidth}{!}{%
\begin{tabular}{l|l|r|r|r}
\toprule
\toprule
\textbf{Variables} & \textbf{Measurement} & \textbf{N} & \textbf{Mean} & \textbf{Standard Deviation} \\
\hline
Gross Outputs (Constant, Million Rupiah) & Ratio &  349,829 	&	88619.23	&	919917.70	\\
Value-Added  (Constant, Million Rupiah) & Ratio & 349,832 	&	39955.99	&	548353.00	\\
Fixed Assets  (Constant, Million Rupiah) & Ratio & 305,893 	&	313266.20	&	66600000.00	\\
Materials  (Constant, Million Rupiah) & Ratio &  349,832 	&	47781.34	&	519040.40	\\
Workers & Workers & 349,834 	&	195.43	&	740.02	\\
TFP - LP 3 Digits & Ratio & 230,526 	&	13.21	&	1.94	\\
TFP - LP 2 Digits & Ratio & 230,526 	&	13.16	&	1.74	\\
TFP - ACF 3 Digits & Ratio & 145,302 	&	6.61	&	4.80	\\
Dummy Superstar & Dummy & 349,834  &	0.03	&	0.18	\\
Dummy Foreign-Owned & Dummy &  349,834 	&	0.09	&	0.29	\\
Imported Material Intensity & Dummy & 334,421 & 0.09 & 0.24 \\
Dummy Exporters & Dummy & 349,834 	&	0.12	&	0.32	\\
Absorptive Capacity (Log) & Ratio &  349,810 	&	16.19	&	1.13	\\
Market Concentration (HHI) & Ratio &  349,834 	&	0.06	&	0.08	\\
Tariff MFN & Ratio & 349,834 	&	9.11	&	7.33	\\
Road Density (Within-Province)  & Ratio &  349,711 	&	0.95	&	1.49	\\
Road Density (Neighboring-Province)  & Ratio &  349,762 	&	1.28	&	1.00	\\
\bottomrule
\bottomrule
\end{tabular}%
}
\label{DescriptiveStatistics}
\\
\footnotesize \justifying \textbf{Note:} Some variables, such as Tariff and Road Density are explained in the Empirical Strategy section.
\end{table}

%% file: Main_Horizontal.tex
\begin{table}[htbp]\centering
\def\sym#1{\ifmmode^{#1}\else\(^{#1}\)\fi}
\caption{Basic Results--Horizontal Spillovers}
\resizebox{\textwidth}{!}{
\begin{tabular}{l*{6}{c}}
\toprule
\toprule
                    &\multicolumn{1}{c}{(1)}&\multicolumn{1}{c}{(2)}&\multicolumn{1}{c}{(3)}&\multicolumn{1}{c}{(4)}&\multicolumn{1}{c}{(5)}&\multicolumn{1}{c}{(6)}\\
                    &\multicolumn{1}{c}{$\varphi$}&\multicolumn{1}{c}{$\Delta \varphi$}&\multicolumn{1}{c}{$\varphi$}&\multicolumn{1}{c}{$\Delta \varphi$}&\multicolumn{1}{c}{$\varphi$}&\multicolumn{1}{c}{$\Delta \varphi$}\\
\midrule
\textit{HSpill}     &       0.007\sym{***}&       0.004\sym{***}&       0.008\sym{***}&       0.004\sym{***}&       0.006\sym{***}&       0.004\sym{***}\\
                    &     (0.001)         &     (0.001)         &     (0.001)         &     (0.001)         &     (0.001)         &     (0.001)         \\
\addlinespace
\textit{Foreign-Owned}&       0.438\sym{***}&      -0.213\sym{***}&       0.633\sym{***}&      -0.303\sym{*}  &       0.437\sym{***}&      -0.213\sym{***}\\
                    &     (0.020)         &     (0.031)         &     (0.127)         &     (0.169)         &     (0.020)         &     (0.031)         \\
\addlinespace
\textit{Exporters}  &       0.176\sym{***}&      -0.136\sym{***}&       0.177\sym{***}&      -0.136\sym{***}&      -0.095         &      -0.164         \\
                    &     (0.011)         &     (0.017)         &     (0.011)         &     (0.017)         &     (0.094)         &     (0.118)         \\
\addlinespace
\textit{Imports}    &       0.400\sym{***}&      -0.061\sym{*}  &       0.400\sym{***}&      -0.061\sym{*}  &       0.392\sym{***}&      -0.062\sym{*}  \\
                    &     (0.023)         &     (0.032)         &     (0.023)         &     (0.032)         &     (0.023)         &     (0.032)         \\
\addlinespace
\textit{Abs}        &       0.502\sym{***}&       0.225\sym{***}&       0.502\sym{***}&       0.225\sym{***}&       0.499\sym{***}&       0.225\sym{***}\\
                    &     (0.004)         &     (0.005)         &     (0.004)         &     (0.005)         &     (0.005)         &     (0.005)         \\
\addlinespace
\textit{HHI}        &      -0.044         &      -0.054         &      -0.036         &      -0.058         &      -0.044         &      -0.055         \\
                    &     (0.053)         &     (0.059)         &     (0.053)         &     (0.059)         &     (0.053)         &     (0.059)         \\
\addlinespace
\textit{HSpill $\times$ Foreign-Owned}&                     &                     &      -0.006         &       0.003         &                     &                     \\
                    &                     &                     &     (0.004)         &     (0.005)         &                     &                     \\
\addlinespace
\textit{HSpill $\times$ Exporters}&                     &                     &                     &                     &       0.009\sym{***}&       0.001         \\
                    &                     &                     &                     &                     &     (0.003)         &     (0.004)         \\
\midrule
Sector-Province-Island-Year FE&         Yes         &         Yes         &         Yes         &         Yes         &         Yes         &         Yes         \\
Kleibergen-Paap Wald F-stat&     643.839         &     633.646         &      44.181         &      55.788         &      24.618         &      22.222         \\
Cragg-Donald Wald F-stat&    9125.236         &    8596.705         &    2143.587         &    2235.552         &    2158.813         &    2051.938         \\
Observations        &  198665         &  180625         &  198665         &  180625         &  198665         &  180625         \\
\bottomrule
\bottomrule
\end{tabular}
}
\label{Main Results Horizontal}
\justifying \footnotesize \textbf{Note:} Clustered-robust standard errors are in parentheses and in firm level. Bartik-IV from labour-based ($LabBartikIV$) is used as IV for $HSpill$. $\varphi$ denotes dependent variable of TFP level, while $\Delta \varphi$ denotes dependent variable for TFP growth. All estimations use Sector-Province-Island-Year Fixed-Effects. Observations only consist of non-superstar firms. $^{***}$, $^{**}$, $^{*}$ denote $\alpha$ at 1\%, 5\%, and 10\%. $Abs$ denotes absorptive capacity. $HHI$ denotes market concentration from Herfindahl-Hirschman Index. $\varphi$ and $\Delta \varphi$ denote total factor productivity in level and growth, respectively. The results with Inverse Probability Weighting are reported in the Appendix in Table \ref{Main Results Horizontal with IPW}.
\end{table}

%% file: Main_Backward.tex
\begin{table}[htbp]\centering
\def\sym#1{\ifmmode^{#1}\else\(^{#1}\)\fi}
\caption{Basic Results--Backward Spillovers}
\resizebox{\textwidth}{!}{
\begin{tabular}{l*{6}{c}}
\toprule
\toprule
                    &\multicolumn{1}{c}{(1)}&\multicolumn{1}{c}{(2)}&\multicolumn{1}{c}{(3)}&\multicolumn{1}{c}{(4)}&\multicolumn{1}{c}{(5)}&\multicolumn{1}{c}{(6)}\\
                    &\multicolumn{1}{c}{$\varphi$}&\multicolumn{1}{c}{$\Delta \varphi$}&\multicolumn{1}{c}{$\varphi$}&\multicolumn{1}{c}{$\Delta \varphi$}&\multicolumn{1}{c}{$\varphi$}&\multicolumn{1}{c}{$\Delta \varphi$}\\
\midrule
\textit{BSpill}     &       0.153\sym{***}&       0.123\sym{***}&       0.140\sym{***}&       0.113\sym{***}&       0.153\sym{***}&       0.122\sym{***}\\
                    &     (0.023)         &     (0.021)         &     (0.022)         &     (0.020)         &     (0.023)         &     (0.022)         \\
\addlinespace
\textit{Foreign-Owned}&       0.131         &      -0.450\sym{***}&       0.141\sym{*}  &      -0.448\sym{***}&       0.131         &      -0.451\sym{***}\\
                    &     (0.083)         &     (0.076)         &     (0.073)         &     (0.068)         &     (0.084)         &     (0.075)         \\
\addlinespace
\textit{Exporters}  &      -0.001         &      -0.277\sym{***}&       0.030         &      -0.246\sym{***}&      -0.274         &      -0.012         \\
                    &     (0.050)         &     (0.046)         &     (0.046)         &     (0.041)         &     (0.797)         &     (0.956)         \\
\addlinespace
\textit{Imports}    &       0.370\sym{***}&      -0.088         &       0.372\sym{***}&      -0.083         &       0.372\sym{***}&      -0.089         \\
                    &     (0.083)         &     (0.073)         &     (0.069)         &     (0.061)         &     (0.085)         &     (0.072)         \\
\addlinespace
\textit{Abs}        &       0.434\sym{***}&       0.173\sym{***}&       0.480\sym{***}&       0.228\sym{***}&       0.429\sym{***}&       0.178\sym{***}\\
                    &     (0.017)         &     (0.015)         &     (0.023)         &     (0.020)         &     (0.023)         &     (0.023)         \\
\addlinespace
\textit{HHI}        &      -1.789\sym{***}&      -1.354\sym{***}&       3.794\sym{*}  &       5.166\sym{**} &      -1.826\sym{***}&      -1.317\sym{***}\\
                    &     (0.337)         &     (0.297)         &     (2.060)         &     (2.060)         &     (0.352)         &     (0.330)         \\
\addlinespace
\textit{BSpill $\times$ High-Concentration}&                     &                     &      -0.089\sym{***}&      -0.103\sym{***}&                     &                     \\
                    &                     &                     &     (0.032)         &     (0.034)         &                     &                     \\
\addlinespace
\textit{BSpill $\times$ Exporters}&                     &                     &                     &                     &       0.012         &      -0.012         \\
                    &                     &                     &                     &                     &     (0.036)         &     (0.043)         \\
\midrule
Sector-Province-Island-Year FE&         Yes         &         Yes         &         Yes         &         Yes         &         Yes         &         Yes         \\
Kleibergen-Paap Wald F-stat&     112.188         &     139.418         &      37.224         &     112.672         &      57.896         &      55.168         \\
Cragg-Donald Wald F-stat&      13.445         &      17.045         &       8.908         &      10.932         &       6.409         &       7.672         \\
Observations        &  200440         &  182093         &  200440         &  182093         &  200440         &  182093         \\
\bottomrule
\bottomrule
\end{tabular}
}
\label{Main Results Backward}
\justifying \footnotesize \textbf{Note:} Clustered-robust standard errors are in parentheses and in firm level. Bartik-IV from output-based ($TarrBartikIV$) is used as IV for $BSpill$. $\varphi$ denotes dependent variable of TFP level, while $\Delta \varphi$ denotes dependent variable for TFP growth. All estimations use Sector-Province-Island-Year Fixed-Effects.  Observations only consist of non-superstar firms. $^{***}$, $^{**}$, $^{*}$ denote $\alpha$ at 1\%, 5\%, and 10\%. $Abs$ denotes absorptive capacity. $HHI$ denotes market concentration from Herfindahl-Hirschman Index. $\varphi$ and $\Delta \varphi$ denote total factor productivity in level and growth, respectively. The results with Inverse Probability Weighting are reported in the Appendix in Table \ref{Main Results Backward with IPW}.
\end{table}

%% file: Main_Forward.tex
\begin{table}[htbp]\centering
\def\sym#1{\ifmmode^{#1}\else\(^{#1}\)\fi}
\caption{Basic Results--Forward Spillovers}
\resizebox{\textwidth}{!}{
\begin{tabular}{l*{6}{c}}
\toprule
\toprule
                    &\multicolumn{1}{c}{(1)}&\multicolumn{1}{c}{(2)}&\multicolumn{1}{c}{(3)}&\multicolumn{1}{c}{(4)}&\multicolumn{1}{c}{(5)}&\multicolumn{1}{c}{(6)}\\
                    &\multicolumn{1}{c}{$\varphi$}&\multicolumn{1}{c}{$\Delta \varphi$}&\multicolumn{1}{c}{$\varphi$}&\multicolumn{1}{c}{$\Delta \varphi$}&\multicolumn{1}{c}{$\varphi$}&\multicolumn{1}{c}{$\Delta \varphi$}\\
\midrule
\textit{FSpill}     &       0.222\sym{***}&       0.212\sym{***}&       0.140\sym{***}&       0.125\sym{***}&       0.222\sym{***}&       0.212\sym{***}\\
                    &     (0.043)         &     (0.049)         &     (0.024)         &     (0.028)         &     (0.043)         &     (0.048)         \\
\addlinespace
\textit{Foreign-Owned}&       0.133         &      -0.491\sym{***}&       0.228\sym{*}  &      -0.389\sym{***}&       0.138         &      -0.482\sym{***}\\
                    &     (0.164)         &     (0.168)         &     (0.117)         &     (0.116)         &     (0.160)         &     (0.161)         \\
\addlinespace
\textit{Exporters}  &       0.021         &      -0.297\sym{***}&       0.071         &      -0.239\sym{***}&       0.483         &       0.518         \\
                    &     (0.080)         &     (0.085)         &     (0.057)         &     (0.059)         &     (1.007)         &     (0.996)         \\
\addlinespace
\textit{Imports}    &       0.291\sym{*}  &      -0.150         &       0.326\sym{***}&      -0.114         &       0.268         &      -0.192         \\
                    &     (0.166)         &     (0.167)         &     (0.122)         &     (0.119)         &     (0.169)         &     (0.169)         \\
\addlinespace
\textit{Abs}        &       0.498\sym{***}&       0.223\sym{***}&       0.485\sym{***}&       0.206\sym{***}&       0.504\sym{***}&       0.232\sym{***}\\
                    &     (0.021)         &     (0.021)         &     (0.017)         &     (0.017)         &     (0.026)         &     (0.025)         \\
\addlinespace
\textit{HHI}        &      -0.224         &      -0.305         &      -4.208\sym{***}&      -4.501\sym{***}&      -0.236         &      -0.310         \\
                    &     (0.219)         &     (0.232)         &     (1.230)         &     (1.270)         &     (0.220)         &     (0.229)         \\
\addlinespace
\textit{FSpill $\times$ High-Concentration}&                     &                     &       0.066\sym{***}&       0.069\sym{***}&                     &                     \\
                    &                     &                     &     (0.020)         &     (0.020)         &                     &                     \\
\addlinespace
\textit{FSpill $\times$ Exporters}    &                     &                     &                     &                     &      -0.027         &      -0.047         \\
                    &                     &                     &                     &                     &     (0.059)         &     (0.058)         \\
\midrule
Sector-Province-Island-Year FE&         Yes         &         Yes         &         Yes         &         Yes         &         Yes         &         Yes         \\
Kleibergen-Paap Wald F-stat&      41.300         &      35.443         &      49.095         &      42.874         &      21.379         &      18.488         \\
Cragg-Donald Wald F-stat&       4.885         &       4.347         &       3.910         &       3.505         &       2.487         &       2.218         \\
Observations        &  200440         &  182093         &  200440         &  182093        &  200440         &  182093         \\
\bottomrule
\bottomrule
\end{tabular}
}
\label{Main Results Forward}
\justifying \footnotesize \textbf{Note:} Clustered-robust standard errors are in parentheses and in firm level. Bartik-IV from output-based ($TarrBartikIV$) is used as IV for $FSpill$. $\varphi$ denotes dependent variable of TFP level, while $\Delta \varphi$ denotes dependent variable for TFP growth. All estimations use Sector-Province-Island-Year Fixed-Effects. Observations only consist of non-superstar firms. $^{***}$, $^{**}$, $^{*}$ denote $\alpha$ at 1\%, 5\%, and 10\%. $Abs$ denotes absorptive capacity. $HHI$ denotes market concentration from Herfindahl-Hirschman Index. $\varphi$ and $\Delta \varphi$ denote total factor productivity in level and growth, respectively. The results with Inverse Probability Weighting are reported in the Appendix in Table \ref{Main Results Forward with IPW}.
\end{table}

%% file: Heterogeneous_Spillovers.tex
\begin{table}[htbp]\centering
\def\sym#1{\ifmmode^{#1}\else\(^{#1}\)\fi}
\caption{Heterogeneous Superstars: Foreign and Domestic}
\resizebox{\textwidth}{!}{
\begin{tabular}{l*{12}{c}}
\toprule
\toprule
                    &\multicolumn{1}{c}{(1)}&\multicolumn{1}{c}{(2)}&\multicolumn{1}{c}{(3)}&\multicolumn{1}{c}{(4)}&\multicolumn{1}{c}{(5)}&\multicolumn{1}{c}{(6)}&\multicolumn{1}{c}{(7)}&\multicolumn{1}{c}{(8)}&\multicolumn{1}{c}{(9)}&\multicolumn{1}{c}{(10)}&\multicolumn{1}{c}{(11)}&\multicolumn{1}{c}{(12)}\\
                    \midrule
                    & \multicolumn{2}{c}{Foreign} &  \multicolumn{2}{c}{Domestic} & \multicolumn{2}{c}{Foreign} &  \multicolumn{2}{c}{Domestic} & \multicolumn{2}{c}{Foreign} &  \multicolumn{2}{c}{Domestic}    \\
                    &\multicolumn{1}{c}{$\varphi$}&\multicolumn{1}{c}{$\Delta\varphi$}&\multicolumn{1}{c}{$\varphi$}&\multicolumn{1}{c}{$\Delta\varphi$}&\multicolumn{1}{c}{$\varphi$}&\multicolumn{1}{c}{$\Delta\varphi$}&\multicolumn{1}{c}{$\varphi$}&\multicolumn{1}{c}{$\Delta\varphi$}&\multicolumn{1}{c}{$\varphi$}&\multicolumn{1}{c}{$\Delta\varphi$}&\multicolumn{1}{c}{$\varphi$}&\multicolumn{1}{c}{$\Delta\varphi$}\\
\midrule
\textit{HSpill}     &       0.004         &       0.004         &     0.019\sym{***}&       0.010\sym{***}&                     &                     &                     &                     &                     &                     &                     &                     \\
                    &     (0.004)         &     (0.005)         &     (0.002)         &     (0.003)         &                     &                     &                     &                     &                     &                     &                     &                     \\
\addlinespace
\textit{BSpill}     &                     &                     &                     &                     &       0.012\sym{**} &       0.007\sym{**} &       0.113\sym{***}&       0.113\sym{***}&                     &                     &                     &                     \\
                    &                     &                     &                     &                     &     (0.005)         &     (0.004)         &     (0.016)         &     (0.019)         &                     &                     &                     &                     \\
\addlinespace
\textit{FSpill}     &                     &                     &                     &                     &                     &                     &                     &                     &       0.037\sym{**} &       0.020\sym{**} &        0.129\sym{***}&       0.125\sym{***}\\
                    &                     &                     &                     &                     &                     &                     &                     &                     &     (0.016)         &     (0.010)         &       (0.017)         &     (0.020)         \\
\addlinespace
\textit{Foreign-Owned}&       0.447\sym{***}&      -0.211\sym{***}&       0.455\sym{***}&      -0.204\sym{***}&       0.429\sym{***}&      -0.214\sym{***}&       0.380\sym{***}&      -0.274\sym{***}&       0.396\sym{***}&      -0.231\sym{***}&       0.464\sym{***}&      -0.182\sym{***}\\
                    &     (0.021)         &     (0.031)         &     (0.021)         &     (0.030)         &     (0.020)         &     (0.030)         &     (0.039)         &     (0.047)         &     (0.029)         &     (0.033)         &     (0.056)         &     (0.060)         \\
\addlinespace
\textit{Exporters}  &       0.176\sym{***}&      -0.135\sym{***}&       0.193\sym{***}&      -0.127\sym{***}&       0.170\sym{***}&      -0.137\sym{***}&       0.157\sym{***}&      -0.149\sym{***}&       0.175\sym{***}&      -0.134\sym{***}&       0.206\sym{***}&      -0.107\sym{***}\\
                    &     (0.011)         &     (0.017)         &     (0.012)         &     (0.017)         &     (0.011)         &     (0.017)         &     (0.022)         &     (0.026)         &     (0.012)         &     (0.017)         &     (0.028)         &     (0.031)         \\
\addlinespace
\textit{Imports}    &       0.399\sym{***}&      -0.065\sym{*}  &       0.422\sym{***}&      -0.049         &       0.388\sym{***}&      -0.059\sym{*}  &       0.514\sym{***}&       0.063         &       0.349\sym{***}&      -0.079\sym{**} &       0.472\sym{***}&       0.020         \\
                    &     (0.023)         &     (0.033)         &     (0.024)         &     (0.032)         &     (0.023)         &     (0.032)         &     (0.046)         &     (0.053)         &     (0.034)         &     (0.035)         &     (0.062)         &     (0.067)         \\
\addlinespace
\textit{Abs}        &       0.503\sym{***}&       0.226\sym{***}&       0.509\sym{***}&       0.229\sym{***}&       0.501\sym{***}&       0.224\sym{***}&       0.491\sym{***}&       0.214\sym{***}&       0.501\sym{***}&       0.225\sym{***}&       0.504\sym{***}&       0.223\sym{***}\\
                    &     (0.004)         &     (0.005)         &     (0.005)         &     (0.005)         &     (0.004)         &     (0.005)         &     (0.008)         &     (0.009)         &     (0.005)         &     (0.005)         &     (0.010)         &     (0.011)         \\
\addlinespace
\textit{HHI}        &       0.056         &      -0.001         &      -0.239\sym{***}&      -0.153\sym{**} &       0.008         &      -0.010         &      -1.741\sym{***}&      -1.662\sym{***}&       0.141\sym{**} &       0.063         &      -1.460\sym{***}&      -1.417\sym{***}\\
                    &     (0.049)         &     (0.056)         &     (0.066)         &     (0.075)         &     (0.051)         &     (0.056)         &     (0.276)         &     (0.318)         &     (0.069)         &     (0.063)         &     (0.221)         &     (0.259)         \\
\midrule
Sector-Province-Island-Year FE&         Yes         &         Yes         &         Yes         &         Yes         &         Yes         &         Yes         &         Yes         &         Yes         &         Yes         &         Yes         &         Yes         &         Yes         \\
Kleibergen-Paap Wald F-stat&     394.891         &     369.157         &     350.564         &     339.332         &     235.571         &     426.651         &     195.089         &     169.255         &      46.902         &     140.173         &     348.729         &     320.436         \\
Cragg-Donald Wald F-stat&     956.510         &     879.831         &    2721.304         &    2426.539         &     179.371         &     461.061         &      39.217         &      33.050         &      29.471         &      95.254         &      20.725         &      19.115         \\
Observations        &  198665         &  180625         &  198665         &  180625         &  200440         &  182093         &  200440         &  182093         &  200440         &  182093         &  200440         &  182093         \\
\bottomrule
\bottomrule
\end{tabular}
}
\label{Heterogeneous Superstar}
\justifying \footnotesize \textbf{Note:} Clustered-robust standard errors are in parentheses and in firm level. Bartik-IV from labour-based and output growth ($LabBartikIV$) is used as IV for $HSpill$, while Bartik-IV from output-based and tariff ($TarrBartikIV$) is used as IV for $BSpill$ adn $FSpill$. $Foreign$ denotes foreign superstar spillovers, while $Domestic$ denotes domestic foreign spillovers. Both foreign and domestic superstars are non-exporters. $\varphi$ denotes dependent variable of TFP level, while $\Delta \varphi$ denotes dependent variable for TFP growth. All estimations use Sector-Province-Island-Year Fixed-Effects.  Observations only consist of non-superstar firms. $^{***}$, $^{**}$, $^{*}$ denote $\alpha$ at 1\%, 5\%, and 10\%. $Abs$ denotes absorptive capacity. $HHI$ denotes market concentration from Herfindahl-Hirschman Index. $\varphi$ and $\Delta \varphi$ denote total factor productivity in level and growth, respectively.
\end{table}

%% file: Static_OP.tex
\begin{table}[htpb]\centering
\caption{TFP Change from Static OP Decomposition}
\begin{tabular}{llccc}
\toprule
\toprule
\textbf{Component} & \textbf{Period} & \textbf{Overall} & \textbf{Superstar} & \textbf{Non-superstar} \\
\midrule
\multirow{3}{*}{Aggregate TFP Change}
&	2001-2015&	1.059	&	0.667	&	1.406	\\
&	2001-2010&	0.780	&	0.430	&	1.113	\\
&	2001-2005&	0.377	&	0.384	&	0.283	\\
\midrule
\multirow{6}{*}{Plant Improvements}
&	2001-2015&	1.294	&	0.865	&	1.313	\\
& 		&(1.222)	&	(1.298)	&	(0.934)	\\
&	2001-2010&	0.665	&	0.716	&	0.671	\\
& 		& (0.852)	&	(1.667)	&	(0.603)	\\
&	2001-2005&	0.318	&	0.363	&	0.305	\\
&   	& (0.844)	&	(0.945)	&	(1.081)	\\
\midrule
\multirow{6}{*}{Reallocation}
&	2001-2015&	-0.235	&	-0.199	&	0.093	\\
&           &		(-0.222)	&	(-0.298)	&	(0.066)	\\
&	2001-2010&	0.116	&	-0.287	&	0.442	\\
&         &		(0.148)	&	(-0.667)	&	(0.397)	\\
&	2001-2005&	0.059	&	0.021	&	-0.023	\\
&         &		(0.156)	&	(0.055)	&	(-0.081)	\\
\bottomrule
\bottomrule
\end{tabular}
\\
\justifying \small
\textbf{Note:} Shares of each components to the aggregate TFP change are in parentheses. Plant improvement refers to the change of unweighted average productivity ($\Delta\bar{\varphi_t}$) while reallocation is the covariance component between market share and productivity. 
\label{Static OP}
\end{table}

%% file: DOPD.tex
\begin{table}[htpb]\centering
\caption{TFP Change from Dynamic OP Decomposition: Survivors, Exiters, and Entrants}
\resizebox{\textwidth}{!}{
\begin{tabular}{llcccccc}
\toprule
\toprule
\multirow{2}{*}{\bfseries Component} & \multirow{2}{*}{\bfseries Period} & \multirow{2}{*}{\bfseries Overall} & \multicolumn{2}{c}{\bfseries General Superstar} & \multicolumn{3}{c}{\bfseries Heterogeneous Superstar} \\
\cmidrule(lr){4-5} \cmidrule(lr){6-8}
 & & & Superstar & Non-superstar & Foreign & Domestic & Non-superstar \\
\midrule
\multirow{3}{*}{Aggregate TFP Change} 
&	2001-2015	&	1.059	&	0.667	&	1.406	&	0.603	&	0.203	&	0.252	\\
&	2001-2010	&	0.780	&	0.430	&	1.113	&	0.838	&	0.068	&	0.825	\\
&	2001-2005	&	0.377	&	0.384	&	0.283	&	1.321	&	-0.187	&	1.492	\\

\midrule
\multirow{6}{*}{Plant Improvements} 
&	2001-2015	&	1.096	&	0.830	&	1.113	&	0.332	&	0.311	&	0.233	\\
	&	&	(1.036)&	(1.245)	&	(0.792)	&	(0.551)	&	(1.534)	&	(0.922)	\\
&	2001-2010	&	0.565	&	0.691	&	0.559	&	0.503	&	0.756	&	0.562	\\
&	              &	(0.724)	&	(1.608)	&	(0.502)	&	(0.600)	&	(11.089)	&	(0.682)	\\
&	2001-2005	&	0.233	&	0.296	&	0.230	&	0.695	&	0.880	&	1.129	\\
&	              &	(0.617)&	(0.771)	&	(0.814)	&	(0.526)	&	(-4.700)	&	(0.756)	\\
\midrule
\multirow{6}{*}{Reallocation within Survivors} 
&	2001-2015	&	-0.446	&	-0.452	&	-0.122	&	-0.056	&	-0.251	&	-0.319	\\
	&	&	(-0.422)	&	(-0.678)	&	(-0.087)	&	(-0.093)	&	(-1.238)	&	(-1.265)	\\
&	2001-2010	&	-0.142	&	-0.441	&	0.120	&	0.651	&	-0.875	&	-0.192	\\
&	              &	(-0.182)	&	(-1.027)	&	(0.108)	&	(0.777)	&	(-12.842)	&	(-0.233)	\\
    &	2001-2005	&	-0.080	&	-0.191	&	-0.043	&	0.274	&	-0.905	&	-0.303	\\
&	              &	(-0.212)	&	(-0.498)	&	(-0.152)	&	(0.207)	&	(4.834)	&	(-0.203)	\\
\midrule
\multirow{6}{*}{Reallocation: Exiters-Entrants} 
&	2001-2015	&	0.409	&	0.289	&	0.415	&	0.326	&	0.143	&	0.339	\\
	&	&	(0.386)&	(0.433	)&	(0.295)&	(0.541)&	(0.704)&	(1.343)\\
&	2001-2010	&	0.357	&	0.180	&	0.434	&	-0.316	&	0.188	&	0.455	\\
&	              &	(0.458)&	(0.419	)&	(0.390)&	(-0.377)	&	(2.753)&	(0.552)\\
    &	2001-2005	&	0.224	&	0.279	&	0.095	&	0.353	&	-0.162	&	0.667	\\
&	              &	(0.595)&	(0.727	)&	(0.338)&	(0.267)&	(0.866)&	(0.447)\\
\bottomrule
\bottomrule
\end{tabular}
}
\\
\justifying \small
\textbf{Note:} The value is in the log-change. Shares of each components to the aggregate TFP change are in parentheses. Plant improvement refers to the change of unweighted average productivity for survivors ($\Delta\bar{\varphi_t}$), while Reallocation within Survivors refers to the change of covariance component within survivors, and Reallocation of Exiters-Entrants is the plant improvements and covariance component across group of survivors with entrants and exiters. 
\label{ContributionDynamicOP}
\end{table}

%% file: Elasticity_2_digits.tex
\begin{table}[htbp]
\centering
\caption{Labour and Capital Elasticities by 2-Digit ISIC Subsector}
\begin{tabular}{lcc}
\toprule
\toprule
& \multicolumn{2}{c}{LP} \\
\cmidrule(lr){2-3}
Subsector & Labour & Capital \\
\midrule
Food & 0.46 & 0.34 \\
Beverages & 0.56 & 0.26 \\
Tobacco & 0.23 & 0.34 \\
Textile & 0.38 & 0.28 \\
Apparel & 0.52 & 0.29 \\
Leather & 0.48 & 0.28 \\
Wood & 0.40 & 0.37 \\
Paper & 0.44 & 0.19 \\
Recording Media & 0.43 & 0.30 \\
Coal Products & 0.38 & 0.45 \\
Chemical & 0.36 & 0.27 \\
Pharmaceutical & 0.49 & 0.21 \\
Rubber & 0.42 & 0.28 \\
Non-Metallic & 0.44 & 0.29 \\
Basic Metal & 0.56 & 0.18 \\
Metals & 0.47 & 0.28 \\
Computer \& Electronics & 0.47 & 0.21 \\
Electrical Equipment & 0.48 & 0.22 \\
Machinery & 0.43 & 0.26 \\
Motor Vehicle & 0.27 & 0.19 \\
Other Transportation & 0.38 & 0.25 \\
Furniture & 0.35 & 0.32 \\
Other Manufacturing & 0.61 & 0.21 \\
Machinery Repairment & 0.44 & 0.29 \\
\bottomrule
\bottomrule
\multicolumn{3}{l}{\footnotesize \textbf{Note:} LP denotes Levinsohn-Petrin approach.}\\
\end{tabular}
\label{Labour and Capital Coefficient 2 Digits}
\end{table}

%% file: Elasticity_3_digits_1.tex
\begin{table}[htbp]
\centering
\caption{Labour and Capital Elasticities by 3-digit ISIC Sector (1)}
\begin{adjustbox}{max width=\textwidth}
\begin{tabular}{lcccccc}
\toprule
\toprule
\textbf{Subsector} & \multicolumn{2}{c}{\textbf{LP}} & \multicolumn{2}{c}{\textbf{ACF}} & \multicolumn{2}{c}{\textbf{WRDG}} \\
 & Labour & Capital & Labour & Capital & Labour & Capital \\
\midrule
Animal Feed & 0.31 & 0.29 & 0.80 & 0.02 & 0.32 & 0.29 \\
Apparel & 0.52 & 0.30 &  &  & 0.62 & 0.30 \\
Beverages & 0.56 & 0.26 &  &  & 0.63 & 0.31 \\
Cables & 0.46 & 0.15 & 0.99 & 0.44 & 0.65 & 0.16 \\
Chemicals & 0.37 & 0.25 & 0.95 & 0.29 & 0.44 & 0.24 \\
Coal and Petroleum & 0.38 & 0.45 & 0.75 & 0.40 & 0.56 & 0.45 \\
Computers \& Communication Equipments & 0.40 & 0.06 & 1.22 & 0.28 & 0.48 & 0.12 \\
Dairy \& Ice Cream & 0.52 & 0.14 &  &  & 0.59 & 0.21 \\
Edible Oils \& Fats & 0.35 & 0.23 &  &  & 0.42 & 0.29 \\
Electric Motors \& Controls & 0.44 & 0.23 & 0.77 & 0.55 & 0.56 & 0.21 \\
Electronic Components & 0.45 & 0.20 & 0.82 & 0.42 & 0.64 & 0.14 \\
Fish \& Aquatic Processing & 0.35 & 0.37 &  &  & 0.52 & 0.32 \\
Footwear & 0.43 & 0.31 & 0.82 & 0.51 & 0.56 & 0.25 \\
Fruit \& Vegetable Processing & 0.37 & 0.29 &  &  & 0.46 & 0.30 \\
Furniture & 0.35 & 0.32 & 0.89 & 0.54 & 0.49 & 0.28 \\
General Machinery & 0.49 & 0.28 &  &  & 0.62 & 0.28 \\
Glass Products & 0.36 & 0.20 & 0.59 & 0.55 & 0.31 & 0.23 \\
Grain \& Flour Milling & 0.50 & 0.38 & 0.69 & 0.46 & 0.52 & 0.39 \\
Household Appliances & 0.50 & 0.30 & 0.56 & 0.46 & 0.60 & 0.32 \\
Iron \& Steel & 0.47 & 0.21 &  &  & 0.57 & 0.32 \\
Jewelries & 0.64 & 0.19 &  &  & 0.76 & 0.18 \\
Knitted \& Embroidered Wear & 0.47 & 0.18 & 0.85 & 0.62 & 0.61 & 0.16 \\
Leather \& Artificial Leather & 0.50 & 0.24 & 1.01 & 0.21 & 0.62 & 0.20 \\
Lighting Equipment & 0.54 & 0.22 &  &  & 0.67 & 0.25 \\
\bottomrule
\bottomrule
\multicolumn{7}{l}{\footnotesize \textbf{Note:} LP denotes Levinsohn-Petrin approach, ACF denotes Ackenbeigh approach, and WRDG denotes Woolridge approach.}\\
\end{tabular}
\end{adjustbox}
\end{table}

%% file: Elasticity_3_digits_2.tex
\begin{table}[htbp]
\centering
\caption{Labour and Capital Elasticities by 3-digit ISIC Sector (2)}
\begin{adjustbox}{max width=\textwidth}
\begin{tabular}{lcccccc}
\toprule
\toprule
\textbf{Subsector} & \multicolumn{2}{c}{\textbf{LP}} & \multicolumn{2}{c}{\textbf{ACF}} & \multicolumn{2}{c}{\textbf{WRDG}} \\
 & Labour & Capital & Labour & Capital & Labour & Capital \\
\midrule
Meat Processing & 0.57 & 0.19 & 0.72 & 0.48 & 0.68 & 0.29 \\
Metal \& Machinery Repair & 0.44 & 0.29 & 9.43 & -0.02 & 0.57 & 0.30 \\
Motor Vehicles & 0.42 & 0.20 &  &  & 0.67 & 0.26 \\
Other Basic Metals & 0.61 & 0.15 & 1.02 & -0.02 & 0.72 & 0.20 \\
Other Chemicals & 0.38 & 0.28 & 0.90 & 0.39 & 0.47 & 0.27 \\
Other Electrical Devices & 0.47 & 0.10 & 1.20 & 0.11 & 0.86 & 0.14 \\
Other Electronic Equipments & 0.48 & 0.27 &  &  & 0.59 & 0.19 \\
Other Food Products & 0.38 & 0.36 & 0.97 & 0.59 & 0.44 & 0.37 \\
Other Manufacturing & 0.67 & 0.21 &  &  & 0.75 & 0.18 \\
Other Metal Goods & 0.46 & 0.28 &  &  & 0.57 & 0.30 \\
Other Non-Metal Minerals & 0.44 & 0.29 & 1.30 & 0.44 & 0.46 & 0.32 \\
Other Textiles & 0.42 & 0.28 & 1.01 & 0.42 & 0.52 & 0.26 \\
Other Transport Equipment & 0.40 & 0.26 &  &  & 0.55 & 0.17 \\
Paper Products & 0.44 & 0.19 & 5.41 & 0.09 & 0.51 & 0.20 \\
Pharmaceuticals Medicine & 0.49 & 0.21 &  &  & 0.59 & 0.23 \\
Prefabricated Metal Goods & 0.56 & 0.27 & 10.60 & 0.06 & 0.74 & 0.29 \\
Printing \& Media Reproduction & 0.43 & 0.30 & 1.03 & 0.45 & 0.56 & 0.25 \\
Rubber \& Plastic Products & 0.42 & 0.28 &  &  & 0.51 & 0.27 \\
Ship \& Boat Building & 0.48 & 0.34 & -26.48 & 0.88 & 0.55 & 0.39 \\
Special Machinery & 0.33 & 0.25 & 2.71 & 0.33 & 0.42 & 0.26 \\
Spinning \& Finishing of Textile & 0.35 & 0.28 & 0.86 & 0.38 & 0.46 & 0.25 \\
Sports Equipment & 0.40 & 0.16 & 0.93 & 0.20 & 0.62 & 0.20 \\
Tobacco Processing & 0.23 & 0.34 &  &  & 0.31 & 0.30 \\
Toys \& Games & 0.40 & 0.18 &  &  & 0.58 & 0.20 \\
Vehicle Bodies \& Trailers & 0.19 & 0.24 & 0.83 & 0.50 & 0.40 & 0.24 \\
Vehicle Parts and Others & 0.23 & 0.14 & 1.01 & 0.50 & 0.32 & 0.20 \\
Woods & 0.40 & 0.37 & 0.78 & 0.50 & 0.53 & 0.32 \\
\bottomrule
\bottomrule
\multicolumn{7}{l}{\footnotesize \textbf{Note:} LP denotes Levinsohn-Petrin approach, ACF denotes Ackenbeigh approach, and WRDG denotes Woolridge approach.}\\
\end{tabular}
\end{adjustbox}
\end{table}

%% file: First_Stage_Horizontal.tex
\begin{table}[htbp]\centering
\def\sym#1{\ifmmode^{#1}\else\(^{#1}\)\fi}
\caption{First-Stage: Basic Results--Horizontal Spillovers}
\begin{tabular}{l*{2}{c}}
\toprule
                    &\multicolumn{1}{c}{(1)}&\multicolumn{1}{c}{(2)}\\
                    &\multicolumn{1}{c}{$\varphi$}&\multicolumn{1}{c}{$\Delta\varphi$}\\
\midrule
\textit{LabBartik-IV}&      18.917\sym{***}&      19.060\sym{***}\\
                    &     (0.746)         &     (0.757)         \\
\addlinespace
\textit{Foreign-Owned}&       1.938\sym{***}&       1.817\sym{***}\\
                    &     (0.469)         &     (0.489)         \\
\addlinespace
\textit{Exporters}  &      -0.673\sym{**} &      -0.611\sym{**} \\
                    &     (0.264)         &     (0.281)         \\
\addlinespace
\textit{Imports}    &       0.878         &       0.959\sym{*}  \\
                    &     (0.559)         &     (0.580)         \\
\addlinespace
\textit{Abs}        &       0.002         &      -0.025         \\
                    &     (0.079)         &     (0.082)         \\
\addlinespace
\textit{HHI}        &      11.083\sym{***}&      10.598\sym{***}\\
                    &     (1.403)         &     (1.487)         \\
\addlinespace
Constant            &      35.392\sym{***}&      35.816\sym{***}\\
                    &     (1.264)         &     (1.326)         \\
\midrule
Sector-Province-Island-Year FE&         Yes         &         Yes         \\
F-statistics        &     130.184         &     125.061         \\
Observations        &  198665         &  180625         \\
\bottomrule
\bottomrule
\end{tabular}
\label{First-Stage Main Horizontal}
\\
\justifying \footnotesize \textbf{Note:} Clustered-robust standard errors are in parentheses and in firm level. Bartik-IV from labour-based ($LabBartikIV$) is used as IV for $HSpill$. $\varphi$ denotes dependent variable of TFP level, while $\Delta \varphi$ denotes dependent variable for TFP growth. All estimations use Sector-Province-Island-Year Fixed-Effects. $^{***}$, $^{**}$, $^{*}$ denote $\alpha$ at 1\%, 5\%, and 10\%. $Abs$ denotes absorptive capacity. $HHI$ denotes market concentration from Herfindahl-Hirschman Index.
\end{table}

%% file: First_Stage_Backward.tex
\begin{table}[htbp]\centering
\def\sym#1{\ifmmode^{#1}\else\(^{#1}\)\fi}
\caption{First-Stage: Basic Results--Backward Spillovers}
\begin{tabular}{l*{2}{c}}
\toprule
                    &\multicolumn{1}{c}{(1)}&\multicolumn{1}{c}{(2)}\\
                    &\multicolumn{1}{c}{$\varphi$}&\multicolumn{1}{c}{$\Delta\varphi$}\\
\midrule
\textit{TarrBartik-IV}        &      -0.365\sym{***}&      -0.421\sym{***}\\
                    &     (0.034)         &     (0.036)         \\
\addlinespace
\textit{Foreign-Owned}&       2.034\sym{***}&       1.978\sym{***}\\
                    &     (0.438)         &     (0.458)         \\
\addlinespace
\textit{Exporters}  &       1.117\sym{***}&       1.138\sym{***}\\
                    &     (0.270)         &     (0.288)         \\
\addlinespace
\textit{Imports}    &       0.190         &       0.304         \\
                    &     (0.517)         &     (0.537)         \\
\addlinespace
\textit{Abs}        &       0.439\sym{***}&       0.410\sym{***}\\
                    &     (0.084)         &     (0.088)         \\
\addlinespace
\textit{HHI}        &      11.944\sym{***}&      11.051\sym{***}\\
                    &     (1.359)         &     (1.444)         \\
\addlinespace
Constant            &      26.263\sym{***}&      26.937\sym{***}\\
                    &     (1.348)         &     (1.417)         \\
\midrule
Sector-Province-Island-Year FE&         Yes         &         Yes         \\
F-statistics        &      42.992         &      42.916         \\
Observations        &  200440         &  182093         \\
\bottomrule
\bottomrule
\end{tabular}
\label{First-Stage Main Backwards}
\\
\justifying \footnotesize \textbf{Note:} Clustered-robust standard errors are in parentheses and in firm level. Bartik-IV from output-based ($TarrBartikIV$) is used as IV for $BSpill$. $\varphi$ denotes dependent variable of TFP level, while $\Delta \varphi$ denotes dependent variable for TFP growth. All estimations use Sector-Province-Island-Year Fixed-Effects. $^{***}$, $^{**}$, $^{*}$ denote $\alpha$ at 1\%, 5\%, and 10\%. $Abs$ denotes absorptive capacity. $HHI$ denotes market concentration from Herfindahl-Hirschman Index.
\end{table}

%% file: First_Stage_Forward.tex
\begin{table}[htbp]\centering
\def\sym#1{\ifmmode^{#1}\else\(^{#1}\)\fi}
\caption{First-Stage: Basic Results--Forward Spillovers}
\begin{tabular}{l*{2}{c}}
\toprule
                    &\multicolumn{1}{c}{(1)}&\multicolumn{1}{c}{(2)}\\
                    &\multicolumn{1}{c}{$\varphi$}&\multicolumn{1}{c}{$\Delta\varphi$}\\
\midrule
\textit{TarrBartik-IV}     &      -0.251\sym{***}&      -0.244\sym{***}\\
                    &     (0.039)         &     (0.041)         \\
\addlinespace
\textit{Foreign-Owned}&       1.393\sym{**} &       1.341\sym{*}  \\
                    &     (0.655)         &     (0.692)         \\
\addlinespace
\textit{Exporters}  &       0.670\sym{**} &       0.752\sym{**} \\
                    &     (0.337)         &     (0.357)         \\
\addlinespace
\textit{Imports}    &       0.487         &       0.467         \\
                    &     (0.730)         &     (0.766)         \\
\addlinespace
\textit{Abs}        &       0.010         &       0.003         \\
                    &     (0.094)         &     (0.098)         \\
\addlinespace
\textit{HHI}        &       1.170         &       1.454         \\
                    &     (0.954)         &     (1.029)         \\
\addlinespace
Constant            &      21.066\sym{***}&      21.238\sym{***}\\
                    &     (1.516)         &     (1.592)         \\
\midrule
Sector-Province-Island-Year FE&         Yes         &         Yes         \\
F-statistics        &      10.423         &       9.059         \\
Observations        &  200440         &  182093         \\
\bottomrule
\bottomrule
\end{tabular}
\label{First-Stage Main Forwards}
\\
\justifying \footnotesize \textbf{Note:} Clustered-robust standard errors are in parentheses and in firm level. Bartik-IV from output-based ($TarrBartikIV$) is used as IV for $FSpill$. $\varphi$ denotes dependent variable of TFP level, while $\Delta \varphi$ denotes dependent variable for TFP growth. All estimations use Sector-Province-Island-Year Fixed-Effects. $^{***}$, $^{**}$, $^{*}$ denote $\alpha$ at 1\%, 5\%, and 10\%. $Abs$ denotes absorptive capacity. $HHI$ denotes market concentration from Herfindahl-Hirschman Index.
\end{table}

%% file: Main_Horizontal_IPW.tex
\begin{table}[htbp]\centering
\def\sym#1{\ifmmode^{#1}\else\(^{#1}\)\fi}
\caption{Basic Results--Horizontal Spillovers with Inverse Probability Weighting (IPW)}
\resizebox{\textwidth}{!}{
\begin{tabular}{l*{6}{c}}
\toprule
\toprule
                    &\multicolumn{1}{c}{(1)}&\multicolumn{1}{c}{(2)}&\multicolumn{1}{c}{(3)}&\multicolumn{1}{c}{(4)}&\multicolumn{1}{c}{(5)}&\multicolumn{1}{c}{(6)}\\
                    &\multicolumn{1}{c}{$\varphi$}&\multicolumn{1}{c}{$\Delta \varphi$}&\multicolumn{1}{c}{$\varphi$}&\multicolumn{1}{c}{$\Delta \varphi$}&\multicolumn{1}{c}{$\varphi$}&\multicolumn{1}{c}{$\Delta \varphi$}\\
\midrule
\textit{HSpill}     &       0.007\sym{***}&       0.004\sym{***}&       0.008\sym{***}&       0.004\sym{***}&       0.006\sym{***}&       0.004\sym{***}\\
                    &     (0.001)         &     (0.001)         &     (0.001)         &     (0.001)         &     (0.001)         &     (0.001)         \\
\addlinespace
\textit{Foreign-Owned}&       0.436\sym{***}&      -0.213\sym{***}&       0.634\sym{***}&      -0.306\sym{*}  &       0.435\sym{***}&      -0.213\sym{***}\\
                    &     (0.020)         &     (0.030)         &     (0.124)         &     (0.168)         &     (0.020)         &     (0.030)         \\
\addlinespace
\textit{Exporters}  &       0.175\sym{***}&      -0.136\sym{***}&       0.175\sym{***}&      -0.136\sym{***}&      -0.094         &      -0.165         \\
                    &     (0.011)         &     (0.017)         &     (0.011)         &     (0.017)         &     (0.093)         &     (0.117)         \\
\addlinespace
\textit{Imports}    &       0.398\sym{***}&      -0.063\sym{**} &       0.398\sym{***}&      -0.063\sym{**} &       0.390\sym{***}&      -0.064\sym{**} \\
                    &     (0.023)         &     (0.032)         &     (0.023)         &     (0.032)         &     (0.023)         &     (0.032)         \\
\addlinespace
\textit{Abs}        &       0.502\sym{***}&       0.225\sym{***}&       0.502\sym{***}&       0.225\sym{***}&       0.499\sym{***}&       0.225\sym{***}\\
                    &     (0.004)         &     (0.005)         &     (0.004)         &     (0.005)         &     (0.005)         &     (0.005)         \\
\addlinespace
\textit{HHI}        &      -0.047         &      -0.056         &      -0.039         &      -0.060         &      -0.048         &      -0.056         \\
                    &     (0.053)         &     (0.059)         &     (0.053)         &     (0.059)         &     (0.053)         &     (0.059)         \\
\addlinespace
\textit{HSpill $\times$ Foreign-Owned}&                     &                     &      -0.006         &       0.003         &                     &                     \\
                    &                     &                     &     (0.004)         &     (0.005)         &                     &                     \\
\addlinespace
\textit{HSpill $\times$ Exporters}&                     &                     &                     &                     &       0.009\sym{***}&       0.001         \\
                    &                     &                     &                     &                     &     (0.003)         &     (0.004)         \\
\midrule
Sector-Province-Island-Year FE&         Yes         &         Yes         &         Yes         &         Yes         &         Yes         &         Yes         \\
Kleibergen-Paap Wald F-stat&     642.907         &     632.546         &      45.756         &      56.611         &      25.076         &      22.615         \\
Cragg-Donald Wald F-stat&    9131.718         &    8604.306         &    2178.942         &    2265.609         &    2173.753         &    2066.671         \\
Observations        &  198665         &  180625         &  198665         &  180625        &  198665         &  180625         \\
\bottomrule
\bottomrule
\end{tabular}
}
\label{Main Results Horizontal with IPW}
\justifying \footnotesize \textbf{Note:} Clustered-robust standard errors are in parentheses and in firm level. Bartik-IV from labour-based ($LabBartikIV$) is used as IV for $HSpill$. $\varphi$ denotes dependent variable of TFP level, while $\Delta \varphi$ denotes dependent variable for TFP growth. All estimations use Sector-Province-Island-Year Fixed-Effects.  Observations only consist of non-superstar firms. $^{***}$, $^{**}$, $^{*}$ denote $\alpha$ at 1\%, 5\%, and 10\%. $Abs$ denotes absorptive capacity. $HHI$ denotes market concentration from Herfindahl-Hirschman Index. $\varphi$ and $\Delta \varphi$ denote total factor productivity in level and growth, respectively.
\end{table}

%% file: Main_Backward_IPW.tex
\begin{table}[htbp]\centering
\def\sym#1{\ifmmode^{#1}\else\(^{#1}\)\fi}
\caption{Basic Results--Backward Spillovers with Inverse Probability Weighting (IPW)}
\resizebox{\textwidth}{!}{
\begin{tabular}{l*{6}{c}}
\toprule
\toprule
                    &\multicolumn{1}{c}{(1)}&\multicolumn{1}{c}{(2)}&\multicolumn{1}{c}{(3)}&\multicolumn{1}{c}{(4)}&\multicolumn{1}{c}{(5)}&\multicolumn{1}{c}{(6)}\\
                    &\multicolumn{1}{c}{$\varphi$}&\multicolumn{1}{c}{$\Delta \varphi$}&\multicolumn{1}{c}{$\varphi$}&\multicolumn{1}{c}{$\Delta \varphi$}&\multicolumn{1}{c}{$\varphi$}&\multicolumn{1}{c}{$\Delta \varphi$}\\
\midrule
\textit{BSpill}     &       0.153\sym{***}&       0.123\sym{***}&       0.140\sym{***}&       0.114\sym{***}&       0.154\sym{***}&       0.123\sym{***}\\
                    &     (0.023)         &     (0.022)         &     (0.022)         &     (0.020)         &     (0.023)         &     (0.022)         \\
\addlinespace
\textit{Foreign-Owned}&       0.129         &      -0.450\sym{***}&       0.138\sym{*}  &      -0.450\sym{***}&       0.130         &      -0.451\sym{***}\\
                    &     (0.083)         &     (0.076)         &     (0.073)         &     (0.068)         &     (0.084)         &     (0.075)         \\
\addlinespace
\textit{Exporters}  &      -0.002         &      -0.276\sym{***}&       0.029         &      -0.246\sym{***}&      -0.273         &      -0.003         \\
                    &     (0.049)         &     (0.046)         &     (0.047)         &     (0.041)         &     (0.794)         &     (0.950)         \\
\addlinespace
\textit{Imports}    &       0.374\sym{***}&      -0.085         &       0.374\sym{***}&      -0.083         &       0.376\sym{***}&      -0.086         \\
                    &     (0.083)         &     (0.073)         &     (0.069)         &     (0.061)         &     (0.084)         &     (0.072)         \\
\addlinespace
\textit{Abs}        &       0.435\sym{***}&       0.173\sym{***}&       0.481\sym{***}&       0.229\sym{***}&       0.429\sym{***}&       0.178\sym{***}\\
                    &     (0.017)         &     (0.015)         &     (0.023)         &     (0.020)         &     (0.023)         &     (0.023)         \\
\addlinespace
\textit{HHI}        &      -1.817\sym{***}&      -1.375\sym{***}&       3.786\sym{*}  &       5.208\sym{**} &      -1.853\sym{***}&      -1.335\sym{***}\\
                    &     (0.341)         &     (0.301)         &     (2.091)         &     (2.089)         &     (0.356)         &     (0.334)         \\
\addlinespace
\textit{BSpill $\times$ High-Concentration}&                     &                     &      -0.089\sym{***}&      -0.104\sym{***}&                     &                     \\
                    &                     &                     &     (0.033)         &     (0.034)         &                     &                     \\
\addlinespace
\textit{BSpill $\times$ Exporters}&                     &                     &                     &                     &       0.012         &      -0.012         \\
                    &                     &                     &                     &                     &     (0.036)         &     (0.043)         \\
\midrule
Sector-Province-Island-Year FE&         Yes         &         Yes         &         Yes         &         Yes         &         Yes         &         Yes         \\
Kleibergen-Paap Wald F-stat&     111.055         &     138.292         &      37.522         &     113.606         &      57.459         &      56.091         \\
Cragg-Donald Wald F-stat&      13.287         &      16.868         &       8.732         &      10.703         &       6.345         &       7.620         \\
Observations        &  200440         &  182093         &  200440         &  182093         &  200440         &  182093         \\
\bottomrule
\bottomrule
\end{tabular}
}
\label{Main Results Backward with IPW}
\justifying \footnotesize \textbf{Note:} Clustered-robust standard errors are in parentheses and in firm level. Bartik-IV from output-based ($TarrBartikIV$) is used as IV for $BSpill$. $\varphi$ denotes dependent variable of TFP level, while $\Delta \varphi$ denotes dependent variable for TFP growth. All estimations use Sector-Province-Island-Year Fixed-Effects.  Observations only consist of non-superstar firms. $^{***}$, $^{**}$, $^{*}$ denote $\alpha$ at 1\%, 5\%, and 10\%. $Abs$ denotes absorptive capacity. $HHI$ denotes market concentration from Herfindahl-Hirschman Index. $\varphi$ and $\Delta \varphi$ denote total factor productivity in level and growth, respectively.
\end{table}

%% file: Main_Forward_IPW.tex
\begin{table}[htbp]\centering
\def\sym#1{\ifmmode^{#1}\else\(^{#1}\)\fi}
\caption{Basic Results--Forward Spillovers with Inverse Probability Weighting (IPW)}
\resizebox{\textwidth}{!}{
\begin{tabular}{l*{6}{c}}
\toprule
\toprule
                    &\multicolumn{1}{c}{(1)}&\multicolumn{1}{c}{(2)}&\multicolumn{1}{c}{(3)}&\multicolumn{1}{c}{(4)}&\multicolumn{1}{c}{(5)}&\multicolumn{1}{c}{(6)}\\
                    &\multicolumn{1}{c}{$\varphi$}&\multicolumn{1}{c}{$\Delta \varphi$}&\multicolumn{1}{c}{$\varphi$}&\multicolumn{1}{c}{$\Delta \varphi$}&\multicolumn{1}{c}{$\varphi$}&\multicolumn{1}{c}{$\Delta \varphi$}\\
\midrule
\textit{FSpill}     &       0.225\sym{***}&       0.215\sym{***}&       0.142\sym{***}&       0.127\sym{***}&       0.226\sym{***}&       0.216\sym{***}\\
                    &     (0.045)         &     (0.050)         &     (0.025)         &     (0.029)         &     (0.045)         &     (0.050)         \\
\addlinespace
\textit{Foreign-Owned}&       0.125         &      -0.497\sym{***}&       0.222\sym{*}  &      -0.393\sym{***}&       0.129         &      -0.488\sym{***}\\
                    &     (0.169)         &     (0.173)         &     (0.120)         &     (0.120)         &     (0.165)         &     (0.165)         \\
\addlinespace
\textit{Exporters}  &       0.016         &      -0.300\sym{***}&       0.068         &      -0.242\sym{***}&       0.483         &       0.518         \\
                    &     (0.081)         &     (0.086)         &     (0.058)         &     (0.060)         &     (1.018)         &     (1.000)         \\
\addlinespace
\textit{Imports}    &       0.291\sym{*}  &      -0.149         &       0.325\sym{***}&      -0.115         &       0.267         &      -0.192         \\
                    &     (0.168)         &     (0.169)         &     (0.123)         &     (0.121)         &     (0.171)         &     (0.171)         \\
\addlinespace
\textit{Abs}        &       0.498\sym{***}&       0.222\sym{***}&       0.485\sym{***}&       0.205\sym{***}&       0.505\sym{***}&       0.232\sym{***}\\
                    &     (0.022)         &     (0.022)         &     (0.018)         &     (0.017)         &     (0.026)         &     (0.025)         \\
\addlinespace
\textit{HHI}        &      -0.221         &      -0.298         &      -4.274\sym{***}&      -4.552\sym{***}&      -0.234         &      -0.305         \\
                    &     (0.222)         &     (0.236)         &     (1.259)         &     (1.299)         &     (0.224)         &     (0.233)         \\
\addlinespace
\textit{FSpill $\times$ High-Concentration}&                     &                     &       0.067\sym{***}&       0.070\sym{***}&                     &                     \\
                    &                     &                     &     (0.020)         &     (0.021)         &                     &                     \\
\addlinespace
\textit{FSpill $\times$ Exporters}     &                     &                     &                     &                     &      -0.027         &      -0.047         \\
                    &                     &                     &                     &                     &     (0.060)         &     (0.058)         \\
\midrule
Sector-Province-Island-Year FE&         Yes         &         Yes         &         Yes         &         Yes         &         Yes         &         Yes         \\
Kleibergen-Paap Wald F-stat&      39.272         &      33.784         &      45.121         &      39.465         &      20.298         &      17.612         \\
Cragg-Donald Wald F-stat&       4.686         &       4.175         &       3.754         &       3.366         &       2.385         &       2.131         \\
Observations        &  200440         &  182093         &  200440         &  182093         &  200440         &  182093         \\
\bottomrule
\bottomrule
\end{tabular}
}
\label{Main Results Forward with IPW}
\justifying \footnotesize \textbf{Note:} Clustered-robust standard errors are in parentheses and in firm level. Bartik-IV from output-based ($TarrBartikIV$) is used as IV for $FSpill$. $\varphi$ denotes dependent variable of TFP level, while $\Delta \varphi$ denotes dependent variable for TFP growth. All estimations use Sector-Province-Island-Year Fixed-Effects.  Observations only consist of non-superstar firms. $^{***}$, $^{**}$, $^{*}$ denote $\alpha$ at 1\%, 5\%, and 10\%. $Abs$ denotes absorptive capacity. $HHI$ denotes market concentration from Herfindahl-Hirschman Index. $\varphi$ and $\Delta \varphi$ denote total factor productivity in level and growth, respectively.
\end{table}

%% file: First_Stage_Horizontal_with_IPW.tex
\begin{table}[htbp]\centering
\def\sym#1{\ifmmode^{#1}\else\(^{#1}\)\fi}
\caption{First-Stage: Basic Results--Horizontal Spillovers with Inverse Probability Weighting (IPW)}
\begin{tabular}{l*{2}{c}}
\toprule
                    &\multicolumn{1}{c}{(1)}&\multicolumn{1}{c}{(2)}\\
                    &\multicolumn{1}{c}{$\varphi$}&\multicolumn{1}{c}{$\Delta\varphi$}\\
\midrule
\textit{LabBartik-IV}&      18.896\sym{***}&      19.042\sym{***}\\
                    &     (0.745)         &     (0.757)         \\
\addlinespace
\textit{Foreign-Owned}&       1.936\sym{***}&       1.811\sym{***}\\
                    &     (0.469)         &     (0.488)         \\
\addlinespace
\textit{Exporters}  &      -0.667\sym{**} &      -0.604\sym{**} \\
                    &     (0.264)         &     (0.281)         \\
\addlinespace
\textit{Imports}    &       0.825         &       0.912         \\
                    &     (0.558)         &     (0.578)         \\
\addlinespace
\textit{Abs}        &      -0.001         &      -0.026         \\
                    &     (0.079)         &     (0.082)         \\
\addlinespace
\textit{HHI}        &      11.188\sym{***}&      10.716\sym{***}\\
                    &     (1.410)         &     (1.495)         \\
\addlinespace
Constant            &      35.361\sym{***}&      35.778\sym{***}\\
                    &     (1.268)         &     (1.330)         \\
\midrule
Sector-Province-Island-Year FE&         Yes         &         Yes         \\
F-statistics        &     130.067         &     124.938         \\
Observations        &  198665         &  180625         \\
\bottomrule
\bottomrule
\end{tabular}
\label{First-Stage Main Horizontal with IPW}
\\
\justifying \footnotesize \textbf{Note:} Clustered-robust standard errors are in parentheses and in firm level. Bartik-IV from labour-based ($LabBartikIV$) is used as IV for $HSpill$. $\varphi$ denotes dependent variable of TFP level, while $\Delta \varphi$ denotes dependent variable for TFP growth. All estimations use Sector-Province-Island-Year Fixed-Effects. $^{***}$, $^{**}$, $^{*}$ denote $\alpha$ at 1\%, 5\%, and 10\%. $Abs$ denotes absorptive capacity. $HHI$ denotes market concentration from Herfindahl-Hirschman Index.
\end{table}

%% file: First_Stage_Backward_with_IPW.tex
\begin{table}[htbp]\centering
\def\sym#1{\ifmmode^{#1}\else\(^{#1}\)\fi}
\caption{First-Stage: Basic Results--Backward Spillovers with Inverse Probability Weighting (IPW)}
\begin{tabular}{l*{2}{c}}
\toprule
                    &\multicolumn{1}{c}{(1)}&\multicolumn{1}{c}{(2)}\\
                    &\multicolumn{1}{c}{$\varphi$}&\multicolumn{1}{c}{$\Delta\varphi$}\\
\midrule
\textit{TarrBartik-IV}        &      -0.363\sym{***}&      -0.420\sym{***}\\
                    &     (0.034)         &     (0.036)         \\
\addlinespace
\textit{Foreign-Owned}&       2.024\sym{***}&       1.968\sym{***}\\
                    &     (0.438)         &     (0.458)         \\
\addlinespace
\textit{Exporters}  &       1.110\sym{***}&       1.132\sym{***}\\
                    &     (0.269)         &     (0.287)         \\
\addlinespace
\textit{Imports}    &       0.147         &       0.267         \\
                    &     (0.515)         &     (0.535)         \\
\addlinespace
\textit{Abs}        &       0.435\sym{***}&       0.407\sym{***}\\
                    &     (0.084)         &     (0.088)         \\
\addlinespace
\textit{HHI}        &      12.071\sym{***}&      11.189\sym{***}\\
                    &     (1.367)         &     (1.453)         \\
\addlinespace
Constant            &      26.212\sym{***}&      26.886\sym{***}\\
                    &     (1.350)         &     (1.420)         \\
\midrule
Sector-Province-Island-Year FE&         Yes         &         Yes         \\
F-statistics        &      42.493         &      42.463         \\
Observations        &  200440         &  182093         \\
\bottomrule
\bottomrule
\end{tabular}
\label{First-Stage Main Backwards with IPW}
\\
\justifying \footnotesize \textbf{Note:} Clustered-robust standard errors are in parentheses and in firm level. Bartik-IV from output-based ($TarrBartikIV$) is used as IV for $BSpill$. $\varphi$ denotes dependent variable of TFP level, while $\Delta \varphi$ denotes dependent variable for TFP growth. All estimations use Sector-Province-Island-Year Fixed-Effects. $^{***}$, $^{**}$, $^{*}$ denote $\alpha$ at 1\%, 5\%, and 10\%. $Abs$ denotes absorptive capacity. $HHI$ denotes market concentration from Herfindahl-Hirschman Index.
\end{table}

%% file: First_Stage_Forward_with_IPW.tex
\begin{table}[htbp]\centering
\def\sym#1{\ifmmode^{#1}\else\(^{#1}\)\fi}
\caption{First-Stage: Basic Results--Forward Spillovers with Inverse Probability Weighting (IPW)}
\begin{tabular}{l*{2}{c}}
\toprule
                    &\multicolumn{1}{c}{(1)}&\multicolumn{1}{c}{(2)}\\
                    &\multicolumn{1}{c}{$\varphi$}&\multicolumn{1}{c}{$\Delta\varphi$}\\
\midrule
\textit{TarrBartik-IV}       &      -0.248\sym{***}&      -0.240\sym{***}\\
                    &     (0.040)         &     (0.041)         \\
\addlinespace
\textit{Foreign-Owned}&       1.397\sym{**} &       1.344\sym{*}  \\
                    &     (0.662)         &     (0.699)         \\
\addlinespace
\textit{Exporters}  &       0.676\sym{**} &       0.758\sym{**} \\
                    &     (0.338)         &     (0.358)         \\
\addlinespace
\textit{Imports}    &       0.472         &       0.450         \\
                    &     (0.730)         &     (0.765)         \\
\addlinespace
\textit{Abs}        &       0.012         &       0.004         \\
                    &     (0.095)         &     (0.099)         \\
\addlinespace
\textit{HHI}        &       1.124         &       1.403         \\
                    &     (0.959)         &     (1.035)         \\
\addlinespace
Constant            &      21.004\sym{***}&      21.195\sym{***}\\
                    &     (1.529)         &     (1.605)         \\
\midrule
Sector-Province-Island-Year FE&         Yes         &         Yes         \\
F-statistics        &      10.086         &       8.769         \\
Observations        &  200440         &  182093         \\
\bottomrule
\bottomrule
\end{tabular}
\label{First-Stage Main Forwards with IPW}
\\
\justifying \footnotesize \textbf{Note:} Clustered-robust standard errors are in parentheses and in firm level. Bartik-IV from output-based ($TarrBartikIV$) is used as IV for $FSpill$. $\varphi$ denotes dependent variable of TFP level, while $\Delta \varphi$ denotes dependent variable for TFP growth. All estimations use Sector-Province-Island-Year Fixed-Effects. $^{***}$, $^{**}$, $^{*}$ denote $\alpha$ at 1\%, 5\%, and 10\%. $Abs$ denotes absorptive capacity. $HHI$ denotes market concentration from Herfindahl-Hirschman Index.
\end{table}

%% file: Heterogeneous_Spillovers_FS.tex
\begin{table}[htbp]\centering
\def\sym#1{\ifmmode^{#1}\else\(^{#1}\)\fi}
\caption{Heterogeneous Superstars: Foreign and Domestic--First Stage}
\resizebox{\textwidth}{!}{
\begin{tabular}{l*{12}{c}}
\toprule
\toprule
                    &\multicolumn{1}{c}{(1)}&\multicolumn{1}{c}{(2)}&\multicolumn{1}{c}{(3)}&\multicolumn{1}{c}{(4)}&\multicolumn{1}{c}{(5)}&\multicolumn{1}{c}{(6)}&\multicolumn{1}{c}{(7)}&\multicolumn{1}{c}{(8)}&\multicolumn{1}{c}{(9)}&\multicolumn{1}{c}{(10)}&\multicolumn{1}{c}{(11)}&\multicolumn{1}{c}{(12)}\\
                    \midrule
                    & \multicolumn{2}{c}{Foreign} &  \multicolumn{2}{c}{Domestic} & \multicolumn{2}{c}{Foreign} &  \multicolumn{2}{c}{Domestic} & \multicolumn{2}{c}{Foreign} &  \multicolumn{2}{c}{Domestic}    \\
                    &\multicolumn{1}{c}{$\varphi$}&\multicolumn{1}{c}{$\Delta\varphi$}&\multicolumn{1}{c}{$\varphi$}&\multicolumn{1}{c}{$\Delta\varphi$}&\multicolumn{1}{c}{$\varphi$}&\multicolumn{1}{c}{$\Delta\varphi$}&\multicolumn{1}{c}{$\varphi$}&\multicolumn{1}{c}{$\Delta\varphi$}&\multicolumn{1}{c}{$\varphi$}&\multicolumn{1}{c}{$\Delta\varphi$}&\multicolumn{1}{c}{$\varphi$}&\multicolumn{1}{c}{$\Delta\varphi$}\\
\midrule
\textit{LabBartik-IV}&       2.952\sym{***}&       2.923\sym{***} &       8.373\sym{***}&       8.128\sym{***}&                     &                     &                     &                     &                     &                     &                     &                     \\
                    &     (0.149)         &     (0.152)         &     (0.447)         &     (0.441)         &                     &                     &                     &                     &                     &                     &                     &                     \\
\addlinespace
\textit{TarrBartik-IV} &                     &                     &                     &                     &       -2.454\sym{***}&      -4.130\sym{***}    &      -0.413\sym{***}&     -0.391\sym{***}    &       -0.770\sym{***}&        -1.472\sym{***}      &         -0.363\sym{***}&      -0.353\sym{***}            \\
                    &                     &                     &                     &                     &       (0.160)         &     (0.200)      &     (0.030)         &      (0.030)         &     (0.112)         &    (0.124)           &        (0.019)         &     (0.020)            \\
\addlinespace
\textit{Foreign-Owned}&       1.509\sym{***}&       1.402\sym{***}&      -0.155         &      -0.112         &       0.977\sym{***}&       0.884\sym{***}&       0.544\sym{*}  &       0.599\sym{*}  &       1.182\sym{***}&       1.166\sym{***}&      -0.176         &      -0.194         \\
                    &     (0.284)         &     (0.293)         &     (0.320)         &     (0.337)         &     (0.282)         &     (0.292)         &     (0.291)         &     (0.307)         &     (0.265)         &     (0.281)         &     (0.407)         &     (0.431)         \\
\addlinespace
\textit{Exporters}  &      -0.483\sym{***}&      -0.490\sym{***}&      -1.140\sym{***}&      -1.112\sym{***}&      -0.031         &      -0.055         &       0.112         &       0.108         &      -0.141         &      -0.127         &      -0.278         &      -0.241         \\
                    &     (0.127)         &     (0.134)         &     (0.168)         &     (0.180)         &     (0.134)         &     (0.143)         &     (0.167)         &     (0.178)         &     (0.112)         &     (0.123)         &     (0.200)         &     (0.208)         \\
\addlinespace
\textit{Imports}    &       1.378\sym{***}&       1.514\sym{***}&      -0.752\sym{**} &      -0.749\sym{**} &       0.960\sym{***}&       1.107\sym{***}&      -1.018\sym{***}&      -1.009\sym{***}&       1.337\sym{***}&       1.358\sym{***}&      -0.568         &      -0.566         \\
                    &     (0.351)         &     (0.365)         &     (0.363)         &     (0.380)         &     (0.350)         &     (0.365)         &     (0.325)         &     (0.341)         &     (0.311)         &     (0.325)         &     (0.448)         &     (0.468)         \\
\addlinespace
\textit{Abs}        &       0.025         &       0.001         &      -0.385\sym{***}&      -0.420\sym{***}&       0.025         &      -0.024         &       0.088         &       0.079         &       0.027         &      -0.019         &      -0.025         &       0.005         \\
                    &     (0.046)         &     (0.049)         &     (0.064)         &     (0.067)         &     (0.049)         &     (0.051)         &     (0.059)         &     (0.062)         &     (0.036)         &     (0.039)         &     (0.072)         &     (0.074)         \\
\addlinespace
\textit{HHI}        &       0.020         &      -0.500         &      14.161\sym{***}&      13.760\sym{***}&       2.299\sym{***}&       1.844\sym{***}&      15.695\sym{***}&      14.767\sym{***}&      -2.834\sym{***}&      -2.900\sym{***}&      11.594\sym{***}&      11.381\sym{***}\\
                    &     (0.641)         &     (0.666)         &     (1.150)         &     (1.210)         &     (0.479)         &     (0.495)         &     (1.185)         &     (1.255)         &     (0.405)         &     (0.411)         &     (0.871)         &     (0.929)         \\
\addlinespace
Constant            &       7.446\sym{***}&       7.807\sym{***}&      22.623\sym{***}&      23.285\sym{***}&       6.777\sym{***}&       7.617\sym{***}&      15.483\sym{***}&      15.819\sym{***}&       3.255\sym{***}&       4.017\sym{***}&      10.690\sym{***}&      10.192\sym{***}\\
                    &     (0.747)         &     (0.786)         &     (1.033)         &     (1.087)         &     (0.790)         &     (0.825)         &     (0.945)         &     (1.001)         &     (0.590)         &     (0.636)         &     (1.168)         &     (1.199)         \\
\midrule
Sector-Province-Island-Year FE&         Yes         &         Yes         &         Yes         &         Yes         &         Yes         &         Yes         &         Yes         &         Yes         &         Yes         &         Yes         &         Yes         &         Yes         \\
F-statistics        &      77.202         &      71.628         &      98.873         &      92.317         &      45.986         &      75.754         &      56.495         &      47.097         &      30.399         &      51.396         &      77.139         &      67.759         \\
Observations        &  198665         &  180625         &  198665         &  180625         &  200440         &  182093         &  200440         &  182093         &  200440         &  182093         &  200440         &  182093         \\
\bottomrule
\bottomrule
\end{tabular}
}
\label{Heterogeneous Superstar FS}
\justifying \footnotesize \textbf{Note:} Clustered-robust standard errors are in parentheses and in firm level. Bartik-IV from labour-based and output growth ($LabBartikIV$) is used as IV for $HSpill$, while Bartik-IV from output-based and tariff ($TarrBartikIV$) is used as IV for $BSpill$ adn $FSpill$. $Foreign$ denotes foreign superstar spillovers, while $Domestic$ denotes domestic foreign spillovers. Both foreign and domestic superstars are non-exporters. $\varphi$ denotes dependent variable of TFP level, while $\Delta \varphi$ denotes dependent variable for TFP growth. All estimations use Sector-Province-Island-Year Fixed-Effects.  Observations only consist of non-superstar firms. $^{***}$, $^{**}$, $^{*}$ denote $\alpha$ at 1\%, 5\%, and 10\%. $Abs$ denotes absorptive capacity. $HHI$ denotes market concentration from Herfindahl-Hirschman Index. $\varphi$ and $\Delta \varphi$ denote total factor productivity in level and growth, respectively.
\end{table}

%% file: Robustness_Split_Large_Medium.tex
\begin{table}[htbp]\centering
\def\sym#1{\ifmmode^{#1}\else\(^{#1}\)\fi}
\caption{Robustness Results--Large and Medium Observations}
\resizebox{\textwidth}{!}{
\begin{tabular}{l*{12}{c}}
\toprule
\toprule
                    &\multicolumn{1}{c}{(1)}&\multicolumn{1}{c}{(2)}&\multicolumn{1}{c}{(3)}&\multicolumn{1}{c}{(4)}&\multicolumn{1}{c}{(5)}&\multicolumn{1}{c}{(6)}&\multicolumn{1}{c}{(7)}&\multicolumn{1}{c}{(8)}&\multicolumn{1}{c}{(9)}&\multicolumn{1}{c}{(10)}&\multicolumn{1}{c}{(11)}&\multicolumn{1}{c}{(12)}\\
\midrule
                     & \multicolumn{2}{c}{$\varphi$} &  \multicolumn{2}{c}{$\Delta\varphi$} & \multicolumn{2}{c}{$\varphi$} &  \multicolumn{2}{c}{$\Delta\varphi$} & \multicolumn{2}{c}{$\varphi$} &  \multicolumn{2}{c}{$\Delta\varphi$}    \\
                    &\multicolumn{1}{c}{M}&\multicolumn{1}{c}{L}&\multicolumn{1}{c}{M}&\multicolumn{1}{c}{L}&\multicolumn{1}{c}{M}&\multicolumn{1}{c}{L}&\multicolumn{1}{c}{M}&\multicolumn{1}{c}{L}&\multicolumn{1}{c}{M}&\multicolumn{1}{c}{L}&\multicolumn{1}{c}{M}&\multicolumn{1}{c}{L}\\
\midrule
\textit{HSpill}     &       0.004\sym{***}&       0.009\sym{***}&       0.002         &       0.006\sym{**} &                     &                     &                     &                     &                     &                     &                     &                     \\
                    &     (0.001)         &     (0.002)         &     (0.001)         &     (0.003)         &                     &                     &                     &                     &                     &                     &                     &                     \\
\addlinespace
\textit{BSpill}     &                     &                     &                     &                     &       0.165\sym{***}&       0.105\sym{**} &       0.120\sym{***}&       0.129\sym{**} &                     &                     &                     &                     \\
                    &                     &                     &                     &                     &     (0.027)         &     (0.042)         &     (0.024)         &     (0.060)         &                     &                     &                     &                     \\
\addlinespace
\textit{FSpill}     &                     &                     &                     &                     &                     &                     &                     &                     &       0.216\sym{***}&       0.174\sym{*}  &       0.195\sym{***}&       0.222\sym{*}  \\
                    &                     &                     &                     &                     &                     &                     &                     &                     &     (0.044)         &     (0.091)         &     (0.049)         &     (0.134)         \\
\addlinespace
\textit{Foreign-Owned}&       0.432\sym{***}&       0.205\sym{***}&      -0.313\sym{***}&      -0.180\sym{***}&      -0.096         &       0.133\sym{**} &      -0.674\sym{***}&      -0.263\sym{***}&      -0.369         &       0.323\sym{**} &      -1.056\sym{***}&       0.007         \\
                    &     (0.030)         &     (0.022)         &     (0.051)         &     (0.037)         &     (0.142)         &     (0.064)         &     (0.121)         &     (0.088)         &     (0.292)         &     (0.144)         &     (0.303)         &     (0.207)         \\
\addlinespace
\textit{Exporters}  &       0.031\sym{**} &      -0.093\sym{***}&      -0.155\sym{***}&      -0.179\sym{***}&      -0.236\sym{***}&      -0.161\sym{***}&      -0.345\sym{***}&      -0.272\sym{***}&      -0.076         &      -0.126         &      -0.278\sym{***}&      -0.223\sym{*}  \\
                    &     (0.014)         &     (0.016)         &     (0.021)         &     (0.024)         &     (0.076)         &     (0.050)         &     (0.064)         &     (0.071)         &     (0.098)         &     (0.085)         &     (0.098)         &     (0.116)         \\
\addlinespace
\textit{Imports}    &       0.233\sym{***}&       0.272\sym{***}&      -0.000         &      -0.156\sym{***}&      -0.004         &       0.329\sym{***}&      -0.175         &      -0.088         &       0.272         &       0.188         &       0.049         &      -0.273         \\
                    &     (0.026)         &     (0.029)         &     (0.042)         &     (0.046)         &     (0.136)         &     (0.078)         &     (0.110)         &     (0.103)         &     (0.202)         &     (0.187)         &     (0.193)         &     (0.256)         \\
\addlinespace
\textit{Abs}        &       0.449\sym{***}&       0.396\sym{***}&       0.244\sym{***}&       0.170\sym{***}&       0.366\sym{***}&       0.376\sym{***}&       0.187\sym{***}&       0.142\sym{***}&       0.452\sym{***}&       0.362\sym{***}&       0.247\sym{***}&       0.128\sym{**} \\
                    &     (0.004)         &     (0.011)         &     (0.005)         &     (0.012)         &     (0.021)         &     (0.022)         &     (0.017)         &     (0.030)         &     (0.024)         &     (0.043)         &     (0.022)         &     (0.058)         \\
\addlinespace
\textit{HHI}        &       0.117\sym{**} &      -0.346\sym{***}&      -0.004         &      -0.170         &      -1.181\sym{***}&      -2.380\sym{**} &      -0.738\sym{***}&      -2.781\sym{**} &       0.007         &      -0.723\sym{*}  &      -0.123         &      -0.757         \\
                    &     (0.055)         &     (0.103)         &     (0.071)         &     (0.123)         &     (0.325)         &     (0.931)         &     (0.254)         &     (1.327)         &     (0.267)         &     (0.420)         &     (0.269)         &     (0.596)         \\
\midrule
Sector-Province-Island-Year FE&         Yes         &         Yes         &         Yes         &         Yes         &         Yes         &         Yes         &         Yes         &         Yes         &         Yes         &         Yes         &         Yes         &         Yes         \\
Kleibergen-Paap Wald F-stat&     745.526         &     181.960         &     717.617         &     176.984         &      80.562         &      42.798         &     104.028         &      37.075         &      38.471         &       7.369         &      31.459         &       5.766         \\
Cragg-Donald Wald F-stat&    6822.894         &    2192.120         &    6392.581         &    2112.548         &      10.763         &       3.439         &      14.330         &       2.907         &       4.861         &       0.872         &       4.219         &       0.684         \\
Observations        &  148091         &   50574         &  133532         &   47093         &  148943         &   51497         &  134243         &   47850         &  148943         &   51497         &  134243         &   47850         \\
\bottomrule
\bottomrule
\end{tabular}
}
\label{Robustness Large Medium}
\justifying \footnotesize \textbf{Note:} Clustered-robust standard errors are in parentheses and in firm level. Bartik-IV from labour-based and output growth ($LabBartikIV$) is used as IV for $HSpill$, while Bartik-IV from output-based and tariff ($TarrBartikIV$) is used as IV for $BSpill$ adn $FSpill$. $Foreign$ denotes foreign superstar spillovers, while $Domestic$ denotes domestic foreign spillovers. Both foreign and domestic superstars are non-exporters. $\varphi$ denotes dependent variable of TFP level, while $\Delta \varphi$ denotes dependent variable for TFP growth. All estimations use Sector-Province-Island-Year Fixed-Effects.  Observations only consist of non-superstar firms. $^{***}$, $^{**}$, $^{*}$ denote $\alpha$ at 1\%, 5\%, and 10\%. $Abs$ denotes absorptive capacity. $HHI$ denotes market concentration from Herfindahl-Hirschman Index. $\varphi$ and $\Delta \varphi$ denote total factor productivity in level and growth, respectively.
\end{table}

%% file: Robustness_Horizontal_-_IV.tex
\begin{table}[htbp]\centering
\def\sym#1{\ifmmode^{#1}\else\(^{#1}\)\fi}
\caption{Robustness Test: Alternative IV--Horizontal Spillovers}
\resizebox{\textwidth}{!}{
\begin{tabular}{l*{6}{c}}
\toprule
\toprule
                    &\multicolumn{1}{c}{(1)}&\multicolumn{1}{c}{(2)}&\multicolumn{1}{c}{(3)}&\multicolumn{1}{c}{(4)}&\multicolumn{1}{c}{(5)}&\multicolumn{1}{c}{(6)}\\
                    &\multicolumn{1}{c}{$\varphi$}&\multicolumn{1}{c}{$\Delta \varphi$}&\multicolumn{1}{c}{$\varphi$}&\multicolumn{1}{c}{$\Delta \varphi$}&\multicolumn{1}{c}{$\varphi$}&\multicolumn{1}{c}{$\Delta \varphi$}\\
\midrule
\textit{HSpill}     &       0.007\sym{***}&       0.004\sym{***}&       0.007\sym{***}&       0.003\sym{**} &       0.006\sym{***}&       0.003\sym{***}\\
                    &     (0.001)         &     (0.001)         &     (0.001)         &     (0.001)         &     (0.001)         &     (0.001)         \\
\addlinespace
\textit{Foreign-Owned}&       0.439\sym{***}&      -0.212\sym{***}&       0.614\sym{***}&      -0.343\sym{**} &       0.438\sym{***}&      -0.212\sym{***}\\
                    &     (0.020)         &     (0.030)         &     (0.125)         &     (0.172)         &     (0.020)         &     (0.030)         \\
\addlinespace
\textit{Exporters}  &       0.176\sym{***}&      -0.136\sym{***}&       0.176\sym{***}&      -0.136\sym{***}&      -0.104         &      -0.174         \\
                    &     (0.011)         &     (0.017)         &     (0.011)         &     (0.017)         &     (0.093)         &     (0.117)         \\
\addlinespace
\textit{Imports}    &       0.400\sym{***}&      -0.061\sym{*}  &       0.400\sym{***}&      -0.061\sym{*}  &       0.391\sym{***}&      -0.062\sym{*}  \\
                    &     (0.023)         &     (0.032)         &     (0.023)         &     (0.032)         &     (0.023)         &     (0.032)         \\
\addlinespace
\textit{Abs}        &       0.502\sym{***}&       0.225\sym{***}&       0.502\sym{***}&       0.225\sym{***}&       0.499\sym{***}&       0.225\sym{***}\\
                    &     (0.004)         &     (0.005)         &     (0.004)         &     (0.005)         &     (0.005)         &     (0.005)         \\
\addlinespace
\textit{HHI}        &      -0.039         &      -0.049         &      -0.032         &      -0.055         &      -0.040         &      -0.050         \\
                    &     (0.052)         &     (0.059)         &     (0.052)         &     (0.059)         &     (0.053)         &     (0.059)         \\
\addlinespace
\textit{HSpill $\times$ Foreign-Owned}&                     &                     &      -0.006         &       0.004         &                     &                     \\
                    &                     &                     &     (0.004)         &     (0.005)         &                     &                     \\
\addlinespace
\textit{HSpill $\times$ Exporters}&                     &                     &                     &                     &       0.010\sym{***}&       0.001         \\
                    &                     &                     &                     &                     &     (0.003)         &     (0.004)         \\
\midrule
Sector-Province-Island-Year FE&         Yes         &         Yes         &         Yes         &         Yes         &         Yes         &         Yes         \\
Kleibergen-Paap Wald F-stat&     620.648         &     610.174         &      40.020         &      49.453         &      23.678         &      21.473         \\
Cragg-Donald Wald F-stat&    9169.872         &    8649.320         &    2188.689         &    2283.008         &    2189.000         &    2083.583         \\
Observations        &  198665         &  180625         &  198665         &  180625        &  198665         &  180625         \\
\bottomrule
\bottomrule
\end{tabular}
}
\label{Robustness IV Horizontal}
\justifying \footnotesize \textbf{Note:} Clustered-robust standard errors are in parentheses and in firm level. Bartik-IV from labour-based (both skilled and unskilled) ($LabBartikIV$) is used as IV for $HSpill$. $\varphi$ denotes dependent variable of TFP level, while $\Delta \varphi$ denotes dependent variable for TFP growth. All estimations use Sector-Province-Island-Year Fixed-Effects. Observations only consist of non-superstar firms. $^{***}$, $^{**}$, $^{*}$ denote $\alpha$ at 1\%, 5\%, and 10\%. $Abs$ denotes absorptive capacity. $HHI$ denotes market concentration from Herfindahl-Hirschman Index. $\varphi$ and $\Delta \varphi$ denote total factor productivity in level and growth, respectively. The results with Inverse Probability Weighting are reported in the Appendix in Table \ref{Main Results Horizontal with IPW}.
\end{table}

%% file: Robustness_Backward_-_IV.tex
\begin{table}[htbp]\centering
\def\sym#1{\ifmmode^{#1}\else\(^{#1}\)\fi}
\caption{Robustness Test: Alternative IV--Backward Spillovers}
\resizebox{\textwidth}{!}{
\begin{tabular}{l*{6}{c}}
\toprule
\toprule
                    &\multicolumn{1}{c}{(1)}&\multicolumn{1}{c}{(2)}&\multicolumn{1}{c}{(3)}&\multicolumn{1}{c}{(4)}&\multicolumn{1}{c}{(5)}&\multicolumn{1}{c}{(6)}\\
                    &\multicolumn{1}{c}{$\varphi$}&\multicolumn{1}{c}{$\Delta \varphi$}&\multicolumn{1}{c}{$\varphi$}&\multicolumn{1}{c}{$\Delta \varphi$}&\multicolumn{1}{c}{$\varphi$}&\multicolumn{1}{c}{$\Delta \varphi$}\\
\midrule
\textit{BSpill}     &       0.025\sym{***}&       0.019\sym{***}&       0.024\sym{***}&       0.019\sym{***}&       0.030\sym{***}&       0.022\sym{***}\\
                    &     (0.006)         &     (0.007)         &     (0.006)         &     (0.006)         &     (0.008)         &     (0.008)         \\
\addlinespace
\textit{Foreign-Owned}&       0.397\sym{***}&      -0.242\sym{***}&       0.398\sym{***}&      -0.241\sym{***}&       0.392\sym{***}&      -0.244\sym{***}\\
                    &     (0.025)         &     (0.033)         &     (0.024)         &     (0.033)         &     (0.029)         &     (0.035)         \\
\addlinespace
\textit{Exporters}  &       0.147\sym{***}&      -0.162\sym{***}&       0.148\sym{***}&      -0.160\sym{***}&      -0.350         &      -0.558\sym{**} \\
                    &     (0.014)         &     (0.019)         &     (0.014)         &     (0.019)         &     (0.243)         &     (0.262)         \\
\addlinespace
\textit{Imports}    &       0.386\sym{***}&      -0.053         &       0.386\sym{***}&      -0.053         &       0.389\sym{***}&      -0.053         \\
                    &     (0.026)         &     (0.033)         &     (0.025)         &     (0.033)         &     (0.029)         &     (0.035)         \\
\addlinespace
\textit{Abs}        &       0.500\sym{***}&       0.217\sym{***}&       0.502\sym{***}&       0.218\sym{***}&       0.489\sym{***}&       0.209\sym{***}\\
                    &     (0.005)         &     (0.006)         &     (0.005)         &     (0.006)         &     (0.009)         &     (0.009)         \\
\addlinespace
\textit{HHI}        &      -0.272\sym{***}&      -0.180\sym{*}  &      -0.127         &      -0.050         &      -0.374\sym{***}&      -0.256\sym{**} \\
                    &     (0.089)         &     (0.092)         &     (0.077)         &     (0.087)         &     (0.129)         &     (0.128)         \\
\addlinespace
\textit{BSpill $\times$ High-Concentration}&                     &                     &      -0.002\sym{*}  &      -0.002         &                     &                     \\
                    &                     &                     &     (0.001)         &     (0.001)         &                     &                     \\
\addlinespace
\textit{BSpill $\times$ Exporters}&                     &                     &                     &                     &       0.023\sym{**} &       0.018         \\
                    &                     &                     &                     &                     &     (0.011)         &     (0.012)         \\
\midrule
Sector-Province-Island-Year FE&         Yes         &         Yes         &         Yes         &         Yes         &         Yes         &         Yes         \\
Kleibergen-Paap Wald F-stat&      78.842         &      83.947         &      48.025         &      50.400         &      14.597         &      15.104         \\
Cragg-Donald Wald F-stat&     112.961         &     124.568         &      67.418         &      73.225         &      26.134         &      30.432         \\
Observations        &  214620         &  183074         &  214620         &  183074         &  214620         &  183074         \\
\bottomrule
\bottomrule
\end{tabular}
}
\label{Robustness IV Backward}
\justifying \footnotesize \textbf{Note:} Clustered-robust standard errors are in parentheses and in firm level. Average Road Density is used as IV for $BSpill$. $\varphi$ denotes dependent variable of TFP level, while $\Delta \varphi$ denotes dependent variable for TFP growth. All estimations use Sector-Province-Island-Year Fixed-Effects.  Observations only consist of non-superstar firms. $^{***}$, $^{**}$, $^{*}$ denote $\alpha$ at 1\%, 5\%, and 10\%. $Abs$ denotes absorptive capacity. $HHI$ denotes market concentration from Herfindahl-Hirschman Index. $\varphi$ and $\Delta \varphi$ denote total factor productivity in level and growth, respectively. The results with Inverse Probability Weighting are reported in the Appendix in Table \ref{Main Results Backward with IPW}.
\end{table}

%% file: Robustness_Forward_-_IV.tex
\begin{table}[htbp]\centering
\def\sym#1{\ifmmode^{#1}\else\(^{#1}\)\fi}
\caption{Robustness Test: Alternative IV--Forward Spillovers}
\resizebox{\textwidth}{!}{
\begin{tabular}{l*{6}{c}}
\toprule
\toprule
                    &\multicolumn{1}{c}{(1)}&\multicolumn{1}{c}{(2)}&\multicolumn{1}{c}{(3)}&\multicolumn{1}{c}{(4)}&\multicolumn{1}{c}{(5)}&\multicolumn{1}{c}{(6)}\\
                    &\multicolumn{1}{c}{$\varphi$}&\multicolumn{1}{c}{$\Delta \varphi$}&\multicolumn{1}{c}{$\varphi$}&\multicolumn{1}{c}{$\Delta \varphi$}&\multicolumn{1}{c}{$\varphi$}&\multicolumn{1}{c}{$\Delta \varphi$}\\
\midrule
\textit{FSpill}     &       0.018\sym{***}&       0.014\sym{***}&       0.018\sym{***}&       0.014\sym{***}&       0.018\sym{***}&       0.015\sym{**} \\
                    &     (0.004)         &     (0.005)         &     (0.004)         &     (0.005)         &     (0.005)         &     (0.006)         \\
\addlinespace
\textit{Foreign-Owned}&       0.422\sym{***}&      -0.223\sym{***}&       0.422\sym{***}&      -0.223\sym{***}&       0.422\sym{***}&      -0.223\sym{***}\\
                    &     (0.023)         &     (0.032)         &     (0.023)         &     (0.032)         &     (0.023)         &     (0.032)         \\
\addlinespace
\textit{Exporters}  &       0.165\sym{***}&      -0.149\sym{***}&       0.165\sym{***}&      -0.149\sym{***}&       0.155         &      -0.104         \\
                    &     (0.012)         &     (0.017)         &     (0.012)         &     (0.018)         &     (0.228)         &     (0.370)         \\
\addlinespace
\textit{Imports}    &       0.380\sym{***}&      -0.056\sym{*}  &       0.380\sym{***}&      -0.056\sym{*}  &       0.381\sym{***}&      -0.058         \\
                    &     (0.025)         &     (0.033)         &     (0.025)         &     (0.033)         &     (0.027)         &     (0.037)         \\
\addlinespace
\textit{Abs}        &       0.510\sym{***}&       0.224\sym{***}&       0.510\sym{***}&       0.224\sym{***}&       0.510\sym{***}&       0.225\sym{***}\\
                    &     (0.005)         &     (0.005)         &     (0.005)         &     (0.005)         &     (0.006)         &     (0.007)         \\
\addlinespace
\textit{HHI}        &      -0.001         &       0.005         &      -0.040         &       0.009         &      -0.001         &       0.005         \\
                    &     (0.051)         &     (0.058)         &     (0.077)         &     (0.094)         &     (0.051)         &     (0.058)         \\
\addlinespace
\textit{FSpill $\times$ High-Concentration}&                     &                     &       0.001         &      -0.000         &                     &                     \\
                    &                     &                     &     (0.001)         &     (0.001)         &                     &                     \\
\addlinespace
\textit{FSpill $\times$ Exporters}   &                     &                     &                     &                     &       0.001         &      -0.003         \\
                    &                     &                     &                     &                     &     (0.014)         &     (0.022)         \\
\midrule
Sector-Province-Island-Year FE&         Yes         &         Yes         &         Yes         &         Yes         &         Yes         &         Yes         \\
Kleibergen-Paap Wald F-stat&     162.107         &     154.517         &      79.508         &      75.566         &       7.041         &       2.881         \\
Cragg-Donald Wald F-stat&     175.538         &     174.592         &      86.419         &      84.925         &      54.279         &      37.561         \\
Observations        &  214620         &  183074         &  214620         &  183074         &  214620.        &  183074         \\
\bottomrule
\bottomrule
\end{tabular}
}
\label{Robustness IV Forward}
\justifying \footnotesize \textbf{Note:} Clustered-robust standard errors are in parentheses and in firm level. Average Road Density is used as IV for $FSpill$. $\varphi$ denotes dependent variable of TFP level, while $\Delta \varphi$ denotes dependent variable for TFP growth. All estimations use Sector-Province-Island-Year Fixed-Effects. Observations only consist of non-superstar firms. $^{***}$, $^{**}$, $^{*}$ denote $\alpha$ at 1\%, 5\%, and 10\%. $Abs$ denotes absorptive capacity. $HHI$ denotes market concentration from Herfindahl-Hirschman Index. $\varphi$ and $\Delta \varphi$ denote total factor productivity in level and growth, respectively. The results with Inverse Probability Weighting are reported in the Appendix in Table \ref{Main Results Forward with IPW}.
\end{table}

%% file: Robustness_Heterogeneous_-_Simple_Productivity.tex
\begin{table}[htbp]\centering
\def\sym#1{\ifmmode^{#1}\else\(^{#1}\)\fi}
\caption{Robustness Test: Heterogeneous Superstar -- Simple Productivity (Value-added per Workers)}
\resizebox{\textwidth}{!}{
\begin{tabular}{l*{12}{c}}
\toprule
\toprule
                    &\multicolumn{1}{c}{(1)}&\multicolumn{1}{c}{(2)}&\multicolumn{1}{c}{(3)}&\multicolumn{1}{c}{(4)}&\multicolumn{1}{c}{(5)}&\multicolumn{1}{c}{(6)}&\multicolumn{1}{c}{(7)}&\multicolumn{1}{c}{(8)}&\multicolumn{1}{c}{(9)}&\multicolumn{1}{c}{(10)}&\multicolumn{1}{c}{(11)}&\multicolumn{1}{c}{(12)}\\
                    \midrule
                    & \multicolumn{2}{c}{Foreign} &  \multicolumn{2}{c}{Domestic} & \multicolumn{2}{c}{Foreign} &  \multicolumn{2}{c}{Domestic} & \multicolumn{2}{c}{Foreign} &  \multicolumn{2}{c}{Domestic}    \\
                    &\multicolumn{1}{c}{$\varphi$}&\multicolumn{1}{c}{$\Delta\varphi$}&\multicolumn{1}{c}{$\varphi$}&\multicolumn{1}{c}{$\Delta\varphi$}&\multicolumn{1}{c}{$\varphi$}&\multicolumn{1}{c}{$\Delta\varphi$}&\multicolumn{1}{c}{$\varphi$}&\multicolumn{1}{c}{$\Delta\varphi$}&\multicolumn{1}{c}{$\varphi$}&\multicolumn{1}{c}{$\Delta\varphi$}&\multicolumn{1}{c}{$\varphi$}&\multicolumn{1}{c}{$\Delta\varphi$}\\
\midrule
\textit{HSpill}     &      -0.001         &       0.009\sym{*}  &       0.020\sym{***}&       0.006\sym{**}       &                     &                     &                     &                     &                     &                     &                     &                     \\
                    &     (0.003)         &     (0.005)         &           (0.002)         &     (0.003)            &                     &                     &                     &                     &                     &                     &                     &                     \\
\addlinespace
\textit{BSpill}     &                     &                     &                     &                     &       0.021\sym{***}&       0.012\sym{**} &         0.109\sym{***}&       0.063\sym{***}&                    &                     &                     &                     \\
                    &                     &                     &                     &                     &     (0.005)         &     (0.006)         &             (0.012)         &     (0.015)         &                        &                     &                     &                     \\
\addlinespace
\textit{FSpill}     &                     &                     &                     &                     &                     &                     &                     &                     &       0.105\sym{***}&       0.064\sym{*}  &            0.126\sym{***}&       0.069\sym{***}          \\
                    &                     &                     &                     &                     &                     &                     &                     &                     &     (0.036)         &     (0.037)         &          (0.012)         &     (0.016)                \\
\addlinespace
\textit{Foreign-Owned}&       0.320\sym{***}&      -0.103\sym{***}&       0.319\sym{***}&      -0.093\sym{***}&       0.295\sym{***}&      -0.107\sym{***}&       0.259\sym{***}&      -0.127\sym{***}&       0.203\sym{***}&      -0.164\sym{***}&       0.346\sym{***}&      -0.075\sym{**} \\
                    &     (0.016)         &     (0.028)         &     (0.016)         &     (0.028)         &     (0.016)         &     (0.028)         &     (0.030)         &     (0.033)         &     (0.048)         &     (0.051)         &     (0.043)         &     (0.036)         \\
\addlinespace
\textit{Exporters}  &      -0.079\sym{***}&      -0.173\sym{***}&      -0.063\sym{***}&      -0.172\sym{***}&      -0.082\sym{***}&      -0.176\sym{***}&      -0.091\sym{***}&      -0.180\sym{***}&      -0.056\sym{***}&      -0.159\sym{***}&      -0.051\sym{**} &      -0.160\sym{***}\\
                    &     (0.010)         &     (0.016)         &     (0.010)         &     (0.016)         &     (0.010)         &     (0.016)         &     (0.019)         &     (0.018)         &     (0.017)         &     (0.020)         &     (0.024)         &     (0.020)         \\
\addlinespace
\textit{Imports}    &       0.415\sym{***}&      -0.181\sym{***}&       0.434\sym{***}&      -0.161\sym{***}&       0.373\sym{***}&      -0.180\sym{***}&       0.517\sym{***}&      -0.100\sym{***}&       0.268\sym{***}&      -0.246\sym{***}&       0.424\sym{***}&      -0.155\sym{***}\\
                    &     (0.018)         &     (0.030)         &     (0.017)         &     (0.029)         &     (0.019)         &     (0.030)         &     (0.035)         &     (0.036)         &     (0.057)         &     (0.058)         &     (0.046)         &     (0.038)         \\
\addlinespace
\textit{Abs}        &       0.630\sym{***}&       0.260\sym{***}&       0.635\sym{***}&       0.262\sym{***}&       0.628\sym{***}&       0.259\sym{***}&       0.619\sym{***}&       0.256\sym{***}&       0.620\sym{***}&       0.257\sym{***}&       0.631\sym{***}&       0.261\sym{***}\\
                    &     (0.004)         &     (0.005)         &     (0.004)         &     (0.005)         &     (0.004)         &     (0.005)         &     (0.007)         &     (0.006)         &     (0.006)         &     (0.006)         &     (0.009)         &     (0.007)         \\
\addlinespace
\textit{HHI}        &      -0.245\sym{***}&      -0.119\sym{**} &      -0.553\sym{***}&      -0.220\sym{***}&      -0.304\sym{***}&      -0.131\sym{***}&      -2.156\sym{***}&      -1.207\sym{***}&       0.045         &       0.096         &      -1.468\sym{***}&      -0.788\sym{***}\\
                    &     (0.040)         &     (0.046)         &     (0.054)         &     (0.065)         &     (0.042)         &     (0.047)         &     (0.228)         &     (0.263)         &     (0.120)         &     (0.130)         &     (0.148)         &     (0.169)         \\
\midrule
Sector-Province-Island-Year FE&         Yes         &         Yes         &         Yes         &         Yes         &         Yes         &         Yes         &         Yes         &         Yes         &         Yes         &         Yes         &         Yes         &         Yes         \\
Kleibergen-Paap Wald F-stat&     499.125         &     476.998         &     491.617         &     476.706         &     224.731         &     236.035         &     260.619         &     244.147         &      13.648         &      12.728         &     483.616         &     487.201         \\
Cragg-Donald Wald F-stat&    1575.424         &    1517.495         &    3922.154         &    3659.963         &     157.055         &     169.346         &      58.145         &      52.525         &      10.076         &       9.410         &      30.307         &      30.987         \\
Observations        &  299497         &  275658         &  299497         &  275658         &  302461         &  278074         &  302461         &  278074         &  302461         &  278074         &  302461         &  278074         \\
\bottomrule
\bottomrule
\end{tabular}
}
\label{Heterogeneous Superstar Simple}
\justifying \footnotesize \textbf{Note:} Clustered-robust standard errors are in parentheses and in firm level. Bartik-IV from labour-based and output growth ($LabBartikIV$) is used as IV for $HSpill$, while Bartik-IV from output-based and tariff ($TarrBartikIV$) is used as IV for $BSpill$ adn $FSpill$. $Foreign$ denotes foreign superstar spillovers, while $Domestic$ denotes domestic foreign spillovers. Both foreign and domestic superstars are non-exporters. $\varphi$ denotes dependent variable of labour productivity level (in log), while $\Delta \varphi$ denotes dependent variable for labour productivity growth. All estimations use Sector-Province-Island-Year Fixed-Effects.  Observations only consist of non-superstar firms. $^{***}$, $^{**}$, $^{*}$ denote $\alpha$ at 1\%, 5\%, and 10\%. $Abs$ denotes absorptive capacity. $HHI$ denotes market concentration from Herfindahl-Hirschman Index. $\varphi$ and $\Delta \varphi$ denote total factor productivity in level and growth, respectively.
\end{table}

%% file: DOPD_Robustness_Amiti.tex
\begin{table}[htpb]\centering
\caption{TFP Change from Dynamic OP Decomposition: Survivors, Exiters, and Entrants (Robustness Test)}
\resizebox{\textwidth}{!}{
\begin{tabular}{llcccc}
\toprule\toprule
\multirow{2}{*}{\textbf{Component}} & \multirow{2}{*}{\textbf{Period}} & \multirow{1}{*}{\textbf{Superstar}} &  \\
\cmidrule(lr){3-5}
   &  & \textbf{Foreign-Owned} & \textbf{Exporter} & \textbf{Large Firm} & \textbf{Non-Superstar}  \\
\midrule
\multirow{3}{*}{Plant Improvements}
& 2001--2015 &1.025	&0.847	&0.930&	1.190 \\
& 2001--2010 &0.664	& 0.551	&0.520	&0.598 \\
& 2001--2005 &0.182	&0.138&	0.167	&0.260 \\
\midrule
\multirow{3}{*}{Reallocation within Survivors}
& 2001--2015 &-0.301	&-0.787&	-0.496&	-0.619 \\
& 2001--2010 &	0.313&	-0.947	&0.520&	-0.497 \\
& 2001--2005 &	-0.014&	-0.634&	0.079	&-0.494 \\
\midrule
\multirow{3}{*}{Reallocation: Exiters--Entrants}
& 2001--2015 &0.527&	0.056	&0.354	&0.503 \\
& 2001--2010 &	-0.658	&0.975&	0.067&	0.111 \\
& 2001--2005 &0.053	&0.103	&-0.032	&0.208 \\
\bottomrule\bottomrule
\end{tabular}
}
\\
\justifying \small
\textbf{Note:} Plant improvement refers to the change of unweighted average productivity for survivors ($\Delta\bar{\varphi_t}$), while Reallocation within Survivors refers to the change of covariance component within survivors, and Reallocation of Exiters-Entrants is the plant improvements and covariance component across group of survivors with entrants and exiters. Robustness test is based on the definition of superstar firms from \citet{amiti2024fdi}.
\label{Robustness DOPD}
\end{table}